\documentclass[11pt]{article}
\usepackage{amsmath, amssymb}
\usepackage{slashed}
\usepackage{verbatim}
\usepackage{latexsym}
\usepackage{euscript}
\usepackage{amsfonts}
\usepackage{verbatim}
\usepackage[]{array}
\usepackage[]{mathrsfs}
\usepackage[]{tensor}
\usepackage{graphicx}
\usepackage{amsmath}
\usepackage{amssymb}
\usepackage{amsxtra}
\usepackage{color}
\def\be{\begin{eqnarray}}
\def\ee{\end{eqnarray}}
\def\0{\nonumber}
\def\d{\partial}
\def\tr{{\rm tr}} 

\def\al{\alpha}
\def\vx{\stackrel{\rightarrow}{x}}

\usepackage{slashed}
\usepackage{verbatim}
\usepackage{latexsym}\usepackage{color}
\usepackage{euscript}
\newcommand\ET{\EuScript{T}}

\newcommand\EA{\EuScript{A}}
\newcommand\EC{\EuScript{C}}
\newcommand\EP{\EuScript{P}}\normalfont\large
\begin{document}
\begin{flushright}
SISSA/10/2017/FISI\\{ZTF-EP-17-02}\\
arXiv:1703.10473
\end{flushright}
\vskip 2cm
\begin{center}

{\LARGE {\bf Axial gravity, massless fermions and trace anomalies  }}

\vskip 1cm

{\large  L.~Bonora$^{a}$, M.~Cvitan$^{b}$, P.~Dominis
Prester$^{c}$, A.~Duarte Pereira$^{d,e}$, S.~Giaccari$^{b}$,   T.~$\rm
\check{S}$temberga$^{b}$
\\\textit{${}^{a}$ International School for Advanced Studies (SISSA),\\Via
Bonomea 265, 34136 Trieste, Italy, and INFN, Sezione di
Trieste\\and KEK Theory Center, KEK, Tsukuba, Japan\\ }%
\textit{${}^{b}$ Department of Physics, Faculty of Science, University of
Zagreb, Bijeni\v{c}ka cesta 32,
10000 Zagreb, Croatia \\}%
\textit{${}^{c}$ Department of Physics, University of Rijeka,\\
Radmile Matej\v{c}i\'{c} 2, 51000 Rijeka, Croatia\\}%
\textit{${}^{d}$UERJ - Universidade Estadual do Rio de Janeiro, Departamento de
F\'isica Te\'orica\\
Rua S$\tilde{\rm a}$o Francisco Xavier 524, 20550-013, Maracan$\tilde{\rm
a}$,Rio de Janeiro, RJ, Brasil\\}
\textit{${}^e$UFF - Universidade Federal Fluminense, Instituto de F\'{\i}sica,\\
Campus da Praia Vermelha, Avenida General Milton Tavares de Souza s/n,\\ 
24210-346, Niter\'oi, RJ, Brasil } }
\vskip 1cm

\end{center}

 \vskip 1cm {\bf Abstract.} This article deals with two main topics. One is odd
parity trace anomalies in
Weyl fermion theories 
in a 4d curved background, the second is the introduction of axial gravity.
The motivation for reconsidering the former is to clarify the theoretical 
background underlying the approach and complete the calculation of the anomaly.
The reference is in particular to the difference between Weyl and massless
Majorana fermions
and to the possible contributions from tadpole and seagull terms in the Feynman
diagram approach. A first, basic, result of this paper is that a more thorough 
treatment, taking account of such additional terms { and using dimensional
regularization}, confirms the earlier result.
The introduction of an axial symmetric tensor besides the usual gravitational
metric
is instrumental to a different derivation of the same result using Dirac
fermions, which are 
coupled not only to the usual metric but also to the additional axial tensor. 
The action of Majorana and Weyl fermions
can be obtained in two different limits of such a general configuration. The
results obtained in this way confirm the previously obtained ones.
\vskip 1cm 
\begin{center}

{\tt Email: bonora@sissa.it,mcvitan@phy.hr,
pprester@phy.uniri.hr,duarte763@gmail.com,sgiaccari@phy.hr,tstember@phy.hr}
\end{center}

\eject

\section{Introduction}\label{s:intro}

This article deals with two main topics. One is odd parity trace anomalies in
chiral fermion theories 
in a 4d curved background, the second is the introduction of axial gravity. The
first 
subject has been already treated in \cite{BGL,BDL}. The second, to our best
knowledge, is new.
The motivation for reconsidering the former is to clarify the theoretical 
background underlying the approach and complete the calculation of the anomaly, 
  {also in view of
more recent results, \cite{Bast}}. For
some aspects of the calculations in \cite{BGL,BDL} were left implicit. We refer
to the possible contributions from tadpole and seagull terms in the Feynman
diagram approach used there. Here we treat them explicitly. 
{ In this paper we use dimensional regularization, deferring to another paper
the
discussion of other regularizations.} A first, basic,
result of this paper is that a more thorough treatment, taking account of such
additional terms, confirms the result of \cite{BGL}.

The second topic is motivated as follows.
It is well known that in anomaly calculations the functional integral measure
plays a basic role. 
In the case of chiral fermions the definition of such a measure is a
long-standing and unsolved problem. 
One can bypass it by using Feynman diagram techniques, where the fermion path
integral measure does not play a direct role. However there is a way to carry
out the same calculation on a theory of Dirac fermions, so that
no fastidious objections can be raised about the fermion functional integral
measure. Here is where the axial metric intervenes.
The idea is to follow the method used in \cite{Bardeen} for chiral gauge
anomalies. It is possible to compute 
covariant and consistent anomalies in a unique model by coupling Dirac fermions
also to an axial potential $A$, beside the familiar vector potential $V$. The
anomalies one obtains in this way satisfy the Wess-Zumino consistency
conditions, but depend on 
two potentials. The covariant anomaly for Dirac fermions coupled to $V$ alone
are obtained by simply setting $A=0$. The consistent 
anomaly of chiral fermions coupled to $V$ are obtained by taking the limit $V\to
V/2, A\to V/2$. Transposing this technique
to the problem of trace anomalies for chiral fermions, requires the introduction
of an axial tensor $f_{\mu\nu}$, which with some abuse of language we call
metric too, besides the usual metric $g_{\mu\nu}$. This second tensor is called
axial because it couples axially to Dirac fermions. 
The second important result of our paper is that we succeed in introducing this
bimetric system, and through it we are able 
to derive the trace anomalies for Dirac, Majorana and Weyl fermions as
particular cases of the general case. Using again dimensional
regularization, 
we obtain in this way a confirmation of the previous, together with new,
results. {We suggest also an explanation for the claimed disagreement with ref.\cite{Bast}}.

The calculations presented here have a more general motivation, stemming from a
more basic question concerning 
massless fermions. More precisely the question we would like to be
able to answer is: is there at present a consistent field  theory of massless
fermions in a curved background? A massless Dirac fermion is not a good
candidate in this sense,
because it admits a mass term that can arise from renormalization, even if
it is not initially present in the action. So the choice must be restricted to
Weyl and Majorana. Also a Majorana fermion can have mass, but if its bare mass
is zero, a (rigid) chiral symmetry could in principle protect this vanishing
mass. However this symmetry is anomalous on a curved background, due to the
Kimura-Delbourgo-Salam anomaly,
\cite{Kimura,DS}. A Weyl fermion is certainly massless and no bare mass term
exists that can threaten this property.
The odd parity trace-anomaly found in \cite{BGL} is a new and perhaps useful
aspect \cite{Shapiro} as long as we consider the theory an effective one.
However, unitarity of the theory might be imperiled in a fully quantized gravity
theory interacting with chiral fermions. 
Perhaps some innovative theory may be necessary in order to describe a truly
massless fermion on a curved background. 

Given the importance of this theme, we intend to return to the analysis of
the odd parity trace anomaly in the presence of a gravitational background by
means of other methods 
and other regularizations, {which we believe will confirm
the results obtained with the dimensional regularization.}  

The paper is organized as follows. In section 2 we review the properties of
massless Weyl and Majorana fermions in 4d. In section 3 the anomaly derivation
of \cite {BGL} is reviewed and integrated with the calculation of the relevant
tadpoles and seagull terms. In section 4 such a revisiting is completed with
the evaluation of the Ward identity for diffeomorphisms. Section 5 contains an
additional discussion on the the odd trace anomaly. In section 6 we
introduce the formalism for a MAT (metric-axial-tensor) gravity, and in section
7 we couple it to Dirac fermions. Then we present a simplified derivation of the
trace anomalies in such a model and, then, we compute in detail the collapsing
limit, which allows us to calculate the trace anomalies in an ordinary gravity
background for Dirac, Weyl and Majorana fermions. Section 8 is devoted to a
justification of the simplifications of the previous section. 
Section 9 is the conclusion. The evaluation of the triangle diagram 
for odd trace anomaly is shown in Appendix \ref{s:trianglediagram}
The derivation of Feynman rules in an ordinary gravity and MAT
background, together with the relevant Ward identities are collected in 
Appendix \ref{s:feynmanrules}. The most encumbering diagram calculations 
can be found in Appendix \ref{s:samples}.

\vskip 1cm

{\bf Notation.} We use a metric $g_{\mu\nu}$ with mostly - signature. The gamma
matrices satisfy $\{\gamma^\mu,\gamma^\nu\}= 2 g^{\mu\nu}$ and
\be
\gamma_{\mu}^\dagger= \gamma_0\gamma_\mu\gamma_0\0
\ee
The generators of the Lorentz group are $\Sigma_{\mu\nu} =\frac 14
[\gamma_\mu,\gamma_\nu]$.
The charge conjugation operator $C$ is defined to satisfy
\be
\gamma_{\mu}^T = -C^{-1} \gamma_\mu C, \quad\quad CC^*=-1, \quad\quad
CC^\dagger=1\label{C}
\ee 
The chiral matrix $\gamma_5=i\gamma^0\gamma^1\gamma^2\gamma^3$ has the
properties
\be
\gamma_5^\dagger=\gamma_5, \quad\quad (\gamma_5)^2=1, \quad\quad C^{-1} \gamma_5
C= \gamma_5^T\0
\ee

\section{Dirac, Majorana and Weyl fermions in 4d. }
\label{s:MW}

{We would like to devote this section to a discussion of the
statement that 
a massless Majorana fermion is the same as a Weyl fermion. The reason is that,
if it is true at both classical and quantum level, there is no chance
for an odd parity trace anomaly to exist and no motivation for this paper. On
the other hand this statement is not undisputed. As one can easily experience,
there is no well defined or generally accepted doctrine about the properties of the quantum theories
of massless Majorana and Weyl fermions. 
Our aim here is to examine various aspects of the problem and bring to light
all the classical and quantum differences between the two types of fermions.  We would like to
convince the reader that there is no a priori uncontroversial evidence that the 
statement is true, and therefore
it is prudent to leave the last word to explicit computations, such as the one
for odd parity trace anomaly.}  

We start from a few basic facts about fermions in 4d. Let us start from a
4-component Dirac fermion $\psi$.
Under Lorentz it transforms as
\be
\psi(x) \rightarrow \psi'(x')=\exp\left[-\frac12
\lambda^{\mu\nu}\Sigma_{\mu\nu}\right]\psi(x)\,,\label{Lorentzpsi}
\ee
for $x'^{\mu}=(e^\lambda)^\mu{}_\nu \, x^\nu$, where $\Sigma_{\mu\nu} =\frac 14
[\gamma_\mu,\gamma_\nu]$. 
The Lagrangian for a free Dirac field is well-known:
\be
i \bar\psi \gamma^\mu \partial_\mu \psi\label{freeDirac}
\ee
What is often forgotten is that, like for the kinetic term of any field theory,
it can be constructed
because, in the spinor space, there exist a Lorentz invariant scalar product
$(\Psi_1,\Psi_2)= \langle \Psi_1^\dagger |\gamma^0| \Psi_2\rangle$. So that
(\ref{freeDirac}) can also be written as
\be
i(\psi, \gamma\! \cdot \!\partial \psi) \label{freeDirac'}
\ee
A Dirac fermion admits a Lorentz invariant mass term $m\bar\psi \psi=
m(\psi,\psi)$.

A Dirac fermion can be seen as the sum of two Weyl fermions
\be
\psi_L = P_L \psi, \quad\quad \psi_R = P_R \psi, \quad\quad {\rm where}
\quad\quad P_L=\frac {1+\gamma_5}2,
\quad\quad  P_R=\frac {1-\gamma_5}2\0
\ee
with opposite chiralities
\be
\gamma_5 \psi_L= \psi_L, \quad\quad \gamma_5 \psi_R= -\psi_R.\0
\ee 
A left-handed Weyl fermion admits a Lagrangian kinetic term
\be
i(\psi_L, \gamma\! \cdot \!\partial \psi_L)=i \overline\psi_L\gamma^\mu
\partial_\mu \psi_L \label{freeWeyl}	 
\ee
but not a mass term, because $(\psi_L,\psi_L)=0$, since
$\gamma_5\gamma^0+\gamma^0\gamma_5=0$. So a Weyl fermion is massless and this
property is protected by its being chiral.

In order to introduce Majorana fermions we need the notion of {\it Lorentz
covariant conjugate} spinor, 
$\hat\psi$:
\be
\hat \psi =  \gamma_0 C \psi^*\label{Lorentzconj}
\ee
It is not hard to show that if $\psi$ transforms like (\ref{Lorentzpsi}), then
\be
\hat\psi(x) \rightarrow \hat\psi'(x')=\exp\left[-\frac12
\lambda^{\mu\nu}\Sigma_{\mu\nu}\right]\hat\psi(x)\,,\label{Lorentzpsihat}
\ee
Therefore it makes sense to impose on $\psi$ the condition 
\be
\psi=\hat\psi\label{Majo}
\ee
because both sides transform in the same way. A spinor satisfying (\ref{Majo})
is, by definition, a Majorana spinor. 

A Majorana spinor admits both kinetic and mass term, which can be written as
$\frac 12\times$ those of a Dirac spinor. 

In terms of Lorentz group representations we can summarize the situation as
follows. $\gamma_5$ commutes with 
Lorentz transformations $\exp\left[-\frac12
\lambda^{\mu\nu}\Sigma_{\mu\nu}\right]$. So do $P_L$ and $P_R$. This means that
the Dirac representation is reducible and multiplying the spinors by  $P_L$ and
$P_R$ identifies irreducible representations, the Weyl ones. To be more precise,
the
Weyl representations
are irreducible representations of the group $SL(2,C)$, which is the covering
group of the {\it proper ortochronous} Lorentz group. They are usually denoted
$(\frac 12,0)$ and $(0,\frac 12)$ in the $SU(2)\times SU(2)$ labeling of the
$SL(2,C)$ irreps. As we have seen in
(\ref{Lorentzpsihat}), Lorentz transformations commute also with the charge
conjugation operation
\be
\EC \psi \EC^{-1}= \eta_C \gamma_0 C \psi^*\label{CC}
\ee
where $\eta_C$ is a phase which, for simplicity, we set equal to 1. This also
says that Dirac spinors are reducible and suggests another way to reduce them:
by imposing (\ref{Majo}) we single out another irreducible representation, the
Majorana one. The Majorana representation is the minimal irreducible
representation of a
(one out of eight) covering of the {\it complete} Lorentz group,
\cite{Racah,Cornwell}. It is evident, and well-known, that Majorana and Weyl
representations are incompatible (in 4d)\footnote{One can express the components
of Majorana fermion as linear combinations of those of a Weyl fermion and
viceversa, see for instance \cite{Pal}. However this does not respect the irrep
decomposition.}. 

Let us recall the properties of a Weyl fermion under charge conjugation and
parity. We have
\be
\EC \psi_L \EC^{-1} =P_L  \EC \psi\EC^{-1} = P_L\hat\psi= 
\hat\psi_L\label{CpsiL}
\ee
The charge conjugate of a Majorana field is itself, by
definition. While the action of a Majorana field is
invariant under charge conjugation, the action of a Weyl fermion is, so to say,
maximally non-invariant, for
\be
{\EC}\left(\int i \overline{\psi_L }\gamma^\mu \partial_\mu
\psi_L\right)  {\EC}^{-1}
= \int i \overline{\hat\psi_L} \gamma^{\mu\dagger} \partial_\mu 
\hat\psi_L=\int i \overline{\psi_R }\gamma^\mu \partial_\mu
\psi_R\label{CactionLH}
\ee

The parity operation is defined by
\be
\EP \psi_L (t, \vx)\EP^{-1} =\eta_P \gamma_0 \psi_R (t,
-\stackrel{\rightarrow}{x})\label{parityop}
\ee
where $\eta_P$ is a phase. In terms of the action we have
\be
\EP\left(\int\overline\psi_L\gamma^\mu \partial_\mu \psi_L  \right) \EP^{-1}=
\int\overline\psi_R\gamma^\mu \partial_\mu \psi_R,\label{parityaction}
\ee
while for a Majorana fermion the action is invariant under parity.

If we consider CP, the action of a Majorana fermion is obviously invariant under
it.
For a Weyl fermion we have
\be
{\EC}{\EP}\psi_L(t,\vx) ( {\EC}{\EP})^{-1} =\gamma_0 \widehat
\psi_L(t,-\vx)= \gamma_0 P_R \hat\psi(t,-\vx)= \gamma_0 \hat\psi_R
(t,-\vx)\label{CP}
\ee  
Applying CP to the Weyl action one gets
\be
{\EC}{\EP}\left(\int i \overline{\psi_L }\gamma^\mu \partial_\mu
\psi_L\right)  ( {\EC}{\EP})^{-1}
= \int i \overline{\hat\psi_R} (t,-\vx) \gamma^{\mu\dagger} \partial_\mu 
\hat\psi_R(t,-\vx)=
\int i \overline{\hat\psi_R} (t,\vx) \gamma^{\mu} \partial_\mu 
\hat\psi_R(t,\vx)
\label{CPactionLH}
\ee
But one can easily prove that
\be
\int i \overline{\hat\psi_R} (t,\vx) \gamma^{\mu} \partial_\mu 
\hat\psi_R(t,\vx)=
\int i \overline {\psi_L}  (x) \gamma^{\mu} \partial_\mu 
\psi_L(x)\label{CPinvariance}
\ee
Therefore the action for a Weyl fermion is CP invariant. It is also, separately,
T invariant, and, so, CPT invariant. 

Now let us go to the quantum interpretation of the field $\psi_L$. It has the
plane wave expansion
\be
\psi_L (x) = \int dp\, \left( a(p) u_L(p) e^{-ipx} + b^\dagger(p) v_L(p) 
e^{ipx} \right) \label{PWexpansion}
\ee
where $u_L,v_L$ are fixed and independent left-handed spinors (there are only
two of them). The interpretation
is: $b^\dagger(p)$ creates a left-handed
particle while $a(p)$ destroys a left-handed particle with negative helicity
(because of the opposite momentum). However 
eqs.(\ref{CP},\ref{CPactionLH}) force us to identify
the latter with a right-handed antiparticle: C maps particles to antiparticles,
while P invert helicities, 
so CP maps left-handed particles to right-handed antiparticles. It goes without
saying that no
right-handed particles or left-handed antiparticles enter the game.

\vskip 1cm

{\bf Remark.} A mass term $\bar \psi \psi$ for a Dirac spinor can be also
rewritten by projecting the latter into its chiral components
\be
\bar\psi \psi= \overline{\psi_L} \psi_R+ \overline {\psi_R}\psi_L
\label{masschiral}
\ee
If $\psi$ is a Majorana spinor this can be written  
\be
\bar\psi \hat\psi= \overline{\hat\psi_L} \psi_R+ \overline {\psi_R}\hat\psi_L
,\label{massMajo}
\ee
which is therefore well defined and Lorentz invariant by construction. Now,
using the Lorentz covariant conjugate we can rewrite (\ref{massMajo}) as
\be
(\psi_L)^T C^{-1} \psi_L + \psi_L^\dagger C(\psi_L)^*,\label{massMajoL}
\ee
which is expressed only in terms of $\psi_L$. (\ref{massMajoL}) may create the
illusion that there exists a mass term also for Weyl fermions. But this is not
the case. If we add this term to the kinetic term (\ref{freeWeyl}) we obtain an
action whose equation of motion is not Lorentz covariant: the kinetic and mass
term in the equation of motion belong to two different representations. To be
more explicit, a massive Dirac 
equation of motion for a Weyl fermion would be
\be
i \gamma^\mu \partial_\mu \psi_L -m \psi_L=0,\label{wrong}
\ee
but this equation breaks Lorentz covariance because the first piece transforms
according to a right-handed representation while the second according to a
left-handed one,
and is not Lagrangian\footnote{{Instead of the second term in
the LHS of \eqref{wrong} one could use $m C \overline{\psi_L}^T$, which has the
right Lorentz 
properties, but the corresponding Lagrangian term would not be self-adjoint and
one would be forced to introduce the adjoint term and end up again with (\ref{massMajoL})
. This implies, in particular, that there does not exist such a
thing as a ``massive Weyl propagator'', that is a massive propagator involving only one chirality, which, in particular, renders the use of the Pauli-Villars regularization problematic. Sometimes a Dirac or Majorana propagator is used in its place. A minimal precaution, in such a case, would be to check the results obtained with this regularization by comparing them with those obtained with others. }},.
The reason is, of course, that (\ref{massMajoL}) is not expressible in the same
canonical form as (\ref{freeWeyl}). This structure is clearly visible in the
four component formalism used
so far, much less recognizable in the two-component formalism.

\vskip 1cm

The fact that a massive Majorana fermion and a Weyl fermion are
different objects is, {in our opinion,}  uncontroversial.
The question whether a massless Majorana fermion is
or is not the same as a Weyl fermion {at both classical and quantum level}, as we pointed out above, is not so clearly
established. Let us
consider the case in which there is no quantum number appended to the fermions.
The reason why they are sometimes considered as a unique object is due, we
think, to the fact that we can establish a one-to-one correspondence between the
components of a Weyl spinor and those of a Majorana spinor
in such a way that the Lagrangian (particularly in two-component notation) looks
the same. In
fact this {is not decisive}, as we will see in a moment. But let us mention
first the evident differences between the two. The first, and most obvious, is
the one we have already mentioned: they belong to two different representations
of the Lorentz group, irreducible to each other (it should be standard lore
that in 4d there cannot exist a spinor that is simultaneously Majorana and
Weyl).
Another macroscopic difference is that the helicity of a Weyl fermion is well
defined and corresponds to its chirality, while the chirality of a Majorana
fermion is undefined, so that the relation with its helicity is also undefined.
Next, a parity operation maps the Majorana action into itself, while it maps the
Weyl action (\ref{freeWeyl}) into the same action for the opposite chirality.
Same for the charge conjugation operator. Finally, going to the quantum
theories, the fermion path integral measures in the two cases are different.
This is the
crucial point as far as the matter discussed in this paper is concerned, i.e.
anomalies. Let us expand on it. The path integral of a free Dirac fermion
(\ref{freeDirac}) is interpreted as the determinant of the massless Dirac
operator ${\slashed D}=i{\slashed \partial+\slashed{V}}$ (where $V$ denotes any
potential), i.e. the (suitably regularized) product of its eigenvalues. A
similar interpretation holds for a massless Majorana fermion, while for a Weyl
fermion it is not so straightforward. Since the Dirac operator anticommutes with
$\gamma_5$, it maps a left-handed spinor to a right-handed one. Therefore the
eigenvalue problem is not even defined for $ {\slashed D}_L = {\slashed D} P_L$.
We may replace the looked for $\det {\slashed D}_L$ with $\left(\det\left(
{\slashed D}^\dagger_L{\slashed D}_L\right)\right)^{\frac 12}$. But in this case
we have to face the problem of an
undetermined overall phase factor. This impasse has been known for a long
time \footnote{It is well-known that in particular this prevents using the
Fujikawa method for chiral theories, because the latter, at least in its original form,
holds when in the theory both
chiralities are present. {This problem
has been discussed in detail in \cite{AB2}, with explicit examples: it is shown
there that 
the original Fujikawa method cannot reproduce the non-Abelian consistent chiral
anomalies, 
but only the covariant ones in chirally symmetric theories.
It follows that one cannot expect to be able to reproduce the odd parity trace
anomaly in a 
left-handed theory, because the latter belongs to the same class as the
non-Abelian consistent 
chiral anomalies, that is the class of anomalies having opposite sign for
opposite chirality.} 
This remark applies to \cite{Bast}, which, following the method of Fujikawa and
using Pauli-Villars regularization,
obtains a vanishing odd trace anomaly and seems to contradict our result below.
Although we intend
to return more punctually on this issue, let us point out for the time being
that using 
a Dirac fermion path integration measure amounts to introducing in the game both
chiralities,
even though formally the action is declared to be the Weyl one. We have seen
that the classical action can take
various forms, but for this anomaly what matters is that only one chirality is
involved through all the steps, including the path integration measure. 
Therefore, we believe, the result of \cite{Bast} applies to Dirac and Majorana
fermions and is in fact consistent with ours.}.  
A few ways have been devised to overcome it. One is to use a
perturbative approach, via the Feynman diagram technique, in a chiral fermion
theory, wherein Lorentz covariance is taken into account via the chiral
vertices. This is the method used in \cite{BGL,BDL}. We will revisit it below.
Later on (in Part II) we will consider another approach, based on Dirac
fermions, \cite{Bardeen,AB1,AB2},
(i.e. with the ordinary Dirac path integral measure), whereby the chiral
fermion theory is recovered as a limiting case. Finally, although we do not use
it here, we should mention the method recently devised in \cite{Grabowska:2016},
where a fifth dimension is introduced as a regulator.

We think  the above arguments are more than enough to conclude that massless
Majorana and Weyl fermions, notwithstanding some similarities, {may really be} 
different objects. However, to conclude, it is worth trying to 
counter a common misconception that comes from what we said above:  we can
establish a one-to-one correspondence between the components of a Weyl spinor
and those of a Majorana spinor
in such a way that the Lagrangians in two-component notation look the same. If,
for instance,
in the chiral representation
we represent $\psi_L$ as $\left(\begin{matrix} \omega\\ 0\end{matrix}\right)$,
where $\omega$ is a two component spinor, then (\ref{freeWeyl}) above becomes
\be
i \omega^\dagger \bar \sigma^\mu \partial_\mu \omega\label{MjoranaWeyl}
\ee
which has the same form as a massless Majorana action. Now, if the action is the
same for both Weyl and Majorana, how can there be differences? This {(problematic)}
syllogism may cause gross misunderstandings. 
Well, first,  in
general, the action of a physical system does not contain all the information
concerning the system, there being specifications that have to be added
separately. Second, even though numerically the actions
coincide, the way the actions respond
to a variation of the Weyl and Majorana field is different. One leads to the
Weyl equation of motion, the other to the Majorana one. The delicate point is
precisely this: when we take the variation of an action 
with respect to a field in order to extract the equations of motion, we have to
make sure that the variations
do not break the symmetries or the properties we wish to be present in the
equations of motion. In general, we do this automatically, without thinking of
it\footnote{For instance, in gravity theories, the metric variation $\delta
g_{\mu\nu}$ is generic while not ceasing to be a symmetric tensor.}. But in this
case more than the normal care has to be used. If we wish the eom to preserve
chirality we must use variations that preserve chirality, i.e. must be
eigenfunctions of $\gamma_5$. If instead we wish the eom to transform in the
Majorana representation we have to use variations that transform suitably, i.e.
must be eigenfunctions of the charge conjugation operation. If we do so we
o
btain two different results, which are irreducible to each other, no matter
what action we use\footnote{It is clear that, eventually, all the components,
both of a Weyl and a massless Majorana fermion, satisfy the massless
Klein-Gordon equation. But this is not a qualifying property in this context,
otherwise, for instance, any two (anticommuting) complex massless scalar fields
would be the
same as a Weyl fermion, and any four real massless scalar fields would be the
same as a massless Majorana fermion.}. Third, and most important, as already
pointed out, in the quantum theory
a crucial role is played by the functional measure, which {is very likely to be} different for Weyl
and Majorana fermions.
 
Concluding this introductory discussion, we think the identification of a Weyl
fermion with a massless Majorana one {should not be taken as
granted as sometimes stated in the literature}. It is prudent to avoid a priori conclusions, but
rather develop both hypotheses (not only one) and compare the end results.
This said, it is important to find out properties that differentiate Weyl and
massless Majorana fermions. In this paper, following \cite{BGL,BDL}, we show
that one such property is the parity odd Weyl anomaly. The latter is 0 for a
massless Majorana fermion, while it equals the Pontryagin density for a Weyl
fermion (the even parity trace anomaly is the same for both).

\vskip 5cm

{\Large Part I}
\vskip 1cm
\section{Odd parity trace anomaly in chiral theories}
\label{s:2}

In this section we reconsider the calculation of the odd trace anomaly in
\cite{BGL}(for an introduction to anomalies see
\cite{Bertlmann,Fujikawa,BastVan}). 
The motivation for this is that in \cite{BGL}, as well as in
\cite{BDL}, tadpoles and seagull diagrams were disregarded. In ordinary 
(non-chiral) theories coupled to gravity such diagrams contribute local 
terms to the effective action, and help restoring conservation, 
which otherwise would be violated by local terms, \cite{Parker}. As we shall see
below, these diagrams are instead ineffective for the parity odd diagrams 
in a chiral theory, and do not change the final result. However a complete
treatment
demands that they should be taken into account and evaluated.

The model we considered in \cite{BGL} was a left-handed Weyl spinor coupled to
external gravity in 4d. The action is
\be
S= \int d^4x \, \sqrt{|g|} \, i\overline {\psi_L} \gamma^m\left(\nabla_m +\frac
12
\omega_m \right)\psi_L \label{action}
\ee
where $\gamma^m = e^m_a \gamma^a$, $\nabla$ ($m,n,...$ are world indices,
$a,b,...$ are flat indices) is the covariant derivative with respect to the
world indices and $\omega_m$ is the spin connection:
\be
\omega_m= \omega_m^{ab} \Sigma_{ab}\0
\ee
where $\Sigma_{ab} = \frac 14 [\gamma_a,\gamma_b]$ are the Lorentz generators.
Finally
$\psi_L= \frac {1+\gamma_5}2 \psi$. Classically the energy-momentum tensor 
\be
T^{\mu\nu}= -\frac i4 \overline {\psi_L} \gamma^\mu
{\stackrel{\leftrightarrow}{\nabla^\nu}}\psi_L+(\mu\leftrightarrow\nu)
\label{emt}
\ee
is both conserved on shell and traceless.

From (\ref{action}) we extracted the (simplified) Feynman rules as follows.
The action (\ref{action}) can be written as
\be 
S= \int d^4x \, \sqrt{|g|} \,\left[ \frac i2\overline {\psi_L} \gamma^\mu
{\stackrel{\leftrightarrow}{\d}}_\mu  \psi_L -\frac 14\epsilon^{\mu a b c}
\omega_{\mu a b} \overline{\psi_L} \gamma_c \gamma_5\psi_L\right]
\label{action1}
\ee
where it is understood that the derivative applies to $\psi_L$ and $\overline
{\psi_L}$ only. We have used the relation $\{\gamma^a, \Sigma^{bc}\}=i \,
\epsilon^{abcd}\gamma_d\gamma_5$.

Expanding
\be
e_\mu^a = \delta_\mu^a +\chi_\mu^a+ ...,\quad\quad e_a^\mu = \delta _a^\mu +\hat
\chi_a^\mu +...,\quad\quad {\rm and}\quad
{g_{\mu\nu}=\eta_{\mu\nu}+h_{\mu\nu}}\label{hmunu}
\ee
and inserting these expansions in the defining relations $e^a_\mu
e^\mu_b=\delta_b^a,\quad g_{\mu\nu}= e_\mu^a e_\nu^b \eta_{ab}$,
one finds
\be
\hat \chi_\nu^\mu =- \chi_\nu^\mu\quad \quad {\rm and}\quad\quad
h_{\mu\nu}=2\,\chi_{\mu\nu}.\label{hatchichi}
\ee

Expanding accordingly the spin connection
\be
\omega_{\mu ab}=e_{\nu a}( \d_\mu e_b^\nu + e^\sigma{}_b
\Gamma_\sigma{}^\nu{}_\mu),\quad\quad
\Gamma_\sigma{}^\nu{}_\mu= \frac 12 \eta^{\nu\lambda}(\d_\sigma
h_{\lambda\mu}+\d_\mu h_{\lambda\sigma}-\d_\lambda
h_{\sigma\mu})+...\0
\ee
after some algebra one gets
\be 
\omega_{\mu a b}\, \epsilon^{\mu a b c}= - \frac 14\epsilon^{\mu a b c}\, 
\d_\mu h_{a\lambda}\,h_b^\lambda+...\label{omega}
\ee

Therefore, up to second order the action can be written (by incorporating
$\sqrt{|g|}$ in a redefinition of the $\psi$ field)
\be 
S\approx \int d^4x \,  \left[\frac i2 (\delta^\mu_a -\frac 12 h^\mu_a )
\overline {\psi_L} \gamma^a {\stackrel{\leftrightarrow}{\d}}_\mu  \psi_L +\frac
1{16}\epsilon^{\mu a b c}\,  \d_\mu h_{a\lambda}\,h_b^\lambda\,  \bar\psi_L
\gamma_c \gamma_5\psi_L\right]\0
\ee
The free action is 
\be
S_{free}=  \int d^4x \,\frac i2 \overline {\psi_L} \gamma^a
{\stackrel{\leftrightarrow}{\d}}_a  \psi_L\label{free}
\ee
and the lowest interaction terms are
\be
S_{int} &=& \int d^4x \left[ -\frac i4 h^\mu_a\,\overline {\psi_L}\gamma^a
{\stackrel{\leftrightarrow}{\d}}_\mu  \psi_L+\frac 1{16} \epsilon^{\mu a b c}\, 
\d_\mu h_{a\lambda}\,h_b^\lambda\,  \bar\psi_L \gamma_c
\gamma_5\psi_L\right]\label{int}
\ee
 
Retaining only the above terms of the action of (\ref{int}), the Feynman rules
are as follows (momenta are ingoing and {\it the external gravitational field is
assumed to be $h_{\mu\nu}$}). The fermion propagator is
\be 
P: \quad\quad \frac i{\slash \!\!\! p+i\epsilon }\label{prop}
\ee
The two-fermion-one-graviton vertex is  
\be
V_{ffh}:\quad\quad- \frac i{8} \left[(p+p')_\mu \gamma_\nu + (p+p')_\nu
\gamma_\mu\right] \frac {1+\gamma_5}2\label{2f1g}
\ee
The two-fermion-two-graviton vertex ($V^{\epsilon}_{ffhh}$) is
\be 
V^{\epsilon}_{ffhh}:\quad\quad \frac 1{64}
t_{\mu\nu\mu'\nu'\kappa\lambda}(k-k')^\lambda\gamma^\kappa\frac
{1+\gamma_5}2\label{2f2g}
\ee 
where
\be
t_{\mu\nu\mu'\nu'\kappa\lambda}=\eta_{\mu\mu'} \epsilon_{\nu\nu'\kappa\lambda}
+\eta_{\nu\nu'} \epsilon_{\mu\mu'\kappa\lambda} +\eta_{\mu\nu'}
\epsilon_{\nu\mu'\kappa\lambda} +\eta_{\nu\mu'}
\epsilon_{\mu\nu'\kappa\lambda}\label{t}
\ee

\subsection{Complete expansion}

The previous action (\ref{action}) is a simplified one. It disregards the
measure $\sqrt{|g|}$, which is incorporated in the fermion field $\psi$. In a more
complete approach one should take into account tadpole and seagull terms and
reinsert $\sqrt{|g|}$ in the action. Some of these, in principle, might be
relevant for the trace anomaly. To this end  we need the complete expansion in
$h_{\mu\nu}$ up to order three of the action,
more precisely, 
\be
g_{\mu\nu}&=& \eta_{\mu\nu}+ h_{\mu\nu}\label{g}\\
g^{\mu\nu}&=& \eta^{\mu\nu}-h^{\mu\nu} +(h^2)^{\mu\nu}+\ldots\0\\
e_a^\mu&=& \delta_a^\mu -\frac 12 h_a^\mu +\frac 38 (h^2)_a^\mu -\frac 5{16}
(h^3)_a^\mu+\ldots\0\\
e^a_\mu&=& \delta^a_\mu +\frac 12 h^a_\mu -\frac 18 (h^2)^a_\mu +\frac 1{16}
(h^3)^a_\mu+\ldots\0\\
\sqrt{|g|}&=&  1+\frac 12 (\tr\, h)+\frac 18 (\tr\,  h)^2 -\frac 14 (\tr\,  h^2)
-\frac 18 (\tr\,  h)(\tr\,  h^2) +\frac 1{48}(\tr h)^3 +\frac 16 (\tr
h^3)+\ldots\0
\ee
and
\be
\Gamma_{\mu\nu}^\lambda =\frac 12 \left( \partial_\mu h_\nu^\lambda
+\partial_\nu h_\mu^\lambda-\partial^\lambda h_{\mu\nu}\right)-\frac 12
(h-h^2)^{\lambda \rho}\left( \partial_\mu h_{\rho\nu} +
\partial_\nu h_{\rho\mu} -\partial_\rho h_{\mu\nu} \right)\label{approxGamma}
\ee
In this approximation the spin connection is
\be 
 \omega_\mu^{ab}&=&\frac 12 \left( \partial^b h_\mu^a-  \partial^a
h_\mu^b\right)+\frac 14 \left( h^{\sigma a} \partial_\sigma h_\mu^b
- h^{\sigma b} \partial_\sigma h_\mu^a+ h^{b\sigma} \partial^a h_{\sigma
\mu}-h^{a\sigma} \partial^b h_{\sigma \mu}\right) \0\\
&&-\frac 18\left( h^{a\sigma} \partial_\mu h_\sigma^b-  h^{b\sigma} \partial_\mu
h_\sigma^a\right) \label{Omegaapprox}\\
&&+\frac 18 \left( (h^2)^{a\lambda}\partial_\mu h^b_\lambda
-(h^2)^{b\lambda}\partial_\mu h^a_\lambda \right)
+\frac 3{16} \left( (h^2)^{a\lambda} \partial^b h_{\mu\lambda}-(h^2)^{b\lambda}
\partial^a h_{\mu\lambda}\right)\0\\
&& - \frac 3{16} \left( (h^2)^{a\lambda} \partial_\lambda
h_{\mu}^b-(h^2)^{b\lambda} \partial_\lambda h_{\mu}^a\right)+\frac
18\left(h^{a\rho} h^{b\lambda} - h^{b\rho} h^{a\lambda}\right)
\partial_{\lambda}h_{\mu\rho}+\ldots\0
\ee
Up to third order in $h$ the action is
\be 
S&=& \int d^4x \, \Big{[} \frac{i}{2} \overline {\psi_L}\gamma^m
{\stackrel{\leftrightarrow}{\partial}}_m  \psi_L   -\frac i4 
 \overline {\psi_L}h^m_a \gamma^a {\stackrel{\leftrightarrow}{\partial}}_m 
\psi_L   +\frac {3i}{16}
 \overline {\psi_L} (h^2)^m_a \gamma^a {\stackrel{\leftrightarrow}{\partial}}_m 
\psi_L  - \frac {5i}{32} \overline {\psi_L} (h^3)^m_a \gamma^a
{\stackrel{\leftrightarrow}{\partial}}_m  \psi_L  \0\\
&&
-\frac 1{16} \epsilon^{m abc} \overline {\psi_L} \gamma_c\gamma_5
\psi_L\left(h_m^\sigma \partial_a h_{b\sigma} +  (h^2)_m^\sigma \partial_b
h_{a\sigma}- h_m^\rho h_a^\sigma \partial_\sigma h_{\rho b}-\frac 12 h_m^\rho
\partial_a h_{\rho\sigma} h_c^\sigma\right) \label{approxaction3rdorder}\\
&&+ \frac 12 (\tr h) \left( \frac{i}{2} \overline {\psi_L}\gamma^m
{\stackrel{\leftrightarrow}{\partial}}_m  \psi_L   -\frac i4 
 \overline {\psi_L}h^m_a \gamma^a {\stackrel{\leftrightarrow}{\partial}}_m 
\psi_L   +\frac {3i}{16}
 \overline {\psi_L} (h^2)^m_a \gamma^a {\stackrel{\leftrightarrow}{\partial}}_m 
\psi_L \right.\0\\
&&\quad\quad -\frac 1{16}\left.  \epsilon^{m abc} \overline {\psi_L}
\gamma_c\gamma_5 \psi_L h_m^\sigma \partial_a h_{b\sigma}\right)\0\\
&&+\left( \frac 18 (\tr\,  h)^2 -\frac 14 (\tr\,  h^2)\right)\left( \frac{i}{2}
\overline {\psi_L}\gamma^m {\stackrel{\leftrightarrow}{\partial}}_m  \psi_L  
-\frac i4 
 \overline {\psi_L}h^m_a \gamma^a {\stackrel{\leftrightarrow}{\partial}}_m 
\psi_L   \right)\0\\
&& + \left(  -\frac 18 (\tr\,  h)(\tr\,  h^2) +\frac 1{48}(\tr h)^3 +\frac 16
(\tr h^3)\right) \frac{i}{2} \overline {\psi_L}\gamma^m
{\stackrel{\leftrightarrow}{\partial}}_m  \psi_L + \ldots\Big{]}\0
\ee
The propagator (\ref{prop}) comes from the first term of the first line in the
RHS of (\ref{approxaction3rdorder}). The vertex
$V_{ffh}$ comes from the second term, while $V^\epsilon_{ffhh}$ originates from
the first term in the second line of (\ref{approxaction3rdorder}). There are
many other vertices of the type  $V_{ffh}, V_{ffhh},V_{ffhhh}$. It is important
to single out which may be relevant to trace anomalies. 

The Ward identity for Weyl invariance, in absence of anomalies, is:
\be
\ET(x)\equiv g_{\mu\nu}(x) \langle\!\langle T^{\mu\nu}(x)\rangle\!\rangle
=\langle\!\langle T^{\mu}_\mu(x)\rangle\!\rangle+
h_{\mu\nu}(x) \langle\!\langle T^{\mu\nu}(x)\rangle\!\rangle=0\label{WWI0}  
\ee
Writing
\be
\langle\!\langle T^{\mu\nu}(x) \rangle \! \rangle&=& \langle
0|T_{(0)}^{\mu\nu}(x)|0\rangle\label{Tseries}\\
&&+\sum_{n=1}^\infty \frac 1{2^n n!}  
\int \prod_{i=0}^n dx_i\, h_{\mu_1\nu_1}(x_1)\ldots h_{\mu_n\nu_n}(x_n) {\cal
T}^{\mu\nu\mu_1\nu_1\ldots \mu_n\nu_n}(x,x_1,\ldots,x_n),\0
\ee
order by order in $h$, 
eq.(\ref{WWI0}) breaks down to
\be
{\ET}^{(0)}(x)&\equiv& \langle 0|T_{(0)\mu}{}^\mu(x)|0\rangle=0\label{Tmumu1}\\
{\ET}^{(1)}(x)&\equiv &  {\cal T}_{\mu}^{\mu\mu_1\nu_1}(x,x_1)+ 2\delta(x-x_1)
\langle 0|T_{(0)}^{\mu_1\nu_1}(x)|0\rangle=0\label{Tmumu2}\\
 {\ET}^{(2)}(x)&\equiv& {\cal T}_{\mu}^{\mu\mu_1\nu_1\mu_2\nu_2}(x,x_1,x_2)+ 2
\delta(x-x_1)  {\cal T}^{\mu_1\nu_1\mu_2\nu_2}(x,x_2)\0\\
&&+2\delta(x-x_2) {\cal T}^{\mu_2\nu_2\mu_1\nu_1}(x,x_1)  =0\label{Tmumu3}
\ee
where
\be
T_{(0)}^{\mu\nu}= 2 \frac {\delta S}{\delta
h_{\mu\nu}(x)}\Bigg{\vert}_{h=0}=-\frac i4 \left(\overline {\psi_L} \gamma^\mu
{\stackrel{\leftrightarrow}{\partial^\nu}}\psi_L
 + \mu\leftrightarrow \nu \right)+ \frac i2 \eta_{\mu\nu}\, \overline
{\psi_L}\gamma^m {\stackrel{\leftrightarrow}{\partial}}_m  \psi_L 
,\label{T0munu} 
\ee
\be
 {\cal T}^{\mu\nu\mu_1\nu_1}(x,x_1) &=& 
i \langle 0|{\cal T}T_{(0)}^{\mu\nu}(x)T_{(0)}^{\mu_1\nu_1}(x_1)|0 \rangle-
\eta^{\mu_1\nu_1}\delta(x-x_1) \langle 0|T_{(0)}^{\mu\nu}(x)|0 \rangle\0\\
&&+4
\langle 0| \frac {\delta^{2} S}{\delta h_{\mu\nu}(x)\delta h_{\mu_1\nu_1}(x_1)
}|0 \rangle\label{TTmunu2}
\ee
and 
\be
&&{\cal T}^{\mu\nu\mu_1\nu_1\mu_2\nu_2}(x,x_1,x_2) \0\\
&&= 
- \langle 0|{\cal
T}T_{(0)}^{\mu\nu}(x)T_{(0)}^{\mu_1\nu_1}(x_1)T_{(0)}^{\mu_2\nu_2}(x_2)|0
\rangle+4
 i  \langle 0|{\cal T}T_{(0)}^{\mu\nu}(x)\frac {\delta^{2} S}{\delta
h_{\mu_1\nu_1}(x_1)\delta h_{\mu_2\nu_2}(x_2) }|0 \rangle\0\\
&&-i  \eta^{\mu_1\nu_1} \delta(x-x_1) \langle 0|{\cal
T}T_{(0)}^{\mu\nu}(x)T_{(0)}^{\mu_2\nu_2}(x_2)|0 \rangle- 
i  \eta^{\mu_2\nu_2} \delta(x-x_2) \langle 0|{\cal
T}T_{(0)}^{\mu\nu}(x)T_{(0)}^{\mu_1\nu_1}(x_1)|0 \rangle\0\\
&&+4
 i  \langle 0|{\cal T}T_{(0)}^{\mu_1\nu_1}(x_1)\frac {\delta^{2} S}{\delta
h_{\mu\nu}(x)\delta h_{\mu_2\nu_2}(x_2) }|0 \rangle+4i  \langle 0|{\cal
T}T_{(0)}^{\mu_2\nu_2}(x_2)\frac {\delta^{2} S}{\delta h_{\mu_1\nu_1}(x_1)\delta
h_{\mu\nu}(x) }|0 \rangle\0\\
&&+\left(\eta^{\mu_1\nu_1}\eta^{\mu_2\nu_2}+\eta^{\mu_1\nu_2}\eta^{\mu_2\nu_1}
+\eta^{\mu_1\mu_2}\eta^{\nu_1\nu_2}\right) \delta(x-x_1)\delta(x-x_2) 
\langle 0|T_{(0)}^{\mu\nu}(x)|0 \rangle\0\\
&& - 4\eta^{\mu_1\nu_1} \delta(x-x_1) \langle 0| \frac {\delta^{2} S}{\delta
h_{\mu\nu}(x)\delta h_{\mu_2\nu_2}(x_2) }|0 \rangle
-4\eta^{\mu_2\nu_2} \delta(x-x_2) \langle 0| \frac {\delta^{2} S}{\delta
h_{\mu\nu}(x)\delta h_{\mu_1\nu_1}(x_1) }|0 \rangle\0\\
&& + 8 \langle 0| \frac {\delta^{3} S}{\delta h_{\mu\nu}(x)\delta
h_{\mu_1\nu_1}(x_1) h_{\mu_2\nu_2}(x_2) }|0 \rangle
\label{TTTmunu3}
\ee
The functional derivatives of $S$ with respect to $h$ are understood to be
evaluated at $h=0$.

In the sequel we will need the explicit expressions of vertices, up to order two
in $h$ (for a derivation of Feynman rules see Appendix \ref{s:feynmanrules}, in
particular \ref{ss:ordgravity} and \ref{ss:oneloop}). Beside (\ref{2f1g}) and
(\ref{2f2g}) we have:
\be
V_{ffh}'&:&\quad\quad \frac i4 \eta_{\mu\nu} (\slashed{p}+ \slashed{p}')
P_L\label{2f1g'}\\
V_{ffhh}'&:& \quad\quad  \frac{ 3i}  {64} \Big{[}\left( (p+p')_\mu
\gamma_{\mu'} \eta_{\nu\nu'} +  (p+p')_{\mu} \gamma_{\nu'} \eta_{\nu\mu'}+
\{\mu\leftrightarrow \nu\}\right)\0\\
&&\quad\quad\quad\quad+ \left(  (p+p')_{\mu'} \gamma_{\mu} \eta_{\nu\nu'}+
(p+p')_{\mu'} \gamma_{\nu} \eta_{\mu\nu'}+ \{\mu'\leftrightarrow
\nu'\}\right)\Big{]} P_L\label{2f2g'}\\
V_{ffhh}''&:&   - \frac{ i}  {16} \Big{[}\eta_{\mu\nu}\left( (p+p')_{\mu'}
\gamma_{\nu'}  
+  (p+p')_{\nu'} \gamma_{\mu'}\right) +\eta_{\mu'\nu'}\left( (p+p')_{\mu}
\gamma_{\nu}  
+  (p+p')_{\nu} \gamma_{\mu}\right)\Big{]} P_L\label{2f2g''} \\
V_{ffhh}'''&:&\quad\quad \frac i{8} (\slashed{p}+ \slashed{p}')
(\eta_{\mu\nu}\eta_{\mu'\nu'} -
\eta_{\mu\nu'}\eta_{\mu'\nu} -\eta_{\mu\mu'}\eta_{\nu\nu'} ) P_L\label{2f2g'''}
\ee 

So far we have been completely general. From now on we consider only odd
correlators, that is only correlators linear in $\epsilon_{\mu\nu\lambda\rho}$.
To start with, to $\langle 0|T_{(0)\mu}{}^\mu(x)|0\rangle$, which is a constant,
only a tadpole can contribute, but its odd part
vanishes because there is no scalar one can construct with $\epsilon$ and
$\eta$. For the same reason also $\langle 0|T_{(0)}^{\mu\nu}(x)|0\rangle$
vanishes. The two-point function $\langle 0|{\cal
T}T_{(0)}^{\mu\nu}(x)T_{(0)}^{\mu_1\nu_1}(x_1)|0 \rangle$ also must vanish,
because in momentum space it must be a 4-tensor linear in $\epsilon$ and formed
with $\eta$ and the momentum $k$: there is no such tensor, symmetric in
$\mu\leftrightarrow \nu$, $\mu_1\leftrightarrow \nu_1$ and $(\mu,\nu)
\leftrightarrow (\mu_1,\nu_1)$. As for the terms
$\langle 0| \frac {\delta^{2} S}{\delta h_{\mu\nu}(x)\delta h_{\mu_1\nu_1}(x_1)
}|0 \rangle$ they might also produce nonvanishing contribution from tadpoles
diagram, but like in the previous case it is impossible to satisfy the
combinatorics. In conclusion
(\ref{Tmumu1}) and (\ref{Tmumu2}) are identically satisfied, while
(\ref{Tmumu3}) becomes
\be
&& {\ET}^{(2)}(x) ={\cal T}_\mu^{\mu\mu_1\nu_1\mu_2\nu_2}(x,x_1,x_2) \0\\
&&= \eta_{\mu\nu}\Big{(}
- \langle 0|{\cal
T}T_{(0)}^{\mu\nu}(x)T_{(0)}^{\mu_1\nu_1}(x_1)T_{(0)}^{\mu_2\nu_2}(x_2)|0
\rangle+
4 i  \langle 0|{\cal T}T_{(0) }^{\mu\nu}(x)\frac {\delta^{2} S}{\delta
h_{\mu_1\nu_1}(x_1)\delta h_{\mu_2\nu_2}(x_2) }|0 \rangle\0\\
 &&+4
 i  \langle 0|{\cal T}T_{(0)}^{\mu_1\nu_1}(x_1)\frac {\delta^{2} S}{\delta
h_{\mu\nu}(x)\delta h_{\mu_2\nu_2}(x_2) }|0 \rangle+4i  \langle 0|{\cal
T}T_{(0)}^{\mu_2\nu_2}(x_2)\frac {\delta^{2} S}{\delta h_{\mu_1\nu_1}(x_1)\delta
h_{\mu\nu}(x) }|0 \rangle\0\\
&& + 8 \langle 0| \frac {\delta^{3} S}{\delta h_{\mu\nu}(x)\delta
h_{\mu_1\nu_1}(x_1) h_{\mu_2\nu_2}(x_2) }|0 \rangle\Big{)}
\label{TTTmunu4}
\ee
To proceed further we focus now on the terms containing the second derivative of
$S$. Looking at (\ref{approxaction3rdorder}) we see that there are several such
terms. We argue now that those among them that do not contain the $\epsilon$
tensor, although the gamma trace algebra may generate an $\epsilon$ tensor,
cannot contribute to the odd trace anomaly. The vertices corresponding to such
terms have two fermion and two graviton legs, that is they are of the type
$V_{ffhh}$. By Fourier transform, we associate an incoming $e^{ipx}$ plane wave
to one fermion and an outgoing $e^{-ip'x}$ one to the other, while we associate
two incoming plane waves $e^{ik_1x}, e^{ik_2x}$ to the two gravitons. Since none
of them contain derivatives of $h$, the vertex will depend at most on
$q=k_1+k_2$, not on $k_1-k_2$, see for instance the vertex coming from the third
term in the first line of (\ref{approxaction3rdorder}), i.e. 
$V^{'}_{ffhh}$.

This being so, the contributions from the terms related to the second derivative
of $S$ in (\ref{TTTmunu4}) via such vertices, and linear in $\epsilon$, must
vanish, because it is impossible to form a 4-tensor  
symmetric in $\mu_1\leftrightarrow \nu_1$, $\mu_2\leftrightarrow \nu_2$ and
$(\mu_1,\nu_1) \leftrightarrow (\mu_2\nu_2)$ with $\epsilon,\eta$ and $q_\mu$.
It follows that only the contribution with the vertex
$V^\epsilon_{ffhh}$ might contribute non trivially to the odd trace anomaly.
Looking at the form
of $V^\epsilon_{ffhh}$,  it is clear that the two terms in the third line
of (\ref{TTTmunu4}) give vanishing contribution because the contraction of $\mu$
with $\nu$ becomes a (vanishing) contraction of the $t$ tensor, (\ref{t}). 

Next let us consider the fourth line of (\ref{TTTmunu4}). These are seagull
terms, with three external graviton lines attached to the same point of a
fermion loop. The gamma trace algebra cannot generate
an $\epsilon$ tensor from all such terms, except of course the second term in
the second line and the one in the fourth line. Therefore we can exclude all the
former from our consideration. As for the latter the relevant vertex has two
fermion legs, with the usual momenta $p$ and $p'$, and three graviton legs, with
incoming momenta $k_1,k_2,k_3$ and labels $\mu_1,\nu_1,\mu_2,\nu_2$ and
$\mu_3,\nu_3$, respectively. Its expression for the second term in the second
line of  (\ref{TTTmunu4}) is
\be
\sim \epsilon_{\mu_2\mu_3\lambda\rho} k_3^\lambda\gamma^\rho \eta_{\mu_1\nu_3}
\eta_{\nu_1\nu_2}\label{Vffhhh}
\ee
symmetrized in $\mu_1\leftrightarrow \nu_1$, $\mu_2\leftrightarrow \nu_2$,
$\mu_3\leftrightarrow \nu_3$,
and with respect to the exchange of any two couples $(\mu_i,\nu_i)$. The seagull
term is therefore proportional to
\be
\int d^4p \, \frac {p^\rho}{p^2}\0
\ee
which vanishes. As for the term in the fourth line of  (\ref{TTTmunu4}), one
comes to similar conclusions.

In summary, the odd trace anomaly receives contributions only from
\be
&& {\ET}^{(2)}(x) ={\cal
T}_\mu^{\mu\mu_1\nu_1\mu_2\nu_2}(x,x_1,x_2)\label{TTTmunu5}\\
&&= \eta_{\mu\nu}\Big{(}
- \langle 0|{\cal
T}T_{(0)}^{\mu\nu}(x)T_{(0)}^{\mu_1\nu_1}(x_1)T_{(0)}^{\mu_2\nu_2}(x_2)|0
\rangle+4
 i  \langle 0|{\cal T}T_{(0) }^{\mu\nu}(x)\frac {\delta^{2} S}{\delta
h_{\mu_1\nu_1}(x_1)\delta h_{\mu_2\nu_2}(x_2) }|0 \rangle \Big{)}\0
\ee
This result looks very much like the starting point of \cite{BGL}, i.e. it seems
to reduce to the same contributions, i.e. the triangle diagram and bubble
diagram (which turned out to vanish), but there is an important modification:
the $T_{(0)}^{\mu\nu}(x)$ is different from the free e.m. tensor in \cite{BGL},
the definition (\ref{T0munu}) contains an additional piece (the second). It is
not hard to show that the
second term in the RHS of  (\ref{TTTmunu5}) vanishes also when taking account of
this modification. As for the three point function in the first term of 
(\ref{TTTmunu5})  we obtain of course the same result as in \cite{BGL} when the
calculation is made with three vertices $V_{ffh}$:
$P-V_{ffh}$-$P$-$V_{ffh}$-$P$-$V_{ffh}$ (for the reader's convenience this
calculation is 
repeated in Appendix \ref{s:trianglediagram}); it is 0 when the second or third
vertices are replaced by $V'_{ffh}$, and it is -4 times the result of \cite{BGL}
if the first vertex is replaced by $V'_{ffh}$, i.e.
$P$-$V'_{ffh}$-$P$-$V_{ffh}$-$P$-$V_{ffh}$  . When we replace more than one
vertex $V_{ffh}$ with $V'_{ffh}$ we get 0. So the overall result of
(\ref{TTTmunu5}) is (1-4=-3) times the end result for the trace anomaly in
\cite{BGL}.

We will see, below, however that this modification of the anomaly must be
canceled in order to guarantee conservation. Let us call the lowest order
integrated anomaly, obtained in \cite {BGL}, ${\cal A}_\omega=-\int \omega \,
{\EA}_0$. Then the new addition equals $-4{\cal A}_\omega$. By adding to the
effective action the term ${\cal C}=-\int \frac 12 \tr h \, {\EA}_0$ we exactly
cancel this additional unwanted piece. We will verify that this counterterm
cancels an analogous anomalous term in the Ward identity of the diffeomorphisms,
anomalous term which is generated by the same diagram
$P$-$V'_{ffh}$-$P$-$V_{ffh}$-$P$-$V_{ffh}$ which is the cause of the additional
term in question in the trace anomaly.

In conclusion, the only relevant term for the odd trace anomaly is the 
$P$-$V_{ffh}$-$P$-$V_{ffh}$-$P$-$V_{ffh}$ one.
This is the term we have computed first in \cite{BGL}, which gives rise to the
Pontryagin anomaly. It should be remarked that in the odd trace anomaly
calculation there are no contributions from tadpole and seagull terms.

\subsection{Odd trace anomaly for Dirac and Majorana fermions}

The action for a Dirac fermion is the same as in (\ref{approxaction3rdorder})
with $\psi_L$ everywhere replaced by the Dirac fermion $\psi$. In order to
evaluate the odd trace anomaly we remark that an odd contribution in
(\ref{TTTmunu3}) can come only from the terms in (\ref{approxaction3rdorder})
that contain the $\epsilon$ tensor. Since these terms contain $\gamma_5$, upon
tracing the gamma matrix part, either they give 0 or another $\epsilon$ tensor.
In the latter case they produce an even contribution to the trace anomaly,
which does not concern us here. In conclusion the odd trace anomaly, in the case
of a Dirac fermion,
vanishes. 

When the fermion are Majorana the conclusion does not change. The simplest way
to see it is to use the Majorana representation for the gamma matrices. Then
$\psi$ has four real components, and the only change with respect to the Dirac
case is that in the path integral we integrate over real fermion fields instead
of complex ones, while all the rest remains unchanged. The conclusion is
obvious.

\section{Conservation of the e.m. tensor}

As already anticipated above, trace anomalies are strictly connected with
diffeomorphism anomalies.
In 4d the so-called Einstein-Lorentz anomalies are absent, but there may appear
other anomalous terms in the Ward identity of the diffeomorphisms. The latter
together with a Weyl anomaly partner form a cocycle of the joint diff+Weyl
cohomology, see \cite{BPT1,BPT2}. Usually, by adding a local counterterm to the
effective action, one can restore diffeomorphism invariance. In the present
case, odd parity trace anomaly, the analysis of such possible anomalies was
carried out in a simplified form in \cite{BDL}. In this section we wish to
complete that analysis by considering also tadpoles and seagull terms.

If we take into account the tadpole and seagull terms in the conservation law 
one has to take into account also the VEV of the e.m. tensor. Let us set 
\be
\langle 0| T_{(0)}^{\mu\nu}(x)|0\rangle = \langle 0|
T_{(0)}^{\mu\nu}(0)|0\rangle= \Theta^{\mu\nu}= A \eta^{\mu\nu}\label{VEV}
\ee
The Ward identity is
\be
\nabla_\mu \langle\!\langle T^{\mu\nu}(x)\rangle\!\rangle=\partial_\mu
\langle\!\langle T^{\mu\nu}(x)\rangle\!\rangle+ \Gamma_{\mu\lambda}^\mu
\langle\!\langle T^{\lambda\nu}(x)\rangle\!\rangle+
\Gamma_{\mu\lambda}^\nu \langle\!\langle
T^{\mu\lambda}(x)\rangle\!\rangle=0\label{nablamuTmunu}
\ee
{ because $\langle\!\langle T^{\mu\nu}(x)\rangle\!\rangle\equiv
\frac{2}{\sqrt{-g}}\frac{\delta W}{\delta g_{\mu\nu}(x)}$.} To first  order in
$h_{\mu\nu}$ we have 
\be
\Gamma_{\mu\lambda}^\nu(x)&\approx &\frac 12 \left( \partial_\mu h^\nu_\lambda +
\partial_\lambda h_\mu^\nu-
\partial^\nu h_{\mu\lambda}\right)\nonumber\\
\Gamma_{\mu\lambda}^\mu(x)&\approx &\frac12 \partial_\lambda h^\mu_\mu 
\label{approxChris}
\ee
Now we use (\ref{Tseries},\ref{T0munu},\ref{TTmunu2}, \ref{TTTmunu3}).
To the 0-th order in $h$ (\ref{nablamuTmunu}) implies
\be
\partial_\mu\langle 0|T^{\mu\nu}(x)|0\rangle =0\label{0thorderWI}
\ee
To get the WI to first order one must differentiate (\ref{nablamuTmunu}) with
respect to $h_{\mu\nu}$. One has
\be
\frac {\delta h_{\mu\nu}(x)}{\delta h_{\lambda\rho}(y)}= \frac 12
\left(\delta^\lambda_\mu \delta^\rho_\nu+\delta^\lambda_\nu
\delta^\rho_\mu\right) \delta(x-y)\label{dhdh}
\ee
Differentiating the first term on the RHS of (\ref{nablamuTmunu}) one gets the
ordinary divergence of the two-point function. Then
\be
\frac {\delta \Gamma^\mu_{\mu\lambda}(x)}{\delta h_{\mu_1\nu_1}(y)}= \frac 12
\eta^{\mu_1 \nu_1} \partial^x_\lambda \delta(x-y)\label{dGammamumu}
\ee
and 
\be
\frac {\delta \Gamma^\nu_{\mu\lambda}(x)}{\delta h_{\mu_1\nu_1}(y)}&=&
\frac 14 \left(\partial_{\mu} \delta(x-y)\,
\left(\delta_\lambda^{\nu_1}\eta^{\mu_1\nu}+
\delta_\lambda^{\mu_1}\eta^{\nu_1\nu}\right)
+ \partial_ {\lambda}\delta(x-y)\, \left(\delta_\mu^{\mu_1}\eta^{\nu\nu_1}
+\delta_\mu^{\nu_1}\eta^{\nu\mu_1}\right)\right.\0\\
&&-\left.\partial^{\nu}
\delta(x-y)\,\left(\delta_\lambda^{\nu_1}\delta^{\mu_1}_{\mu}+ 
\delta_\lambda^{\mu_1}\delta^{\nu_1}_{\mu}\right)  \right)\label{dGammamunu}
\ee
Putting everything together one finds
\be
&&\partial_{\mu}^x {\cal T}^{\mu\nu\mu_1\nu_1}(x,y) +  \frac 12 \eta^{\mu_1
\nu_1} \partial^x_\lambda \delta(x-y) \Theta^{\lambda\nu}\label{newWI}\\
&&+ \frac 12 \left( \partial_\lambda^x \delta(x-y) \eta^{\mu_1\nu}
\Theta^{\lambda\nu_1}+\partial_\lambda^x \delta(x-y)
 \eta^{\nu_1\nu} \Theta^{\lambda\mu_1} - \partial^{x\,\nu} \delta(x-y)
\Theta^{\mu_1\nu_1}\right)\0\\
 &&= i \partial_{\mu}^x  \langle 0| {\cal T} T_{(0)}^{\mu\nu}(x) 
T_{(0)}^{\mu_1\nu_1}(y)|\rangle +4\partial_{\mu}^x \langle 0| \frac {\delta^{2}
S}{\delta h_{\mu\nu}(x)\delta h_{\mu_1\nu_1}(y) }|0 \rangle\0\\
 &&+   \partial_\lambda^x \delta(x-y) \eta^{\mu_1\nu}
\Theta^{\lambda\nu_1}+\partial_\lambda^x \delta(x-y)
 \eta^{\nu_1\nu} \Theta^{\lambda\mu_1} - \partial^{x\,\nu} \delta(x-y)
\Theta^{\mu_1\nu_1} =0.\0
\ee
We have already noted that, for what concerns the odd part, all the terms in the
RHS vanish. Therefore conservation is guaranteed up to second order in $h$.

The order three Ward identity has a rather cumbersome expression, in particular
it contains various terms
linear in $\Theta^{\mu\nu}$. Since they do not contribute to the odd part of the
identity we drop them altogether. The remaining terms are:{\small
\be
&&-\partial^x_\mu \langle 0| {\cal T}T_{(0)}^{\mu\nu}(x) 
T_{(0)}^{\mu_1\nu_1}(x_1) T_{(0)}^{\mu_2\nu_2}(x_2)|0\rangle+ 4i \partial^x_\mu
\langle 0| {\cal T}T_{(0)}^{\mu\nu}(x) \frac {\delta^{2} S}{\delta
h_{\mu_1\nu_1}(x_1)\delta h_{\mu_2\nu_2}(x_2) } |0\rangle \label{3rdWI}\\
&& + 4i \partial^x_\mu \langle 0| {\cal T}T_{(0)}^{\mu_2\nu_2}(x_2) \frac
{\delta^{2} S}{\delta h_{\mu\nu}(x)\delta h_{\mu_1\nu_1}(x_1) } |0\rangle + 4i
\partial^x_\mu \langle 0| {\cal T}T_{(0)}^{\mu_1\nu_1}(x_1) \frac {\delta^{2}
S}{\delta h_{\mu\nu}(x)\delta h_{\mu_2\nu_2}(x_2) } |0\rangle\0 \\
&&-4\eta^{\mu_1\nu_1}\partial^x_\mu \left( \delta(x-x_1)  \langle 0| \frac
{\delta^{2} S}{\delta h_{\mu\nu}(x)\delta h_{\mu_2\nu_2}(x_2) } |0\rangle\right)
-4\eta^{\mu_2\nu_2}\partial^x_\mu \left( \delta(x-x_2)  \langle 0| \frac
{\delta^{2} S}{\delta h_{\mu\nu}(x)\delta h_{\mu_1\nu_1}(x_1) }
|0\rangle\right)\0\\
&&-i \partial^x_\mu \left( \delta(x-x_1) \eta^{\mu_1\nu_1}\langle 0| {\cal
T}T_{(0)}^{\mu\nu}(x)  T_{(0)}^{\mu_2\nu_2}(x_2)\rangle + \delta(x-x_2)
\eta^{\mu_2\nu_2}\langle 0| {\cal T}T_{(0)}^{\mu\nu}(x) 
T_{(0)}^{\mu_1\nu_1}(x_1)\rangle\right)\0\\
&&+\partial^x_\lambda \delta(x-x_1)\eta^{\mu_1\nu_1}\left(i\langle 0| {\cal
T}T_{(0)}^{\lambda\nu}(x)  T_{(0)}^{\mu_2\nu_2}(x_2)|0\rangle+4  \langle 0|
\frac {\delta^{2} S}{\delta h_{\lambda\nu}(x)\delta h_{\mu_2\nu_2}(x_2) }
|0\rangle\right) \0\\
&&+\partial^x_\lambda \delta(x-x_2)\eta^{\mu_2\nu_2}\left(i\langle 0| {\cal
T}T_{(0)}^{\lambda\nu}(x)  T_{(0)}^{\mu_1\nu_1}(x_1)|0\rangle+4  \langle 0|
\frac {\delta^{2} S}{\delta h_{\lambda\nu}(x)\delta h_{\mu_1\nu_1}(x_1) }
|0\rangle\right) \0\\
&& + \left( \partial^x_\mu   \delta(x-x_1) \left( \delta_\lambda^{\mu_1}
\eta^{\nu\nu_1}+  \delta_\lambda^{\nu_1} \eta^{\nu\mu_1}\right) -\partial^{x\nu}
\delta(x-x_1) \delta_\mu^{\mu_1}\delta_\lambda^{\nu_1}\right) \0\\
&&\quad\quad\cdot \left(i\langle 0| {\cal T}T_{(0)}^{\mu\lambda}(x) 
T_{(0)}^{\mu_2\nu_2}(x_2)\rangle+4  \langle 0| \frac {\delta^{2} S}{\delta
h_{\mu\lambda}(x)\delta h_{\mu_2\nu_2}(x_2) } |0\rangle\right) \0\\
&& + \left( \partial^x_\mu  \delta(x-x_2) \left( \delta_\lambda^{\mu_2}
\eta^{\nu\nu_2}+  \delta_\lambda^{\nu_2} \eta^{\nu\mu_2}\right) -\partial^{x\nu}
\delta(x-x_2) \delta_\mu^{\mu_2}\delta_\lambda^{\nu_2}\right)\0\\
&&\quad\quad\cdot \left(i\langle 0| {\cal T}T_{(0)}^{\mu\lambda}(x) 
T_{(0)}^{\mu_2\nu_2}(x_2)\rangle+4  \langle 0| \frac {\delta^{2} S}{\delta
h_{\mu\lambda}(x)\delta h_{\mu_1\nu_1}(x_1) } |0\rangle\right)\0\\
&&+8 \partial^x_\mu \langle 0| \frac {\delta^{3} S}{\delta
h_{\lambda\nu}(x)\delta h_{\mu_1\nu_1}(x_1) \delta h_{\mu_2\nu_2}(x_2)}
|0\rangle=0\0
\ee}
In the above discussion concerning the odd trace anomaly we have already met
some of the terms appearing in this formula. As already noted there, the two
point functions $\langle 0| {\cal T}T_{(0)}^{\mu\nu}(x) 
T_{(0)}^{\lambda\rho}(y)|0\rangle$ cannot contribute to the odd part because the
combinatorics of the $\epsilon$ and $\eta$ tensor plus an external momentum does
not allow it. Next the VEV's of second and third derivative of $S$ with respect
to $h$ cannot contribute with a tadpole term: if we look at
(\ref{approxaction3rdorder}) and focus on the vertices that can give an odd
parity contribution, i.e. those containing the $\epsilon$ tensor, we notice that
they depend linearly on the external momenta (not on the fermion momenta);
therefore, in a tadpole term, the momentum integrand can only be linear in the
internal momentum $p^\mu$, and thus vanishes.

Therefore (\ref{3rdWI}), as far as the odd part is concerned, reduces to
\be
&&-\partial^x_\mu \langle 0| {\cal T}T_{(0)}^{\mu\nu}(x) 
T_{(0)}^{\mu_1\nu_1}(x_1) T_{(0)}^{\mu_2\nu_2}(x_2)|0\rangle+ 4i \partial^x_\mu
\langle 0| {\cal T}T_{(0)}^{\mu\nu}(x) \frac {\delta^{2} S}{\delta
h_{\mu_1\nu_1}(x_1)\delta h_{\mu_2\nu_2}(x_2) } |0\rangle \0\\
&&+ 4i \partial^x_\mu \langle 0| {\cal T}T_{(0)}^{\mu_2\nu_2}(x_2) \frac
{\delta^{2} S}{\delta h_{\mu\nu}(x)\delta h_{\mu_1\nu_1}(x_1) } |0\rangle + 4i
\partial^x_\mu \langle 0| {\cal T}T_{(0)}^{\mu_1\nu_1}(x_1) \frac {\delta^{2}
S}{\delta h_{\mu\nu}(x)\delta h_{\mu_2\nu_2}(x_2) } |0\rangle\0\\
&&=0.\label{3rdWIb}
\ee
The last three terms on the LHS can be shown to vanish. The proof is not as
simple as the previous ones. One has to push the calculations one step further,
introduce a dimensional regulator and use Feynman parametrization (see Appendix
B). The
integration over the relevant parameter can easily be shown to vanish. What
remains to be verified is therefore 
\be
&&\partial^x_\mu \langle 0| {\cal T}T_{(0)}^{\mu\nu}(x) 
T_{(0)}^{\mu_1\nu_1}(x_1) T_{(0)}^{\mu_2\nu_2}(x_2)|0\rangle=0.\label{3rdWIc}
\ee

Let us consider the term generated by the diagram
$P-V'_{ffh}-P-V_{ffh}-P-V_{ffh}$. We have already calculated it above, it equals
$-\partial^x_\nu \EA(x)$, where $\EA(x)$ is the unintegrated Weyl anomaly
calculated in \cite{BGL}. So conservation is violated by this term. Adding to
the action the term ${\cal C}=-\int \frac 12 \tr h \omega {\EA}_0$, as we have
anticipated above, we get the diff variation
\be
\delta_\xi {\cal C}=- \int \partial_\nu \xi^\nu\, \EA=    \int
\xi^\nu\partial_\nu\EA\label{canc1}
\ee
which exactly cancels this anomaly\footnote{Concerning the signs remember that
there is a relative - sign between the unintegrated Diff and trace anomalies}. 

Next we have to consider the diagram $P-V_{ffh}-P-V'_{ffh}-P-V_{ffh}$ and
$P-V_{ffh}-P-V_{ffh}-P-V'_{ffh}$. In the on-shell case, $k_1^2=0=k_2^2$, these
contributions can be shown to vanish. It is enough to take formula (3.18) of
\cite{BGL}. The first diagram corresponds to contracting this formula with
$k_1^\mu$ or $k_1^\nu$. It is easy to see that such a contraction vanishes. The
second diagram corresponds to contracting the same formula with $k_2^{\mu'}$ or
$k_2^{\nu'}$, which again vanishes. Therefore, at least in the on-shell case
these diagrams do not contribute.

In  conclusion we have to verify (\ref{3rdWIc}) for the triangle diagram
$P-V_{ffh}-P-V_{ffh}-P-V_{ffh}$ (and the crossed one). This is what we have
already done in \cite{BGL,BDL}.

\subsection{On-shell, off-shell and locality}

In \cite{BGL,BDL} the following integrals were used in order to compute the
relevant Feynman diagram
\be
&&   \int\!\frac{d^4p}{(2\pi)^4}\int\!\frac{d^\delta \ell}{(2\pi)^\delta}\frac
{p^2}{({p}^2+\ell^2+\Delta)^3}=
\frac 1{(4\pi)^2}\left(- \frac 2{\delta}-\gamma +\log (4\pi)-
\log{\Delta}\right)\0\\
&&  \int\!\frac{d^4p}{(2\pi)^4}\int\!\frac{d^\delta \ell}{(2\pi)^\delta}\frac
{p^4}{({p}^2+\ell^2+\Delta)^3}=
\frac {\Delta}{2(4\pi)^2}\left(- \frac 2{\delta}-\gamma +4+\log
(4\pi)-\log{\Delta}\right)\label{C4a}
\ee
and
\be
&&  \int\!\frac{d^4p}{(2\pi)^4}\int\!\frac{d^\delta \ell}{(2\pi)^\delta}\frac
{\ell^2}{({p}^2+\ell^2+\Delta)^3}=-\frac 1{2(4\pi)^2}\0\\
&&  \int\!\frac{d^4p}{(2\pi)^4}\int\!\frac{d^\delta \ell}{(2\pi)^\delta}\frac
{\ell^2 p^2}{({p}^2+\ell^2+\Delta)^3}=
\frac 1{(4\pi)^2}\Delta\label{C4b}
\ee
where $\Delta = u(1-u) k^2_1+v(1-v) k^2_2 + 2 uv \,k_1k_2$, $\, u,v$ are Feynman
parameters, and $\delta$ is the dimensional regulator: $d=4+\delta$. 

The odd trace anomaly is due to the term \cite{BGL,BDL}
\be
&&  -\frac 1 {128} \int \frac {d^4p}{(2\pi)^4}
\int \frac {d^{\delta}\ell}{(2\pi)^{\delta}} \,\tr\left( \frac
{\slashed{p}+\slashed{\ell}}{p^2-\ell^2} (2p-k_1)_\lambda\gamma_\rho
\,\right.\0\\
&&\times \left.\,\frac
{\slashed{p}+\slashed{\ell}-\slashed{k}_1}{(p-k_1)^2-\ell^2}
\,(2p-2k_1-k_2)_{\al}\gamma_{\beta}  \frac
{\slashed{p}+\slashed{\ell}-\slashed{q}}{(p-q)^2-\ell^2}\,   \slashed{\ell} \, 
\frac {\gamma_5}2 \right).\label{DeltaT}
\ee
This requires the two integrals (\ref{C4b}), which must be further integrated on
$v$ from 0 to $1-u$ and on $u$ from 0 to 1. The integrations over the Feynman
parameters are elementary and lead to the result
\be 
\ET^{\mu}_{\mu\alpha\beta\lambda\rho }(k_1,k_2)=\frac 1{192(4\pi)^2} k_1^\sigma
k_2^\tau\left(t_{\lambda\rho\al\beta\sigma\tau}
(k_1^2+k_2^2+k_1k_2)-t^{(21)}_{\lambda\rho\al\beta\sigma\tau}\right)\label{
oddtrace}
\ee
We report this result here to stress the fact that the terms contained in it are
contact terms and thus lead to a local anomaly. In \cite{BDL} we remarked that
the piece proportional to $(k_1^2+k_2^2)$ disappears on shell, and off-shell
corresponds to a trivial anomaly.

To compute the conservation law (\ref{3rdWIc}) we need also the integrals
(\ref{C4a}). It is evident from the form of their RHS's that integrating on $u$
and $v$ will lead to non-contact terms, and non-local expressions
for the odd diff anomaly. However if we put $k_1$ and $k_2$ on shell things
change. The contact terms have been discussed in \cite{BDL}. They can be
eliminated by subtracting local counterterms without spoiling the trace anomaly.
As for the noncontact terms they
are polynomials of $k_1$ and $k_2$ multiplied by $\log k_1 \!\cdot\!k_2$. All
such terms are listed in Appendix E of \cite{BDL}. They look non-local. However,
using the Fourier transform
\be
&&\int \frac{d^4k_1}{(2\pi)^4} \frac  {d^4k_2}{(2\pi)^4} e^{i (k_1(x-z)
+k_2(y-z))} \log{(k_1+k_2)^2}\0\\
&&\quad\quad=- \frac 1{4\pi^2}  \delta^{(4)}(x-y) \square_z \left(\frac
1{(x-z)^2}\log
\frac{(x-z)^2}4\right),\label{C4}
\ee
one can show that they give a vanishing contribution when inserted into the
effective action, because
of the on shell condition $\square h_{\mu\nu}=0$ (De Donder gauge). On the other
hand, when $k_1$ and $k_2$ are off shell, the anomaly looks nonlocal. This is a
surprise because
we are used to think of anomalies as local expressions. But we have learned from
\cite{BCDGLS}
and from the higher spins analysis that when higher spins are involved
(including
the metric) covariance generally requires to sacrifice locality. However the
ensuing  
non-locality is a gauge artifact. By imposing a suitable gauge choice, locality
can be
restored. As an example see eq.(8.21) and others in \cite{BCDGLS}.

\section{Additional remarks on Weyl and Majorana}

Before leaving Part I of this paper let us add some comments on the Pontryagin
trace anomaly.
{ A non-trivial property is that it belongs to the family of chiral anomalies
characterized by having opposite coefficients 
for opposite chiralities\footnote{ This family includes in particular the
consistent chiral anomalies in gauge theories. Thus in selecting the 
regularization to compute the odd trace anomaly, a necessary criterion is the
ability to reproduce such well-known consistent gauge anomalies.}}.
This anomaly did not appear for the first time in \cite{BGL}. The possibility
of its existence due to its Wess-Zumino consistency was pointed out in
\cite{BBP} and, although
somewhat implicitly, its existence was implied by \cite{ChD2}. A similar anomaly
was found 
in a different contest (originating from an antisymmetric tensor field)
in the framework of an AdS/CFT in \cite{Nakayama}, where a possible conflict
with unitarity was pointed out.
The same risk has been pointed out, from a different viewpoint, in the
introduction of the 
present paper and in \cite{BGL}. 
In general it seems that its presence signals some kind of difficulty in
properly defining the theory. 
Very likely for this reason the existence of the Pontryagin trace
anomaly for chiral fermions is still considered controversial and
objections have been raised against it. Such objections are often reducible to
the credence
that Weyl fermions are equivalent to massless Majorana fermions. We have already
answered this naive
objection and will not come back to it. There are more serious issues however.
One is the following. In conformal field theory 
in 4d the three-point functions of the energy-momentum tensor cannot have an odd
part, so how can an anomaly arise from the regularization of a vanishing bare
correlator? We have already answered this question in \cite{BDL}: an anomaly
can arise as a simple quantum effect; we have shown other examples of
correlators which do not arise from the regularization of nonvanishing bare
correlators,
\cite{BL}. The crucial criterion is consistency. 

A frequent  prejudice is based on the lore that
anomalies appear only in 
connection with complex representations of the gauge group in question. This is
actually true in many cases for 
consistent chiral gauge anomalies. The latter are linear in the completely
symmetric ad-invariant tensors
of order $n$ in even $d=2n-2$ dimension. For instance in $d=4$ the tensor in
question is the
symmetric tensor $d^{abc} = \frac 12 \tr (T^a \{T^b,T^c\})$ with $T^a$ being the
anti-hermitean generators 
of the Lie algebra. It is clear that if the representation is real, i.e. $T^a$
is antisymmetric, $d^{abc}$ vanishes.
For instance in 4d there are no Einstein-Lorentz (a.k.a. diffeomorphisms and/or
local Lorentz anomalies), because the corresponding representation
of the Lorentz group is real.
{However one cannot blindly tranfer the above criterion to the
case of trace anomalies.
A hint that in such a case it may not apply is the following: as we have
explained
in section 2, the fermionic functional determinant in a left-handed theory can
be thought as a square root.
This square root is likely to give rise to a phase, which in turn would explain
the imaginary anomaly. To our best knowledge it is impossible to decide this
{\it a priori}.
Therefore we can only rely on the explicit computation.}

Another objection may arise from the following consideration. 
Let us split the Dirac fermion into two Weyl fermions:
$ \psi= P_L \psi + P_R \hat \psi=\psi_L+\psi_R$, 
The terms that appear in (\ref{approxaction3rdorder}), in the Dirac case, are of
the form $ \overline {\psi}\gamma^a {\stackrel{\leftrightarrow}{\partial}}_m 
\psi $ and
$\overline {\psi} \gamma_c\gamma_5 \psi$. They both split into the sum of the
left and right-handed part. With simple manipulations we have
\be
&&\overline {\psi_R}\gamma^a {\stackrel{\leftrightarrow}{\partial}}_m  \psi_R =
\psi^\dagger \gamma^0 \gamma^a P_R{\stackrel{\leftrightarrow}{\partial}}_m  \psi
=\psi^T C^{-1} \gamma^a \gamma^0 P_R C  {\stackrel{\leftrightarrow}{\partial}}_m
\psi^*\0\\
&&=
\overline{\hat\psi} P_R \gamma^a P_L {\stackrel{\leftrightarrow}{\partial}}_m
\hat \psi= \overline {\hat\psi_L} \gamma^a
{\stackrel{\leftrightarrow}{\partial}}_m \hat{\psi}_L\label{chiralsplit2}
\ee
where $\hat\psi= \gamma^0 C\psi^*$, and a similar expression for $\overline {\psi}
\gamma_c\gamma_5 \psi$. Thus, for instance, we can write 
\be
\overline {\psi}\gamma^a {\stackrel{\leftrightarrow}{\partial}}_m  \psi=
\overline {\psi_L}\gamma^a {\stackrel{\leftrightarrow}{\partial}}_m  \psi_L + 
\overline {\hat\psi_L} \gamma^a {\stackrel{\leftrightarrow}{\partial}}_m
\hat{\psi}_L\label{Diracsplit}
\ee
Recall that $\widehat {\psi_L} = {\hat \psi}_R$. A Majorana fermion satisfies
the reality condition
$\psi= \widehat \psi$, so we can split it, according to the chiralities, 
$\psi= P_L \psi + P_R \hat \psi$. Then, looking at (\ref{Diracsplit}), we have,
for instance
\be
\overline {\psi}\gamma^a {\stackrel{\leftrightarrow}{\partial}}_m  \psi= 2 \,
\overline {\psi_L}\gamma^a {\stackrel{\leftrightarrow}{\partial}}_m  \psi_L
\label{Majo2}
\ee
It would seem that the full Majorana action can be expressed as twice the action
for its left-handed part.
Then one would be led to conclude that there is an odd trace anomaly also for a
Majorana fermion. This is
another possible pitfall induced by a careless use of formal manipulations. The
answer is the same as in section \ref{s:MW}: one cannot consider the passage
from $\psi$ to
$\psi_L$ in \eqref{Majo2} as an allowed field redefinition, because it changes
the integration measure.
Majorana and Weyl fermions have
their own appropriate actions, which faithfully represent their properties; in
each
case one must refer to the appropriate action, in particular, for Majorana
fermions one should avoid using the RHS of (\ref{Majo2}).

Finally there is one subtle issue that has been somehow understood so far. We
have stressed above that
the crucial ingredient in the calculation of anomalies is the functional
integral measure. We have also explained the problems connected with the latter
when chiral fermions are involved. In section 3 we
have employed a Feynman diagram technique, tacitly assuming that it reproduces
the correct path integration measure. Although this must be the case, because
the relevant Feynman diagrams (with chiral propagators and chiral vertices) are
different from those
for Dirac or Majorana fermions, it
is fair to say that we do not have a direct proof of it. There is however
a way to spell out any residual doubts concerning the path integration measure.
It relies in the analog of the method used by Bardeen, \cite{Bardeen}, for
chiral gauge anomalies, see also \cite{AB1}. In such an approach one employs
Dirac fermions (and,
consequently, the ordinary Dirac measure) and recovers the chiral
fermion theory as a subcase, by taking a specific limit. To this approach is
devoted the second part of the paper.

\vskip 3cm
{\Large Part II}
\vskip 1cm
In this second part we consider another approach to the odd trace anomaly,
similar to Bardeen's method to chiral gauge anomalies in
gauge theories, \cite{Bardeen,AB1,AB2}. The latter consists in introducing both
a
vector and an axial potential as external sources of a free Dirac fermion
theories in 4d. The usual consistent and covariant anomalies are obtained as
specific limits of this model. In order to transfer to gravity such a model we
need a
second metric, an axial metric, beside the usual one. We will call such a model 
Metric-Axial-Tensor (MAT) Gravity.

\section{Metric-Axial-Tensor Gravity}

\subsection{Axial metric}

We use the symbols $g_{\mu\nu}, g^{\mu\nu}$ and $e_\mu^a, e_a^\mu$ in the usual
sense of metric and vierbein and their inverses. Then we introduce the formal
writing{\footnote{{We use at times the
suggestive terminology axial-complex for an expression like $G_{\mu\nu}$,
axial-real for $g_{\mu\nu}$ and axial-imaginary for  $f_{\mu\nu}$. This alludes
to a geometrical interpretation, which is however not necessary to expand on in
this paper.}}} 
\be
G_{\mu\nu}=g_{\mu\nu}+\gamma_5 f_{\mu\nu}\label{G}
\ee
where $f$ is a symmetric tensor. Their background values are $\eta_{\mu\nu}$ and
0, respectively. So, to first order
\be
g_{\mu\nu}= \eta_{\mu\nu}+ h_{\mu\nu},\quad\quad f_{\mu\nu}=
k_{\mu\nu}\label{Gapprox}
\ee

In matrix notation the inverse of $G$, $G^{-1}$, is defined by
\be
G^{-1} = \hat g +\gamma_5\hat f,\quad\quad G^{-1} G=1,\quad\quad \hat
G^{\mu\lambda}G_{\lambda\nu}= \delta^\mu_\nu\label{GG-1}
\ee
which implies
\be 
\hat g f + \hat f g=0, \quad\quad \hat g g+\hat f f =1.\label{GG-12} 
\ee
That is
\be
\hat f=-\hat g f g^{-1},\quad\quad \hat g =
\left({g-fg^{-1}f}\right)^{-1}\label{GG-13} 
\ee
So
\be
\hat g=(1-g^{-1}\,f g^{-1}f)^{-1} g^{-1},\quad\quad  {\hat f}=-(1-g^{-1}f
\,g^{-1}f)^{-1} g^{-1}f\, g^{-1}\label{G-12}
\ee
Keeping up to second order terms:
\be 
g^{\mu\nu} &=& \eta^{\mu\nu} -h^{\mu\nu}+ h^\mu_\lambda h^{\lambda\nu}+\ldots
\0\\
\hat g^{\mu\nu} &=&  \eta^{\mu\nu} -h^{\mu\nu}+ h^\mu_\lambda h^{\lambda\nu}+
k^\mu_\lambda k^{\lambda\nu}+\ldots\0\\
\hat f^{\mu\nu} &=& -k^{\mu\nu} +  h^\mu_\lambda k^{\lambda\nu}+ k^\mu_\lambda
h^{\lambda\nu}+\ldots\label{G1G2approx}
\ee

\subsection{MAT vierbein} 

Likewise for the vierbein one writes
\be
E^a_\mu = e^a_\mu+ \gamma_5 c^a_\mu,\quad\quad \hat E_a^\mu =\hat e_a^\mu+
\gamma_5\hat c_a^\mu\label{vier1}
\ee
This implies
\be
\eta_{ab}\left(e^a_\mu e^b_\nu + c^a_\mu c^b_\nu\right)= g_{\mu\nu},
\quad\quad \eta_{ab}\left(e^a_\mu c^b_\nu + e^a_\nu c^b_\mu\right)=
f_{\mu\nu}\label{vier2}
\ee
Moreover, from $\hat E_a^\mu E_\nu^a=\delta^\mu_\nu$,
\be
\hat e_a^\mu c^a_\nu+\hat c_a^\mu e^a_\nu =0,\quad\quad \hat e_a^\mu
e^a_\nu+\hat c_a^\mu c^a_\nu =\delta^\mu_\nu,\label{vier3}
\ee
one gets
\be
\hat e_a^\mu = \left(\frac 1{1-e^{-1}c \,e^{-1} c}e^{-1}\right)_a^\mu
\label{vier4}
\ee
and
\be
\hat c_a^\mu = -\left(e^{-1}c \frac 1{1-e^{-1}c \,e^{-1} c} e^{-1}c
e^{-1}\right)_a^\mu 
\label{vier5}
\ee

In accord with (\ref{Gapprox}) we have
\be
&&e_\mu^a=\delta_\mu^a +\frac 12 h_\mu^a-\frac 18 \left(h
h+kk\right)^a_\mu+\frac 1{16}(h^3+khk+hk^2+k^2h)^a_\mu+\ldots\label{Eapprox}\\
&&\hat e^\mu_a=\delta^\mu_a -\frac 12 h^\mu_a+ \frac 38 \left(hh+ k
k\right)_a^\mu-\frac 5{16}(h^3+khk+hk^2+k^2h)_a^\mu+ \ldots \0\\
&& c_\mu^a= \frac 12 k_\mu^a-\frac 18 \left(h k   +k h\right)^a_\mu +\frac 1{16}
(k^3+hkh+h^2k+kh^2)^a_\mu+\ldots \0\\
&&\hat c^\mu_a=- \frac 12 k^\mu_a+\frac 1{16}\left(h  k   +k
h\right)^\mu_a-\frac
5{16} (k^3+hkh+h^2k+kh^2)_a^\mu+ \ldots\0
\ee
or
\be
E^a_\mu &=& \delta_\mu^a +\frac 12 h_\mu^a-\frac 18 \left(h
h+kk\right)^a_\mu+\gamma_5 \left(\frac 12  k_\mu^a
-\frac 18 \left(h  k   +k  h\right)^a_\mu \right)+\ldots\label{Eapprox2}\\
\hat E_a^\mu &=& \delta^\mu_a -\frac 12 h^\mu_a+ \frac 38 \left(h h+ k
k\right)_a^\mu-  
\gamma_5\left(\frac 12  k^\mu_a-\frac 38\left(h  k   +k   h \right)_a^\mu\right)
+\ldots\0
\ee

\subsection{Christoffel and Riemann}

The ordinary Christoffel symbols are
\be
\gamma_{\mu\nu}^\lambda= \frac 12 g^{\lambda\rho}\left( \partial_\mu
g_{\rho\nu}+\partial_\nu g_{\rho\mu} -\partial_\rho g_{\mu\nu}\right)
\label{chris}
\ee
The MAT Christoffel symbols are  defined in a similar way
\be
\Gamma_{\mu\nu}^\lambda&=& \frac 12 \hat G^{\lambda\rho}\left( \partial_\mu
G_{\rho\nu}+\partial_\nu G_{\rho\mu} -\partial_\rho G_{\mu\nu}\right)
\label{Chris}\\
&=&\frac 12 \left( \hat g^{\lambda\rho}\left( \partial_\mu
g_{\rho\nu}+\partial_\nu g_{\rho\mu} -\partial_\rho g_{\mu\nu}\right)+
 \hat f^{\lambda\rho}\left( \partial_\mu f_{\rho\nu}+\partial_\nu f_{\rho\mu}
-\partial_\rho f_{\mu\nu}\right)\right)\0\\
&&+\frac 12 \gamma_5\left( \hat g^{\lambda\rho}\left( \partial_\mu
f_{\rho\nu}+\partial_\nu f_{\rho\mu} -\partial_\rho f_{\mu\nu}\right)+
 \hat f^{\lambda\rho}\left( \partial_\mu g_{\rho\nu}+\partial_\nu g_{\rho\mu}
-\partial_\rho g_{\mu\nu}\right)\right)\0\\
&\equiv& \Gamma^{(1)\lambda}_{\mu\nu} + \gamma_5 \Gamma^{(2)\lambda}_{\mu\nu}\0
\ee
Up to  order two in $h$ and $k$ these become
\be
\Gamma^{(1)\lambda}_{\mu\nu}\!&=&\! \frac 12 \Big{(} \partial_\mu h_\nu^\lambda+
\partial_\nu h_\mu^\lambda- \partial^\lambda h_{\mu\nu}\0\\
&&-
h^{\lambda\rho}\left(\partial_\mu h_{\nu\rho}+ \partial_\nu h_{\mu\rho}-
\partial_\rho h_{\mu\nu} \right)-
k^{\lambda\rho}\left(\partial_\mu k_{\nu\rho}+ \partial_\nu k_{\mu\rho}-
\partial_\rho k_{\mu\nu} \right)\Big{)}+\ldots\label{Gamma1}\\
\Gamma^{(2)\lambda}_{\mu\nu}\!&=&\! \frac 12 \Big{(} \partial_\mu k_\nu^\lambda+
\partial_\nu k_\mu^\lambda- \partial^\lambda k_{\mu\nu}\0\\
&&-
h^{\lambda\rho}\left(\partial_\mu k_{\nu\rho}+ \partial_\nu k_{\mu\rho}-
\partial_\rho k_{\mu\nu} \right)-
k^{\lambda\rho}\left(\partial_\mu h_{\nu\rho}+ \partial_\nu h_{\mu\rho}-
\partial_\rho h_{\mu\nu} \right)\Big{)}+\ldots\label{Gamma2}
\ee

Proceeding the same way one can define the MAT Riemann tensor via ${\cal
R}_{\mu\nu\lambda}{}^\rho $:
\be
{\cal R}_{\mu\nu\lambda}{}^\rho &=& -\partial_\mu \Gamma_{\nu\lambda}^\rho +
\partial_\nu \Gamma_{\mu\lambda}^\rho- \Gamma_{\mu\sigma}^\rho
\Gamma_{\nu\lambda}^\sigma +
\Gamma_{\nu\sigma}^\rho\Gamma_{\mu\lambda}^\sigma\label{Riem}\\
&=& -\partial_\mu \Gamma_{\nu\lambda}^{(1)\rho } + \partial_\nu
\Gamma_{\mu\lambda}^{(1)\rho }- \Gamma_{\mu\sigma}^{(1)\rho}
\Gamma_{\nu\lambda}^{(1)\sigma} +
\Gamma_{\nu\sigma}^{(1)\rho}\Gamma_{\mu\lambda}^{(1)\sigma} 
-\Gamma_{\mu\sigma}^{(2)\rho} \Gamma_{\nu\lambda}^{(2)\sigma} +
\Gamma_{\nu\sigma}^{(2)\rho}\Gamma_{\mu\lambda}^{(2)\sigma}\0\\
&&+\gamma_5 \Big{(}  -\partial_\mu \Gamma_{\nu\lambda}^{(2)\rho } + \partial_\nu
\Gamma_{\mu\lambda}^{(2)\rho }- \Gamma_{\mu\sigma}^{(1)\rho}
\Gamma_{\nu\lambda}^{(2)\sigma} +
\Gamma_{\nu\sigma}^{(1)\rho}\Gamma_{\mu\lambda}^{(2)\sigma} 
-\Gamma_{\mu\sigma}^{(2)\rho} \Gamma_{\nu\lambda}^{(1)\sigma} +
\Gamma_{\nu\sigma}^{(2)\rho}\Gamma_{\mu\lambda}^{(1)\sigma}\Big{)}\0\\
&\equiv& {\cal R}^{(1)}_{\mu\nu\lambda}{}^\rho+\gamma_5 {\cal
R}^{(2)}_{\mu\nu\lambda}{}^\rho\0
\ee
The MAT spin connection is introduced in analogy 
\be
\Omega_\mu^{ab} &=& E_\nu^a\left(\partial_\mu\hat  E^{\nu b}+\hat  E^{\sigma b}
\Gamma^\nu_{\sigma \mu}\right) \label{spinconn1}\\
&=& \Omega_\mu^{(1)ab} +\gamma_5 \Omega_\mu^{(2)ab}\0
\ee
where
\be
\Omega_\mu^{(1)ab}&=& e^a_\nu \left(\partial_\mu \hat e^{\nu b}+ \hat e^{\sigma
b} \Gamma^{(1)\nu}_{\sigma \mu}+ \hat c^{b\sigma} 
\Gamma^{(2)\nu}_{\sigma \mu}\right)
+ c^a_\nu \left(\partial_\mu \hat c^{\nu b}+ \hat e^{\sigma b}
\Gamma^{(2)\nu}_{\sigma \mu}+ \hat c^{b\sigma} 
\Gamma^{(1)\nu}_{\sigma \mu}\right)\label{Omega1}\\
\Omega_\mu^{(2)ab}&=& e^a_\nu \left(\partial_\mu \hat c^{\nu b}+ \hat e^{\sigma
b} \Gamma^{(2)\nu}_{\sigma \mu}+ \hat c^{b\sigma} 
\Gamma^{(1)\nu}_{\sigma \mu}\right)
+ c^a_\nu \left(\partial_\mu \hat e^{\nu b}+ \hat e^{\sigma b}
\Gamma^{(1)\nu}_{\sigma \mu}+ \hat c^{b\sigma} 
\Gamma^{(2)\nu}_{\sigma \mu}\right)\label{Omega2}
\ee

\subsection{Transformations. Diffeomorphisms}

Under diffeomorphisms, $\delta x^\mu=\xi^\mu$, the Christoffel symbols transform
as tensors except for one non-covariant piece
\be
\delta_\xi^{(n.c.)}\gamma_{\mu\nu}^\lambda=
\partial_\mu\partial_\nu\xi^\lambda\label{noncovgamma}
\ee
The same happens for the MAT Christoffel symbols
\be
\delta_\xi^{(n.c.)}\Gamma_{\mu\nu}^\lambda=
\partial_\mu\partial_\nu\xi^\lambda\label{noncovGamma}
\ee
This means in particular that $\Gamma^{(2)\lambda}_{\mu\nu}$ is a tensor. 

{It is more convenient to introduces also axial diffeomorphisms
and use the following compact notation. The axially-extended (AE)
diffeomorphisms are defined by
\be
x^\mu\rightarrow x^\mu+\Xi^\mu, \quad\quad \Xi^\mu=\xi^\mu+\gamma_5
\zeta^\mu\label{axialdiff}
\ee
Since operationally these transformations act the same way as the usual
diffeomorphisms, it is easy to obtain
for the non-covariant part
\be
\delta^{(n.c.)}\Gamma_{\mu\nu}^\lambda=
\partial_\mu\partial_\nu\Xi^\lambda\label{deltancGamma}
\ee
We can also write 
\be
\delta_\Xi G_{\mu\nu} ={\cal D}_\mu \Xi_\nu+ {\cal D}_\nu \Xi_\mu
\ee}
where $\Xi_\mu = G_{\mu\nu} \Xi^\nu$.

In components one easily finds
\be
\delta_\xi g_{\mu\nu} &=&\xi^\lambda \partial_\lambda  g_{\mu\nu} + \partial_\mu
\xi^\lambda g_{\lambda\nu}+
\partial_\nu \xi^\lambda g_{\lambda\mu}\label{deltaxi}\\
\delta_\xi f_{\mu\nu} &=&\xi^\lambda \partial_\lambda  f_{\mu\nu} + \partial_\mu
\xi^\lambda f_{\lambda\nu}+
\partial_\nu \xi^\lambda f_{\lambda\mu}\0\\
\delta_\zeta g_{\mu\nu} &=&\zeta^\lambda \partial_\lambda  f_{\mu\nu} +
\partial_\mu \zeta^\lambda f_{\lambda\nu}+
\partial_\nu \zeta^\lambda f_{\lambda\mu}\label{deltazeta}\\
\delta_\zeta f_{\mu\nu} &=&\zeta^\lambda \partial_\lambda  g_{\mu\nu} +
\partial_\mu \zeta^\lambda g_{\lambda\nu}+
\partial_\nu \zeta^\lambda g_{\lambda\mu}\0
\ee

Summarizing
\be
&&\delta_\xi^{(n.c.)}\Gamma^{(1)\lambda}_{\mu\nu}=
\partial_\mu\partial_\nu\xi^\lambda,
\quad\quad\delta_\xi^{(n.c.)}\Gamma^{(2)\lambda}_{\mu\nu}=0\label{noncovGamma1}
\\
&&\delta_\zeta^{(n.c.)}\Gamma^{(1)\lambda}_{\mu\nu}=0, \quad\quad 
\delta_\zeta^{(n.c.)}\Gamma^{(2)\lambda}_{\mu\nu} =
\partial_\mu\partial_\nu\zeta^\lambda\0
\ee
and the overall Riemann and Ricci tensors are tensor, and the Ricci
scalar ${\cal R}$ is a scalar. But also ${\cal R}^{(1)}$ and ${\cal R}^{(2)}$,
separately, have the same tensorial properties.

\subsection{Transformations. Weyl transformations}

There are two types of Weyl transformations. The first is the obvious one
\be
G_{\mu\nu} \longrightarrow e^{2\omega}  G_{\mu\nu}, \quad\quad \hat G^{\mu\nu}
\to e^{-2\omega}\hat  G^{\mu\nu}\label{Weyl1G}
\ee
and 
\be
E_{\mu}^a \longrightarrow e^{\omega}  E_{\mu}^a, \quad\quad\hat  E^{\mu}_a \to
e^{-\omega}\hat  E^{\mu}_a\label{Weyl1E}
\ee
This leads to the usual relations
\be
\Gamma_{\mu\nu}^\lambda \longrightarrow \Gamma_{\mu\nu}^\lambda+ \partial_\mu
\omega\, \delta_\nu^\lambda + \partial_\nu \omega\, \delta_\mu^\lambda -
\partial_\rho \omega \,\hat  G^{\lambda\rho}G_{\mu\nu} \label{Weyl1Gamma}
\ee
and
\be
\Omega_\mu^{ab} \longrightarrow\Omega_\mu^{ab} + \left( E_\mu^a \hat E^{\sigma
b} - E_\mu^b\hat  E^{\sigma a} \right) \partial_\sigma \omega
\label{Weyl1Omega}
\ee
For infinitesimal $\omega$ this implies
\be
&&\delta_\omega g_{\mu\nu} = 2\omega \, g_{\mu\nu},\quad\quad \delta_\omega
f_{\mu\nu} = 2\omega \, f_{\mu\nu}\label{Weyl1gf}\\
&& \delta_\omega^{(0)} h_{\mu\nu} = 2 \omega \eta_{\mu\nu},\quad\quad
\delta_\omega^{(1)} h_{\mu\nu} = 2 \omega h_{\mu\nu},\ldots\0\\
&& \delta_\omega^{(0)} k_{\mu\nu} = 0,\quad\quad \delta_\omega^{(1)} k_{\mu\nu}
= 2 \omega k_{\mu\nu},\ldots\0
\ee

The second type of Weyl transformation is the axial one
\be
G_{\mu\nu} \longrightarrow e^{2\gamma_5 \eta}  G_{\mu\nu}, \quad\quad\hat 
G^{\mu\nu} \to e^{-2\gamma_5 \eta}\hat  G^{\mu\nu}\label{Weyl2G}
\ee
and 
\be
E_{\mu}^a \longrightarrow e^{\gamma_5 \eta}  E_{\mu}^a, \quad\quad\hat 
E^{\mu}_a \to e^{-\gamma_5 \eta} \hat E^{\mu}_a\label{Weyl2E}
\ee
This leads to
\be
\Gamma_{\mu\nu}^\lambda \longrightarrow \Gamma_{\mu\nu}^\lambda+
\gamma_5\left(\partial_\mu \eta\, \delta_\nu^\lambda + 
\partial_\nu \eta\, \delta_\mu^\lambda -
\partial_\rho \eta \,\hat G^{\lambda\rho}G_{\mu\nu}\right) \label{Weyl2Gamma}
\ee
and
\be
\Omega_\mu^{ab} \longrightarrow\Omega_\mu^{ab} + \gamma_5 \left( E_\mu^a \hat
E^{\sigma b} - E_\mu^b\hat  E^{\sigma a} \right) \partial_\sigma \eta
\label{Weyl2Omega}
\ee
Eq.(\ref{Weyl2G}) implies 
\be
g_{\mu\nu} \longrightarrow \cosh (2\eta) \, g_{\mu\nu} + \sinh(2\eta) \,
f_{\mu\nu},\quad\quad 
f_{\mu\nu} \longrightarrow \cosh (2\eta) \, f_{\mu\nu} + \sinh(2\eta) \,
g_{\mu\nu}\label{getafeta}
\ee
which, for infinitesimal $\eta$ becomes
\be
&&\delta_\eta g_{\mu\nu} = 2\eta\, f_{\mu\nu},\quad\quad \delta_\eta^{(0)}
h_{\mu\nu} =0, 
\quad\quad \delta_\eta^{(1)} h_{\mu\nu} =2\eta\, k_{\mu\nu},\quad \ldots\0\\ 
&&\delta_\eta f_{\mu\nu} = 2\eta\, g_{\mu\nu},\quad\quad \delta_\eta^{(0)}
k_{\mu\nu} =2\eta \,\eta_{\mu\nu}, 
\quad\quad \delta_\eta^{(1)} k_{\mu\nu} =2\eta\, h_{\mu\nu},\quad
\ldots\label{Weyl2fg}
\ee
 
\subsection{Volume density}

{The ordinary density $\sqrt{|g|}$ is replaced by 
\be
{\sqrt{|G|} =\sqrt{\det (G)} = \sqrt{\det (g+\gamma_5 f)}} \label{volume1}
\ee
The expression in the RHS has to be understood as a formal Taylor expansion in
terms of the
axial-complex variable $g+\gamma_5 f$. This means
\be
 \tr\ln ( g+\gamma_5 f) &=&\tr \ln g +\tr \ln\left(1 + \gamma_5 (g^{-1}f)\right)\0\\
&=& \tr \ln g 
+\frac 12\,  \tr\ln \left(1- (g^{-1}f)^2 \right) +\gamma_5 \, \tr  \, {\rm arcth}    (g^{-1}f)
\label{volume2}\\
 &=& \frac {1+\gamma_5}2\,\tr \ln (g+f)  +\frac {1-\gamma_5}2\,  \tr\ln (g-f)\0 
\ee 
It follows that
\be 
\sqrt{|G|} &=& e^{\frac 12 \tr\ln ( g+\gamma_5 f)} = e^{\frac 12\left(\frac
{1+\gamma_5}2 \tr\ln (g+f)  +\frac {1-\gamma_5}2\tr \ln (g-f)\right)}\0\\
&=& \frac 12 \left(\sqrt{\det(g+f)} +\sqrt{\det(g-f)}\right) + \frac {\gamma_5}2
   \left(\sqrt{\det(g+f)} -\sqrt{\det(g-f)}\right) 
\label{volume3}
\ee
$\sqrt{|G|}$ has the basic property that, under diffeomorphisms,
\be
\delta_\xi \sqrt{|G|}= \xi^\lambda \partial_\lambda \sqrt{|G|} +
\sqrt{|G|}\,\partial_\lambda \xi^\lambda\label{volume4}
\ee
This is a volume density, and has the following properties 
\be
\sqrt{|G|}\rightarrow e^{4\omega}\sqrt{|G|}, \sqrt{|G|}\rightarrow e^{4\eta
\gamma_5}\sqrt{|G|},\label{volume5}
\ee
under Weyl and axial-Weyl transformations, respectively. Moreover
\be 
\frac 1{\sqrt{|G|} }\partial_\nu  \sqrt{|G|} =\frac 12  \hat G^{\mu\lambda}
\partial_\nu G_{\mu\lambda}=\Gamma_{\mu\nu}^\mu\label{Gammamumunu}
\ee}

\section{Axial fermion theories}

{From the above it is evident that the action for fermion a
fermion field in interaction with MAT cannot be written in the classical form
$\int d^4x\, \sqrt{|g|} \overline \psi {\cal O} \psi$, as in the case of
ordinary gravity, where ${\cal O}$ is the usual operatorial kinetic operator in
the presence of gravity, because in the MAT case $\sqrt |G|$ contains the
$\gamma_5$ matrix. Instead, $\sqrt |G|$ must be inserted between $\overline\psi$
and $\psi$. Moreover we have to take into account that the that the kinetic
operator contains a $\gamma$ matrix that anticommutes with $\gamma_5$. Thus, for
instance, using ${\cal D}_\lambda G_{\mu\nu}=0$ and $({\cal D}_\lambda +\frac 12
\Omega_\lambda) E=0$, where ${\cal D}=\partial+\Gamma$, one gets
 \be
{\overline {\psi}} \gamma^a\hat E_a^m
\left(\partial_m+\frac 12 \Omega_m \right)\psi = \overline {\psi} (\bar {D}_\mu
+\frac 12  \bar  {\Omega}_m) \gamma^a\hat E_a^m\psi\label{equation}
\ee
where a bar denotes axial-complex conjugation, i.e. a sign reversal in front of
each $\gamma_5$ contained in the expression, for instance   $ \bar  \Omega_m=
\Omega_m^{(1)}-\gamma_5 \Omega_m^{(2)}$. 
The reader should be aware that, in particular, a concise notation like ${\cal
D}_\mu \gamma^\lambda$ is ambiguous. 
The MAT fermion action is now
\be
S&=& \int d^4x \, i\overline {\psi} \sqrt{|\bar G|}\gamma^a\hat E_a^m
\left(\partial_m+\frac 12 \Omega_m \right)\psi  \label{axialaction}\\
&{=}&  \int d^4x \,  i\overline {\psi} \sqrt{|\bar G|}\gamma^a(\hat e_a^m+\gamma_5
\hat c_a^m)  \left(\partial_m +\frac 12 \left(\Omega^{(1)}_m+\gamma_5
\Omega^{(2)}_m\right) \right)\psi \0\\
&=&\int d^4x \, \overline {\psi} \sqrt{|\bar G|}(\hat e_a^m-\gamma_5 \hat
c_a^m)\left[\frac  i2 \gamma^a {\stackrel{\leftrightarrow} 
{\partial}}_m + {\frac i4} \left( \gamma^a \Omega_m + \bar  \Omega_m
\gamma^a\right) \right]\psi\0\\
&=&\int d^4x \, \overline {\psi} \sqrt{|\bar G|}(\hat e_a^m-\gamma_5 \hat
c_a^m)\left[\frac  i2 \gamma^a {\stackrel{\leftrightarrow} 
{\partial}}_m  -\frac 14 \epsilon^{abcd}
\left( \Omega^{(1)}_{m bc} \gamma_d\gamma_5+ \Omega^{(2)}_{m bc} \gamma_d
\right)\right]\psi\0
\ee
where it is understood that $\partial_m$ applies only to $\psi$ or $\overline
\psi$, as indicated, and $\bar G$ denotes the axial-complex conjugate. To obtain
this one must use \eqref{Gammamumunu} and \eqref{equation}.}

\subsection{{Classical Ward identities}}

{Let us consider AE diffeomorphisms first, \eqref{axialdiff}.
It is not hard to prove that the action \eqref{axialaction} is invariant under
these transformations. 
Now, define the full MAT e.m. tensor by means of
\be
{\bf T}^{\mu\nu} = \frac 2{\sqrt{|G|}} \frac {\stackrel{\leftarrow} {\delta} S}
{\delta G_{\mu\nu}} \label{fullem}
\ee
This formula needs a comment, since $\sqrt{|G|}$ contains $\gamma_5$. To give a
meaning to it we understand that
the operator $\frac 2{\sqrt{|G|}} \frac {\stackrel{\leftarrow} {\delta} }
{\delta G_{\mu\nu}}$  in the RHS acts on the operatorial expression, say 
${\cal O}{\sqrt{|G|}}$, which is inside the scalar product, i.e. $\overline \psi
{\cal O}\sqrt{|G|} \psi$. 
Moreover the functional derivative acts from the right of the action. 
Now the conservation law under diffeorphisms is
\be
0=\delta_\Xi S &=& \int \overline \psi \frac { \stackrel{\leftarrow} {\delta}
{\cal O}} {\delta G_{\mu\nu}}\delta G_{\mu\nu}\psi  
=\int \overline \psi\frac {\stackrel{\leftarrow} {\delta} {\cal O}} {\delta
G_{\mu\nu}} \left({\cal D}_\mu \Xi_\nu+ {\cal D}_\nu \Xi_\mu\right)\psi
\0\\
&=& -2 \int \overline \psi \frac {\stackrel{\leftarrow} {\delta} {\cal O}}
{\delta G_{\mu\nu}}  {\stackrel{\leftarrow} {\cal D}}_\mu
\Xi_\nu\psi\label{deltaXi}
\ee
where ${\cal D}$ acts (from the right) on everything except the parameter
$\Xi_\nu$. Differentiating with respect to the arbitrary
parameters $\xi^\mu$ and $\zeta^\nu$ we obtain two conservation laws involving
the two tensors
\be
T^{\mu\nu}&=&{2}  \overline \psi \frac {\stackrel{\leftarrow} {\delta} {\cal O}}
{\delta G_{\mu\nu}}\psi\label{Tmunu}\\
T_5^{\mu\nu} &=& {2} \overline \psi \frac {\stackrel{\leftarrow} {\delta} {\cal O}}
{\delta G_{\mu\nu}}\gamma_5\psi\label{T5munu}
\ee
At the lowest order the latter are given by eqs.\eqref{Tmunu0},\eqref{T5munu0}
below.}

{
Repeating the same derivation for the axial complex Weyl trasformation one can
prove that, assumimg for the fermion field the trasnformation rule
\be
\psi \rightarrow e^{-\frac 32 (\omega+\gamma_5 \eta)} \psi,\label{Weylpsi}
\ee
\eqref{axialaction} is invariant and obtain the Ward identity
\be
0= \int \overline \psi \frac {\stackrel{\leftarrow} {\delta} {\cal O}} {\delta
G_{\mu\nu}} G_{\mu\nu} \,(\omega+\gamma_5 \eta)\psi \label{WeylWI}
\ee 
We obtain in this way two WI's
\be
&&T^{\mu\nu} g_{\mu\nu} + T_{5}^{\mu\nu} f_{\mu\nu}=0, \label{WardWeyl1}\\
&&T^{\mu\nu} f_{\mu\nu} + T_{5}^{\mu\nu} g_{\mu\nu}=0, \label{WardWeyl2}
\ee}

\subsection{A simplified version}

{A simplified approach to
the trace anomaly calculation consists first in absorbing $\sqrt{|G|}$ in $\psi$
by setting $\Psi =
|G|^{\frac 14} \psi$ and thereby assuming the transformation properties 
\be
\delta_\Xi \Psi= \Xi^\mu\partial_\mu \Psi +\frac 12 D_\mu \Xi^\mu
\Psi\label{Psidiff}
\ee
for AE diffeomorphisms, and
\be
 \delta_{\omega+\gamma_5} \Psi=e^{\frac 12 {\omega +\gamma_5 \eta}} \Psi, 
\label{PsiWeyl}
\ee
for axial-complex Weyl transformations.}

To arrive at an expanded action one uses (\ref{Gapprox},\ref{Eapprox}), up to
second order, and
finds
\be 
 \Omega_\mu^{(1)ab}&=&\frac 12 \left( \partial^b h_\mu^a-  \partial^a
h_\mu^b\right)+\frac 14 \left( h^{\sigma a} \partial_\sigma h_\mu^b
- h^{\sigma b} \partial_\sigma h_\mu^a+ h^{b\sigma} \partial^a h_{\sigma
\mu}-h^{a\sigma} \partial^b h_{\sigma \mu}\right)\0\\
&&-\frac 18\left( h^{a\sigma} \partial_\mu h_\sigma^b-  h^{b\sigma} \partial_\mu
h_\sigma^a\right) 
-\frac 18 \left( k^{a\sigma} \partial_\mu k_\sigma^b-  k^{b\sigma} \partial_\mu
k_\sigma^a\right)\0\\ 
&& +\frac 14 \left( k^{\sigma a} \partial_\sigma k_\mu^b
- k^{\sigma b} \partial_\sigma k_\mu^a+ k^{b\sigma} \partial^a k_{\sigma
\mu}-k^{a\sigma} \partial^b k_{\sigma \mu}\right)+\ldots\label{Omega1approx}
\ee
and 
\be 
\Omega_\mu^{(2)ab}&=&\frac 12 \left( \partial^b k_\mu^a-  \partial^a
k_\mu^b\right)+\frac 14 \left( h^{\sigma a} \partial_\sigma k_\mu^b
- h^{\sigma b} \partial_\sigma k_\mu^a+ h^{b\sigma} \partial^a k_{\sigma
\mu}-h^{a\sigma} \partial^b k_{\sigma \mu}\right)\0\\
&&-\frac 18\left( h^{a\sigma} \partial_\mu k_\sigma^b- h^{b\sigma} \partial_\mu
k_\sigma^a\right) 
-\frac 18 \left( k^{a\sigma} \partial_\mu h_\sigma^b-  k^{b\sigma}
\partial_\mu h_\sigma^a\right)\0\\
&&  +\frac 14 \left( k^{\sigma a} \partial_\sigma h_{\mu}^b
-  k^{\sigma b} \partial_\sigma h_{\mu}^b+ k^{b\sigma} \partial^a h_{\sigma
\mu}-k^{a\sigma} \partial^b h_{\sigma \mu}\right)+\ldots
\label{Omega2approx}
\ee
In particular
\be
\epsilon^{\mu abc} \Omega^{(1)}_{\mu ab} &=& -\frac 14 \epsilon^{\mu abc }
\left( h_a^\sigma \partial_b h_{\mu\sigma}
+k_a^\sigma \partial_b k_{\mu\sigma}\right)+\ldots \label{epsilonOmega1}\\
\epsilon^{\mu abc} \Omega^{(2)}_{\mu ab} &=& -\frac 14 \epsilon^{\mu abc} \,(
h_a^\sigma \partial_b k_{\mu\sigma} 
+k_a^\sigma \partial_b h_{\mu\sigma})+\ldots\label{epsilonOmega2}
\ee

Up to order two in $h$ and $k$ we have
\be 
S&=&{ \int d^4x \, \overline {\psi}|\bar G|^{\frac 14}(\hat e_a^m-\gamma_5 \hat
c_a^m) \left[\frac  i2 \gamma^a {\stackrel{\leftrightarrow} 
{\partial}}_m  -\frac 14 \epsilon^{abcd}
\left( \Omega^{(1)}_{m bc} \gamma_d\gamma_5+ \Omega^{(2)}_{m bc} \gamma_d
\right)\right]|G|^{\frac 14}\psi} \0\\ 
&=& \int d^4x \, \Big{[} \frac{i}{2} \overline {\Psi}\gamma^m
{\stackrel{\leftrightarrow}{\partial}}_m  \Psi   -\frac i4 
 \overline {\Psi}(h^m_a-\gamma_5 k^m_a) 
\gamma^a {\stackrel{\leftrightarrow}{\partial}}_m  \Psi 
\label{2ndorderaction}\\
&& +\frac {3i}{16}
 \overline {\Psi}\left( (k^2)^m_a+(h^2)^m_a- \gamma_5  (hk+kh)^m_a\right) 
\gamma^a {\stackrel{\leftrightarrow}{\partial}}_m  \Psi \0\\
&& +\frac 1{16} \epsilon^{m abc} \overline {\Psi}\left(\left( h_a^\sigma
\partial_b h_{m\sigma} + 
k_a^\sigma \partial_b k_{m\sigma}\right)\gamma_c\gamma_5 +(h_a^\sigma\partial_b
k_{m\sigma} +k_a^\sigma\partial_b h_{m\sigma} )\gamma_c\right)\Psi\0\\
&&+ \frac 1{8} \epsilon^{ abcd}\overline {\Psi}(h_a^m-\gamma_5 k_a^m) \left(
\partial_c h_{bm} \gamma_d \gamma_5 +
\partial_c k_{bm} \gamma_d\right)\Psi\Big{]}+\ldots\0\\
&=& \int d^4x \, \Big{[} \frac{i}{2} \overline {\Psi}\gamma^m
{\stackrel{\leftrightarrow}{\partial}}_m  \Psi   -\frac i4 
 \overline {\Psi}(h^m_a{-}\gamma_5 k^m_a) 
\gamma^a {\stackrel{\leftrightarrow}{\partial}}_m  \Psi  \0\\
&& +\frac {3i}{16}
 \overline {\Psi}\left( (k^2)^m_a+(h^2)^m_a- \gamma_5 
(hk+kh)^m_a\right) 
\gamma^a {\stackrel{\leftrightarrow}{\partial}}_m  \Psi \0\\
&& -\frac 1{16} \epsilon^{m abc} \overline {\Psi}\left(\left( h_a^\sigma
\partial_b h_{m\sigma} + 
k_a^\sigma \partial_b k_{m\sigma}\right)\gamma_c\gamma_5 +(h_a^\sigma\partial_b
k_{m\sigma} +k_a^\sigma\partial_b h_{m\sigma}
)\gamma_c\right)\Psi\Big{]}+\ldots\0
\ee
Here we do not report explicitly the terms cubic in $h$ and $k$: they contains
three powers of $h$ and/or $k$  multiplied by $\overline \Psi \gamma_\mu \Psi$
or $\overline \Psi \gamma_\mu \gamma_5\Psi$ and possibly by the $\epsilon$
tensor. They contain one single derivative, applied to either $h,k$ or $\Psi$.
These cubic terms will not affect our results.

\subsection{Feynman rules}

For a derivation of the Feynman rules in this case see \ref{ss:MATbackground}
and \ref{ss:oneloopaxial}.
The fermion propagator is
\be 
\frac i{\slash \!\!\! p+i\epsilon }\label{chprop}
\ee
The two-fermion-h-graviton vertex is ($V_{ffh}$):
\be
-\frac i{8} \left[(p+p')_\mu \gamma_\nu + (p+p')_\nu \gamma_\mu\right]
\label{Vffh}
\ee
The axial two-fermion-k-graviton vertex is ($V_{ffk}$):
\be
-\frac i{8} \left[(p+p')_\mu \gamma_\nu + (p+p')_\nu \gamma_\mu \right]
\gamma_5\label{Vffk}
\ee
($p$ incoming, $p'$ outgoing).
There are 6 2-fermion-2-graviton vertices:

1) $V^{(1)}_{ffhh}$:  

\be
&&\frac {3i}{64} \bigg{[}\left( (p+p')_\mu \gamma_{\mu'} \eta_{\nu\nu'} + 
(p+p')_{\mu} \gamma_{\nu'} \eta_{\nu\mu'}+ \{\mu\leftrightarrow \nu\}\right)\0\\
&&+ \left(  (p+p')_{\mu'} \gamma_{\mu} \eta_{\nu\nu'}+ (p+p')_{\mu'}
\gamma_{\nu} \eta_{\mu\nu'}+ \{\mu'\leftrightarrow
\nu'\}\right)\bigg{]}\label{Vffhh}
\ee

2) $V^{(2)}_{ffkk}$:

\be
&&\frac {3i}{64} \bigg{[}\left( (p+p')_\mu \gamma_{\mu'} \eta_{\nu\nu'} + 
(p+p')_{\mu} \gamma_{\nu'} \eta_{\nu\mu'}+ \{\mu\leftrightarrow \nu\}\right)\0\\
&&+ \left(  (p+p')_{\mu'} \gamma_{\mu} \eta_{\nu\nu'}+ (p+p')_{\mu'}
\gamma_{\nu} \eta_{\mu\nu'}+ \{\mu'\leftrightarrow
\nu'\}\right)\bigg{]}\label{Vffkk}
\ee

3) $V^{(3)}_{ffhk}$:

\be
&&\frac {3i}{64} \bigg{[}\left( (p+p')_\mu \gamma_{\mu'} \eta_{\nu\nu'} + 
(p+p')_{\mu} \gamma_{\nu'} \eta_{\nu\mu'}+ \{\mu\leftrightarrow
\nu\}\right)\0\\\
&&+  \left((p+p')_{\mu'} \gamma_{\mu} \eta_{\nu\nu'}+ (p+p')_{\mu'} \gamma_{\nu}
\eta_{\mu\nu'}+ \{\mu'\leftrightarrow \nu'\}\right)\bigg{]}\gamma_5\label{Vffhk}
\ee

4) $V^{(1)\epsilon}_{ffhh}$:

\be
\frac 1{64} \, t_{\mu\nu\mu'\nu'\kappa\lambda}\,(k-k')^\lambda\gamma^\kappa\,
\gamma_5\label{Veffhh}
\ee 
where $t$ is the tensor (\ref{t}).

5) $V^{(2)\epsilon}_{ffkk}$:

\be
\frac 1{64} \,
t_{\mu\nu\mu'\nu'\kappa\lambda}\,(k-k')^\lambda\gamma^\kappa\,\gamma_5\label{
Veffkk}
\ee 

6) $V^{(3)\epsilon}_{ffhk}$:

\be
\frac 1{64}\,  t_{\mu\nu\mu'\nu'\kappa\lambda}\,(k-k')^\lambda\gamma^\kappa
\label{Veffhk}
\ee 

The graviton momenta $k,k'$ are incoming.

As anticipated above, we dispense from writing down the vertices with three $h,
k$ legs. For the purposes of this paper it is possible to dispose of them with a
general argument, without entering detailed calculations.

\subsection{Trace anomalies - a simplified derivation}
\label{s:backenvelop}

We will now derive the odd parity trace anomalies in the model
(\ref{2ndorderaction}), by considering only the triangle diagram contributions
{and disregarding tadpoles and seagull terms.}
We will justify later on this simplified procedure.
 
The overall effective action is
\be
W[h,k]&=&W[0]+ \sum_{n,m=0} ^\infty \frac {i^{m+n-1}}{2^{n+m} n!m!} \int
\prod_{i=1}^n
dx_i h_{\mu_i\nu_i}(x_i)  \prod_{j=1}^m dy_j k_{\lambda_j\rho_j}(y_j)\0  \\
&&\cdot \langle0| {\cal T}T^{\mu_1\nu_1}(x_1)\ldots T^{\mu_n\nu_n}(x_n)
T_5^{\lambda_1\rho_1}(y_1) \ldots  T_5^{\lambda_m\rho_m
}(y_m)|0\rangle\label{Whk}
\ee
where, in the simplified version of this section, the $T$ operator in the
time-ordered amplitudes refer
to the classical ones, i.e.
\be
T^{\mu\nu}\equiv T_{(0,0)}^{\mu\nu}=-\frac i4 \left(\overline {\psi } \gamma^\mu
{\stackrel{\leftrightarrow}{\partial^\nu}}\psi
 + \mu\leftrightarrow \nu \right),\label{Tmunu0} 
\ee
and 
\be
T_5^{\mu\nu}\equiv T_{5(0,0)}^{\mu\nu}=\frac i4 \left(\overline {\psi}\gamma_5
\gamma^\mu  {\stackrel{\leftrightarrow}{\partial^\nu}}\psi
 + \mu\leftrightarrow \nu \right),\label{T5munu0} 
\ee

The quantum Ward identities for the Weyl and axial Weyl symmetry are obtained by
replacing the classical   e.m. tensor expressions with the one-loop one-point
functions in (\ref{WardWeyl1}) {and (\ref{WardWeyl2})} 
\be
\ET(x)\equiv\langle\!\langle T^{\mu\nu}\rangle\!\rangle g_{\mu\nu} +
\langle\!\langle T_{5}^{\mu\nu}\rangle\!\rangle f_{\mu\nu}=0,\quad\quad {\rm
i.e.} \quad\quad  \langle\!\langle T^\mu_\mu\rangle\!\rangle+ \ldots
=0\label{quantumWardWeyl1}
\ee
 and 
\be
\ET_5(x)\equiv\langle\!\langle T^{\mu\nu}\rangle\!\rangle f_{\mu\nu} +
\langle\!\langle T_{5}^{\mu\nu}\rangle\!\rangle g_{\mu\nu}=0,\quad\quad {\rm
i.e.} \quad\quad \langle\!\langle T_{5\mu}^\mu\rangle\!\rangle+ \ldots
=0\label{quantumWardWeyl2}
\ee
In the present simplified setup the relevant one-loop one-point functions are
\be
\langle\!\langle T^{\mu\nu}(x)\rangle \!\rangle &=& \sum_{n,m=0} ^\infty \frac
{i^{m+n}}{2^{n+m}n!m!} \int \prod_{i=1}^n dx_i h_{\mu_i\nu_i}(x_i)  
\prod_{j=1}^m dy_j k_{\lambda_j\rho_j}(y_j)  \0\\
&&\cdot \langle0| {\cal T}T^{\mu\nu}(x)T^{\mu_1\nu_1}(x_1)\ldots
T^{\mu_n\nu_n}(x_n) T_5^{\lambda_1\rho_1} (y_1)\ldots  T_5^{\lambda_m\rho_m
}(y_m)|0\rangle\label{1pointhk}
\ee
and
\be
\langle\!\langle T_5^{\mu\nu}(x)\rangle \!\rangle &=& \sum_{n,m=0} ^\infty \frac
{i^{m+n}}{2^{n+m}n!m!} \int \prod_{i=1}^n dx_i h_{\mu_i\nu_i}(x_i) 
 \prod_{j=1}^m dy_j k_{\lambda_j\rho_j}(y_j)  \0\\
&&\cdot \langle0| {\cal T}T_5^{\mu\nu}(x)T^{\mu_1\nu_1}(x_1)\ldots
T^{\mu_n\nu_n}(x_n)T_5^{\lambda_1\rho_1}(y_1) \ldots  T_5^{\lambda_m\rho_m
}(y_m)|0\rangle\label{1pointhk5}
\ee
In particular for the trace anomalies, at level ${\cal O}(h^2, hk, k^2)$, we
have 
\be
 \langle\!\langle T^{\mu}_\mu(x)\rangle \!\rangle^{(2)}  &=& -\frac 18 \int
dx_1dx_2
h_{\mu_1\nu_1}(x_1) h_{\mu_2\nu_2}(x_2)  \langle 0| {\cal
T}T_{\mu}^\mu(x)T^{\mu_1\nu_1}(x_1)T^{\mu_2\nu_2}(x_2)
|0\rangle \0\\
&& -\frac 14\int dx_1 dy h_{\mu_1\nu_1}(x_1) k_{\lambda\rho}(y)  \langle 0|
{\cal
T}T_{\mu}^\mu(x)T^{\mu_1\nu_1}(x_1)T_5^{\lambda\rho}(y)|0\rangle\label{Tmumu} \\
&& -\frac 18 \int dy_1 dy_2  k_{\lambda_1\rho_1}(y_1)  k_{\lambda_2\rho_2}(y_2) 
\langle 0| {\cal
T}T_{\mu}^\mu(x)T_5^{\lambda_1\rho_1}(y_1)T_5^{\lambda_2\rho_2}(y_2)|0\rangle\0
\ee
and
\be
 \langle\!\langle T_{5\mu}{}^{\mu}(x)\rangle \!\rangle^{(2)} &=& -\frac 18 \int
dx_1dx_2 h_{\mu_1\nu_1}(x_1) h_{\mu_2\nu_2}(x_2)  
\langle 0| {\cal T}T_{5\mu}{}^{\mu}(x)T^{\mu_1\nu_1}(x_1)T^{\mu_2\nu_2}(x_2)
|0\rangle \0\\
&& -\frac 14\int dxdy h_{\mu_1\nu_1}(x_1) k_{\lambda\rho}(y)  \langle 0| {\cal
T}T_{5\mu}{}^\mu(x)T^{\mu_1\nu_1}(x_1)T_5^{\lambda\rho}(y)|0\rangle 
\label{T5mumu}\\
&& -\frac 18 \int dy_1 dy_2  k_{\lambda_1\rho_1}(y_1)  k_{\lambda_2\rho_2}(y_2) 
\langle 0| {\cal
T}T_{5\mu}{}^{\mu}(x)T_5^{\lambda_1\rho_1}(y_1)T_5^{\lambda_2\rho_2}
(y_2)|0\rangle\0
\ee
It is clear that only the terms containing an odd number of $T_5$ will
contribute to the odd parity trace anomaly.

The three-point functions (\ref{Tmumu},\ref{T5mumu}) are given by the ordinary
triangle diagrams. All such diagrams give the same contribution
\be
\sim \left(k_1\!\cdot \! k_2 \, t_{\mu\nu\mu'\nu' \lambda \rho} -
t^{(21)}_{\mu\nu\mu'\nu' \lambda \rho}\right) k_1^\lambda k_2^\rho
\label{trianglediagram}
\ee
where 
\be 
t^{(21)}_{\mu\nu\mu'\nu'\kappa\lambda}=k_{2\mu}k_{1\mu'}
\epsilon_{\nu\nu'\kappa\lambda} 
+ k_{2\nu}k_{1\nu'}\epsilon_{\mu\mu'\kappa\lambda} +k_{2\mu}k_{1\nu'}
\epsilon_{\nu\mu'\kappa\lambda} 
+k_{2\nu}k_{1\mu'} \epsilon_{\mu\nu'\kappa\lambda}\label{t21}
\ee
Upon Fourier-anti-transforming and replacing in (\ref{Tmumu}) we get:
\be 
\langle\!\langle T^{ \mu}_{\mu}(x)\rangle\!\rangle^{(2)} =-2N
\epsilon^{\mu\nu\lambda \rho} \left(\partial_\mu\partial_\sigma h^\tau_\nu \,
\partial_\lambda\partial_\tau k_{\rho}^\sigma-
\partial_\mu\partial_\sigma h^\tau_\nu \, \partial_\lambda\partial^\sigma
k_{\tau\rho}\right)\label{final}
\ee
and in (\ref{T5mumu}) we get:
\be 
\langle\!\langle T_{5\mu}{}^{\mu}(x)\rangle\!\rangle^{(2)} &=&-2N\Big{[}\frac 12
\epsilon^{\mu\nu\lambda \rho} \left(\partial_\mu\partial_\sigma h^\tau_\nu \,
\partial_\lambda\partial_\tau h_{\rho}^\sigma-
\partial_\mu\partial_\sigma h^\tau_\nu \, \partial_\lambda\partial^\sigma
h_{\tau\rho}\right)\label{final5}\\
&&+\frac 12 \epsilon^{\mu\nu\lambda \rho} \left(\partial_\mu\partial_\sigma
k^\tau_\nu \, \partial_\lambda\partial_\tau k_{\rho}^\sigma-
\partial_\mu\partial_\sigma k^\tau_\nu \, \partial_\lambda\partial^\sigma
k_{\tau\rho}\right)\Big{]}\0
\ee
where $N$ is the constant that appears in front of the Pontryagin anomaly in 
\cite{BGL}, i.e. $N= \frac i{768 \pi^2}$. 

Covariantizing these expressions we get
\be
\Theta_\mu^\mu&\equiv&\int \omega \langle\!\langle T^{
\mu}_{\mu}(x)\rangle\!\rangle= N \int \omega \epsilon^{\mu\nu\lambda \rho}  
{\cal
R}^{(1)}_{\mu\nu}{}^{\sigma \tau} 
{\cal R}^{(2)}_{{\lambda\rho}\sigma \tau}\label{Weylanomaly1}
\ee
 {
\be
\Theta_{5\mu}{}^{\mu}&\equiv&\int\eta \langle\!\langle
T_{5\mu}{}^{\mu}(x)\rangle\!\rangle= \frac N2  \int \eta
\,\epsilon^{\mu\nu\lambda \rho}\left(  {\cal
R}^{(1)}_{\mu\nu}{}^{\sigma \tau} 
{\cal R}^{(1)}_{\lambda \rho\sigma \tau}+  {\cal R}^{(2)}_{\mu\nu}{}^{\sigma
\tau} 
{\cal R}^{(2)}_{\lambda \rho \sigma \tau}\right) \label{Weylanomaly2}
\ee}

The important remark is now that the odd parity trace anomaly, in an ordinary
theory of Weyl fermions, can be calculated using the above theory of Dirac
fermions coupled to MAT gravity and setting at the end 
$h_{\mu\nu}\to \frac {h_{\mu\nu}}2,k_{\mu\nu} \to \frac {h_{\mu\nu}}2$ and
$\omega=\eta$, for left-handed Weyl fermions, and
$h_{\mu\nu}\to \frac {h_{\mu\nu}}2,k_{\mu\nu} \to -\frac {h_{\mu\nu}}2$ 
for right-handed ones. We will refer to these as {\it collapsing limits}.

\subsection{What happens when $h_{\mu\nu}\to \frac {h_{\mu\nu}}2,k_{\mu\nu} \to
\frac {h_{\mu\nu}}2$.}

Let us show that in the collapsing limit $h_{\mu\nu}\to \frac
{h_{\mu\nu}}2,k_{\mu\nu} \to \frac {h_{\mu\nu}}2$ we have the following results:
\be
\Gamma_{\mu\nu}^{(1)\lambda} \to \frac 12
\gamma_{\mu\nu}^\lambda,\quad\quad\Gamma_{\mu\nu}^{(2)\lambda} \to \frac 12
\gamma_{\mu\nu}^\lambda\label{Gamma12limit}
\ee
This is evident in the approximate expressions (\ref{Gamma1},\ref{Gamma2}), but
it can be proved in general.
To order $n$ in the expansion of $h$ and $k$ of $\Gamma_{\mu\nu}^{(1)\lambda}$
we are going to have a first term of order $n$ in $h$ alone, then
$\left(\begin{matrix}  n \\2   \end{matrix} \right)$ of order $n-2$ in $h$ and
order 2 in $k$, then $\left(\begin{matrix}  n \\4 \end{matrix} \right)$ of order
$n-4$ in $h$ and order 4 in $k$, and so on, up to order $[n/2]$ in $h$. In the
collapsing limit, all these terms collapse to the first term of order $n$ in $h$
divided by $2^n$. In total they are
\be
\sum_{k=0}^{[n/2]}\left(\begin{matrix} n \\2k \end{matrix} \right)=
2^{n-1}\label{conto1}
\ee
Therefore they give the order $n$ term in $h$ of $\gamma_{\mu\nu}^{\lambda}$
divided by 2. A similar proof holds for $\Gamma_{\mu\nu}^{(2)\lambda}$. 

Looking at the definition (\ref{Riem}) of the curvatures ${\cal
R}^{(1)}_{\mu\nu\lambda}{}^\rho$ and ${\cal R}^{(2)}_{\mu\nu\lambda}{}^\rho$ one
easily sees that in the collapsing limit
\be
{\cal R}^{(1)}_{\mu\nu\lambda}{}^\rho\to \frac 12 R_{\mu\nu\lambda}{}^\rho,
\quad\quad
{\cal R}^{(2)}_{\mu\nu\lambda}{}^\rho\to \frac 12
R_{\mu\nu\lambda}{}^\rho,\label{R12limit}
\ee
where $R_{\mu\nu\lambda}{}^\rho$ is the curvature of $g_{\mu\nu}$.

In a similar way, using (\ref{Omega1approx},\ref{Omega2approx}), one can show
that
\be
 \Omega_\mu^{(1)ab}\to \frac 12  \omega_\mu^{ab},\quad\quad 
\Omega_\mu^{(2)ab}\to \frac 12  \omega_\mu^{ab}\label{Omega12limit}
\ee

Notice also that in the collapsing limit
\be
&&g_{\mu\nu}+f_{\mu\nu}= \eta_{\mu\nu}+h_{\mu\nu}+k_{\mu\nu}\rightarrow
g_{\mu\nu} \0\\
&& g_{\mu\nu}-f_{\mu\nu}= \eta_{\mu\nu}+h_{\mu\nu}-k_{\mu\nu}\rightarrow
\eta_{\mu\nu}\label{detgflimit}
\ee
so that
\be
\sqrt{|G|}\rightarrow{ \frac {1-\gamma_5}2+\frac {1+\gamma_5}2 \sqrt{|g|}},\label{detGlimit}
\ee
and
\be
E^a_m \to \delta^a_m\frac {1-\gamma_5}2 + e^a_m\, \frac {1+\gamma_5}2,\quad\quad
\hat E^m_a \to  \delta_a^m \frac {1-\gamma_5}2 +\hat e^m_a \, \frac
{1+\gamma_5}2. \label{vierbeinlimit}
\ee

>From the above follows that the action (\ref{2ndorderaction}) tends to
\be
S&=&\int d^4x \,   i \overline {\Psi}\gamma^a \hat E_a^m (\partial_m +\frac 12
\Omega_m )\Psi\label{actionlimit}
\\
&& \longrightarrow \int d^4x \, \left[ i\overline {\Psi}\gamma^m
\frac{1-\gamma_5}2\partial_m \Psi+ i\overline {\Psi}\gamma^a\hat e_a^m
\left(\partial_m +\frac 12 \omega_m \right)\frac{1+\gamma_5}2\Psi\right]  \0
\ee

As for the opposite handedness one notices that, if $h_{\mu\nu}\to \frac
{h_{\mu\nu}}2,k_{\mu\nu} \to -\frac {h_{\mu\nu}}2$, we have  
\be
 \Omega_\mu^{(1)ab}\to \frac 12  \omega_\mu^{ab},\quad\quad 
\Omega_\mu^{(2)ab}\to -\frac 12  \omega_\mu^{ab}\label{Omega12limit-}
\ee
and in (\ref{vierbeinlimit}) the sign in front of $\gamma_5$ is reversed.
Therefore the limiting action is 
\be
S'=\int d^4x \, \left[ i\overline {\Psi}\gamma^a \frac{1+\gamma_5}2\partial_a
\Psi+ i\overline {\Psi}\gamma^a\hat e_a^m \left(\partial_m +\frac 12 \omega_m
\right)\frac{1-\gamma_5}2\Psi\right]  \label{actionlimit-}
\ee
We recall that $\gamma^a$ is the flat (non-dynamical) gamma matrix.

Concerning the energy-momentum tensor, from the definitions
(\ref{Tmunu},\ref{T5munu}), in the
collapsing limit both
$T^{\mu\nu}$ and $T_5^{\mu\nu}$ become
\be
T^{'\mu\nu}(x) = {4} \frac {\delta S'}{\delta
h_{\mu\nu}(x)}\label{T'munu}
\ee
As a consequence (\ref{quantumWardWeyl1}) {and (\ref{quantumWardWeyl2})}
collapse to the same expression
\be
\ET(x)\rightarrow  \langle\!\langle T^{'\mu\nu}\rangle\!\rangle g_{\mu\nu}
\equiv  \ET'(x) \label{quantumWardWeyl1'}
\ee
and 
\be
\ET_5(x) \rightarrow  \langle\!\langle T^{'\mu\nu}\rangle\!\rangle g_{\mu\nu}
\equiv \ET'(x) \label{quantumWardWeyl2'}
\ee
that is, there is only one trace Ward identity.

\subsection{The Pontryagin anomaly}

As pointed out above the odd parity trace anomaly in an ordinary theory of Weyl
fermions can be calculated, to first order, using the above theory of Dirac
fermions coupled to MAT gravity and calculating the collapsing limit of the Weyl
anomaly for a Dirac fermion coupled to MAT gravity. The collapsing limit of the
relevant action reproduces the action for Weyl fermions
\be 
S'&=& \int d^4x \sqrt{|g|} \, \Big{[} \frac{i}{2} \overline {\psi}_L\gamma^m
{\stackrel{\leftrightarrow}{\partial}}_m  \psi_L   -\frac i4 \omega^{\mu abc}
 \overline {\psi_L} 
\gamma_c \gamma_5 \psi_L \Big{]} \label{Weylfermionaction}
\ee
up to a right-handed kinetic term, which is however harmless
due to the presence of the $P_L$ projector in the vertices. Inserting the
replacements into either (\ref{Weylanomaly1}) or (\ref{Weylanomaly2})we find
\be 
\ET'(x)= \frac N
4 \,
\epsilon^{\mu\nu\lambda \rho}{ R} _{\mu\nu}{}^{\sigma \tau} 
{ R}_{\lambda \rho\sigma \tau} \label{Pontryagin0}
\ee
This is not yet the correct result for one must take into account the different
combinatorics in (\ref{Whk}) and
in 
\be
W[h]&=&W[0]+ \sum_{n=0} ^\infty \frac {i^{n-1}}{2^n n!} \int \prod_{i=1}^n
dx_i h_{\mu_i\nu_i}(x_i)   \langle0| {\cal T}T^{\mu_1\nu_1}(x_1)\ldots
T^{\mu_n\nu_n}(x_n)
|0\rangle\label{Wh}
\ee
which is appropriate for (\ref{Weylfermionaction})\footnote{The factor $\frac
1{2^n}$ in the RHS must be properly interpreted. 
When inserting the results for the n-point functions in (\ref{Wh}), one should
recall that the vertex (\ref{Vffh}) contains 
already a $\frac 12$ factor in it with respect to the e.m. tensor: symbolically
we could write $V_{ffh}= \frac 12 \tilde T$, 
where $\tilde T$ is the Fourier transform of the e.m. tensor with fields
replaced by corresponding plane waves. 
A simple practical recipe is to just forget factor $\frac 1{2^n}$ 
in (\ref{Wh}), as was done, somewhat sloppily, in \cite{BGL}. The same holds
also for the formula (\ref{Whk}).}. This amounts to multiplying
(\ref{Pontryagin0}) by a factor of 2.
Therefore, finally the anomaly is
\be 
\ET(x)= \frac N
2 \,
\epsilon^{\mu\nu\lambda \rho}{ R} _{\mu\nu}{}^{\sigma \tau} 
{ R}_{\lambda \rho\sigma \tau} \label{Pontryagin}
\ee
which is the already found Pontrygin anomaly.

In the case of right-handed fermions the anomaly is the same, but with reversed
sign. Thus the odd trace anomaly for Dirac fermions vanishes. This is confirmed
by the following subsection.

\subsection{Odd trace anomaly in the Dirac and Majorana case}

>From the results (\ref{Weylanomaly1}{,\ref{Weylanomaly2}}) we can draw other
conclusions. The action (\ref{axialaction})
reduces to the usual Dirac action if we set $f_{\mu\nu}=0$, and to the Majorana
action if $\psi$ satisfies the Majorana condition. From (\ref{Weylanomaly1}) we
have the confirmation that the odd trace anomaly of these theories vanishes. 
But we also see that in both cases there is an anomaly in the axial
energy-momentum
tensor.
\be
\Theta_{5\mu}{}^{\mu}
=\frac N2  \int \eta \,\epsilon^{\mu\nu\lambda \rho}\,{
R}_{\mu\nu}{}^{\sigma\tau} 
{ R}_{\lambda \rho\sigma \tau} \label{axialWeylanomaly}
\ee
for the Dirac case and $\frac 12$ of it in the Majorana case.
This is a new result. This anomaly is the analog in the trace case of the
Kimura-Delbourgo-Salam anomaly for the axial current.

\section{Odd trace anomalies (the complete calculation)}
\label{s:complete}

Now we would like to justify the assumption made above, according to which only
triangle diagrams provide a nonvanishing contribution to the odd trace anomaly.
The complete calculation requires taking into account all the  tadpoles and
seagull terms that arise from the action (\ref{axialaction}). We start with the
quantum Ward identity (\ref{quantumWardWeyl1},\ref{quantumWardWeyl2})

\subsection{Trace Ward indentity}

We need to expand this Ward identity in series of $h$ and $k$. The expanded
versions is written down in appendix \ref{ss:WIaxial}. Since eventually we are
interested only in the odd terms we will drop all the terms that we already know
are even or vanish (the vev of $T_{(0,0)}^{\mu\nu}(x)$ and 
$T_{5(0,0)}^{\mu\nu}(x)$, the two-point functions of the em and axial em tensor,
as well as the vev of the second and third derivatives of $S$).
In this way the WI's get simplified as follows 
\be
 \ET_{(1,1)}(x,x_1,y_1) &\equiv&
{\cal T}_{(1,1)}{}_{\mu}^{\mu\mu_1\nu_1\lambda_1\rho_1}(x,x_1,y_1) =0
\label{ET11sim}\\
{\ET}_{(2,0)}(x,x_1,x_2) &\equiv& {\cal T}_{(2,0)}{}_\mu^{\mu
\mu_1\nu_1\mu_2\nu_2}(x,x_1,x_2)=0\label{ET20sim}\\
\ET_{(0,2)}(x,y_1,y_2) &\equiv& {\cal T}_{(0,2)}{}_\mu^{\mu
\lambda_1\rho_1\lambda_2\rho_2}(x,y_1,y_2) =0\label{ET02sim}\\
&&\ldots\0
\ee
and
\be
\ET_{5(1,1)}(x,x_1,y_1) &\equiv&
{\cal T}_{5(1,1)}{}_\mu^{\mu\mu_1\nu_1\lambda_1\rho_1}(x,x_1,y_1)
=0\label{ET511sim}\\
 \ET_{5(2,0)}(x,x_1,x_2) &\equiv& {\cal T}_{5(2,0)}{}_\mu^{\mu
\mu_1\nu_1\mu_2\nu_2}(x,x_1,x_2)=0\label{ET520sim}\\
\ET_{5(0,2)}(x,y_1,y_2) &\equiv& {\cal T}_{5(0,2)\mu}{}^{\mu
\lambda_1\rho_1\lambda_2\rho_2}(x,y_1,y_2) =0\label{ET502sim}\\
&&\ldots\0
\ee
These are the Ward identities in the absence of anomalies, but we expect the
rhs's of all these identities to 
be in fact different from zero at one-loop. The odd parity anomaly can be
present only in the rhs of (\ref{ET11sim} ,\ref{ET520sim}) and (\ref{ET502sim}
):
the remaining two cannot contain the $\epsilon$ tensor linearly. After such a
repeated trimming, the relevant WI for our purposes are
(\ref{ET11sim} {,\ref{ET520sim}) and (\ref{ET502sim} ), and the terms that need
to
be closely scrutinized are
\be
&&{\cal T}_{(1,1)}^{\mu\nu\mu_1\nu_1\lambda_1\rho_1}(x,x_1,y_1) = 
- \langle 0|{\cal
T}T_{(0,0)}^{\mu\nu}(x)T_{(0,0)}^{\mu_1\nu_1}(x_1)T_{5(0,0)}^{\lambda_1\rho_1}
(y_1)|0 \rangle\0\\
&&+4
 i  \langle 0|{\cal T}T_{5(0,0)}^{\lambda\rho_1}(y_1)\frac {\delta^{2} S}{\delta
h_{\mu\nu}(x)\delta h_{\mu_1\nu_1}(x_1) }|0 \rangle+4i  \langle 0|{\cal
T}T_{(0,0)}^{\mu_1\nu_1}(x_1)\frac {\delta^{2} S}{\delta
k_{\lambda_1\rho_1}(y_1)\delta h_{\mu\nu}(x) }|0 \rangle\0\\
&&+4
 i  \langle 0|{\cal T}T_{(0,0)}^{\mu\nu}(x)\frac {\delta^{2} S}{\delta
k_{\lambda_1\rho_1}(y_1)\delta k_{\mu_1\nu_1}(x_1) }|0 \rangle ,
\label{TTTmunu11sim}
\ee
 together with
\be
&&{\cal T}_{5(2,0)}^{\lambda\rho\mu_1\nu_1\mu_2\nu_2}(x,x_1,x_2) = 
- \langle 0|{\cal
T}T_{5(0,0)}^{\lambda\rho}(x)T_{(0,0)}^{\mu_1\nu_1}(x_1)T_{(0,0)}^{\mu_2\nu_2}
(x_2)|0 \rangle \0\\
&&+4
 i  \langle 0|{\cal T}T_{(0,0)}^{\mu_1\nu_1}(x_1)\frac {\delta^{2} S}{\delta
k_{\lambda\rho}(x)\delta h_{\mu_2\nu_2}(x_2) }|0 \rangle+4i  \langle 0|{\cal
T}T_{(0,0)}^{\mu_2\nu_2}(x_2)\frac {\delta^{2} S}{\delta
h_{\mu_1\nu_1}(x_1)\delta k_{\lambda\rho}(x) }|0 \rangle\0\\
&&+4
 i  \langle 0|{\cal T}T_{5(0,0)}^{\lambda\rho}(x)\frac {\delta^{2} S}{\delta
h_{\mu_1\nu_1}(x_1)\delta h_{\mu_2\nu_2}(x_2) }|0 \rangle  \label{TTT5munu20sim}
\ee
and
\be
&&{\cal T}_{5(0,2)}^{\lambda\rho\lambda_1\rho_1\lambda_2\rho_2}(x,y_1,y_2) = 
- \langle 0|{\cal
T}T_{5(0,0)}^{\lambda\rho}(x)T_{5(0,0)}^{\lambda_1\rho_1}(y_1)T_{5(0,0)}^{
\lambda_2\rho_2}(y_2)|0 \rangle\0\\
&&+4
 i  \langle 0|{\cal T}T_{5(0,0)}^{\lambda_1\rho_1}(y_1)\frac {\delta^{2}
S}{\delta k_{\lambda\rho}(x)\delta k_{\lambda_2\rho_2}(y_2) }|0 \rangle+4i 
\langle 0|{\cal T}T_{5(0,0)}^{\lambda_2\rho_2}(y_2)\frac {\delta^{2} S}{\delta
k_{\lambda_1\rho_1}(y_1)\delta k_{\lambda\rho}(x) }|0 \rangle\0\\
&&+4
 i  \langle 0|{\cal T}T_{5(0,0)}^{\lambda\rho}(x)\frac {\delta^{2} S}{\delta
k_{\lambda_1\rho_1}(y_1)\delta k_{\lambda_2\rho_2}(y_2) }|0 \rangle
\label{TTT5munu02sim}
\ee

The terms above that contain the second derivative of $S$ are bubble diagrams
where one vertex has two external $h$ and/or $k$ graviton lines. These diagrams
are similar to those already met above and in \cite{BGL}, and can be shown to
similarly vanish, see appendix \ref{ss:PVPV'} and \ref{ss:PVPVe}. Therefore we
are left with 
\be
 \ET_{(1,1)}(x,x_1,y_1) &=& - \langle 0|{\cal
T}T_{(0,0)\mu}{}^\mu(x)T_{(0,0)}^{\mu_1\nu_1}(x_1)T_{5(0,0)}^{\lambda_1\rho_1}
(y_1)|0 \rangle
\label{ET11simfin}
\ee
\be
\ET_{5(2,0)}(x,x_1,x_2) & =& - \langle 0|{\cal
T}T_{5(0,0)}{}_\lambda^\lambda(x)T_{(0,0)}^{\mu_1\nu_1}(x_1)T_{(0,0)}^{
\mu_2\nu_2}(x_2)|0 \rangle \label{ET520simfin}\\
\ET_{5(0,2)}(x,y_1,y_2) & =&- \langle 0|{\cal
T}T_{5(0,0)}{}_\lambda^{\lambda}(x)T_{5(0,0)}^{\lambda_1\rho_1}(y_1)T_{5(0,0)}^{
\lambda_2\rho_2}(y_2)|0 \rangle\label{ET502simfin}
\ee
which are the intermediate results already obtained above. From this point on
the
calculation proceeds as in section {\ref{s:backenvelop}.

\section{Conclusion}

In this paper we have dealt with two subjects: the odd parity trace anomaly in
chiral fermion theories 
in a 4d curved background and the introduction of an axial 'metric' beside the
familiar gravity metric.
We have recalculated the first with the Feynman diagram method in a more
complete way, by including in the computation also tadpole and seagull terms. We
have verified that the latter do not modify the result of \cite{BGL}. To do so
we have also recalculated the Ward identity for diffeomorphims. { In 
this paper we have constantly been using DR, leaving to a future investigation 
the discussion of other regularizations. The other important topic of this paper
is the
introduction of MAT (metric-axial-metric) gravity and the relevant
formalism}. MAT gravity may have of course an autonomous development and could
be
studied as a new bimetric model, with the new characteristics that it interacts
also axially with fermions. We postpone this analysis to a future work. In this
paper we have utilized MAT gravity in order to disentangle the thorny issue of
the path integral measure in a theory of chiral fermions. In fact MAT gravity
interact naturally with Dirac fermions. We have shown that one can compute the
trace anomalies of a theory of Dirac fermions coupled to a background MAT
gravity and, then,
recover the results for a chiral fermion theory coupled to ordinary gravity by
simply taking a (smooth) limit.
We have shown that in this way one obtains the same results as in \cite{BGL}.

Finally, let us remark that in this paper we did not verify the Ward identity
for two types of diffeomorphisms 
in MAT background, much as was done in section 4. From consolidated experience
we believe that this will not 
modify the trace anomalies of the model, but the problem is interesting in
itself.
Can there be anomalies of the Einstein-Lorentz type in one of the Ward
identities?
This is an intriguing problem we leave for the future.
\vskip 1cm
{\bf Acknowledgements.}
L.B. would like to thank the Yukawa Institute for Theoretical Physics, Kyoto and
the KEK Theory Center, KEK, Tsukuba, where he carried out 
most of this research, for their kind hospitality and support.  We would like to
thank Fiorenzo Bastianelli for a useful exchange of messages.
This research has been supported by the Croatian
Science Foundation under the project No.~8946 and by the University of
Rijeka under the research support No.~13.12.1.4.05. Finally, A.D.P. is grateful
to CAPES and CNPq for support.

\vskip 1cm

\noindent{\bf \large Appendices}

\appendix

\section{The triangle diagram}
\label{s:trianglediagram}

In this Appendix we derive in more detail the result of \cite{BGL}.
Employing the Feynman rules of the free chiral fermion coupled to an external
gravitational field, the contribution from the triangle diagram is expressed as

\begin{eqnarray}
T_{\mu\nu\mu'\nu'}(k_1,k_2) &=& \int
\frac{d^4p}{(2\pi)^4}\mathrm{Tr}\left\{\frac{i}{8}\left[(2p-k_1)_\mu\gamma_\nu +
(\mu \leftrightarrow
\nu)\right]\left(\frac{1+\gamma_5}{2}\right)\frac{i}{(\slashed{p}-\slashed{k}
_1)+i\epsilon}\right.\nonumber\\
&\times&\left.\frac{i}{8}\left[(2p-2k_1-k_2)_{\mu'}\gamma_{\nu'}+(\mu'
\leftrightarrow
\nu')\right]\left(\frac{1+\gamma_5}{2}\right)\frac{i}{(\slashed{p}-\slashed{k}
_1-\slashed{k}_2)+i\epsilon}\right.\nonumber\\
&\times&\left.\frac{i}{4}(2\slashed{p}-\slashed{k}_1-\slashed{k}_2)\left(\frac{
1+\gamma_5}{2}\right)\frac{i}{\slashed{p}+i\epsilon}\right\}\,.
\label{triangle1}
\end{eqnarray}
Using the properties of the gamma matrices, one obtains\footnote{We have dropped
the $i\epsilon$ factor in the denominators, for convenience.} 

\begin{eqnarray}
T_{\mu\nu\mu'\nu'}(k_1,k_2) &=& - \frac{1}{256}\int
\frac{d^4p}{(2\pi)^4}\mathrm{Tr}\left\{\left[\frac{\slashed{p}}{p^2}
(2p-k_1)_\mu\gamma_\nu + (\mu
\leftrightarrow
\nu)\right]\frac{(\slashed{p}-\slashed{k}_1)}{(p-k_1)^2}\left[(2p-2k_1-k_2)_{
\mu'}\gamma_{\nu'}\right.\right.\nonumber\\
&+&\left.\left.(\mu' \leftrightarrow
\nu')\right]\frac{(\slashed{p}-\slashed{k}_1-\slashed{k}_2)}{(p-k_1-k_2)^2}
(2\slashed{p}-\slashed{k}_1-\slashed{k}_2)\left(\frac{
1+\gamma_5}{2}\right)\right\}\,.
\label{triangle2}
\end{eqnarray}
Clearly, such an integral is ultraviolet divergent. In order to proceed with the
computation, we employ dimensional regularization, where additional components
are added to the momentum, namely, $p\rightarrow p+\ell$, where $\ell =
(\ell_4,\ldots,\ell_{n-4})$. This implies, in particular, 

\begin{equation}
\gamma^\mu p_\mu\,\longrightarrow\,\gamma^{\mu}p_\mu +
\gamma^{\bar{\mu}}\ell_{\bar{\mu}}\,,
\label{triangle3}
\end{equation}
with $\bar{\mu}\in\left\{4,\ldots,n-4\right\}$. Hence, eq.\eqref{triangle2} is
replaced by
\begin{eqnarray}
T_{\mu\nu\mu'\nu'}(k_1,k_2) &=& -\frac{1}{256}\int \frac{d^4p}{(2\pi)^4} \int
\frac{d^{n-4}\ell}{(2\pi)^{n-4}}\mathrm{Tr}\left\{\left[\frac{\slashed{p}
+\slashed{\ell}}{p^2-\ell^2}(2p-k_1)_\mu\gamma_\nu +
(\mu \leftrightarrow
\nu)\right]\frac{(\slashed{p}+\slashed{\ell}-\slashed{k}_1)}{(p-k_1)^2-\ell^2}
\right.\nonumber\\
&\times&\left.\left[(2p-2k_1-k_2)_{\mu'}\gamma_{\nu'}+(\mu' \leftrightarrow
\nu')\right]\underbrace{\frac{(\slashed{p}+\slashed{\ell}-\slashed{k}_1-\slashed
{k}_2)}{(p-k_1-k_2)^2-\ell^2}(2\slashed{p}+2\slashed{\ell}-\slashed{k}
_1-\slashed{k}_2)}_{(\ast)}\right.\nonumber\\
&\times&\left.\left(\frac{
1+\gamma_5}{2}\right)\right\}\,.
\label{triangle4}
\end{eqnarray}
Expression \eqref{triangle4} is now regularized and we can continue with the
computation of the diagram. In order to simplify our analysis a bit, we ignore
the identity in the projector $(1+\gamma_5)/2$ since we are concerned with the
parity odd part contribution of the diagram, which is encoded in the $\gamma_5$
sector. Also, we omit the symmetrizations in $(\mu \leftrightarrow \nu)$ and in
$(\mu' \leftrightarrow \nu')$ for the time being and reintroduce them later on. 

Let us take the term $(\ast)$ and define $q=k_1+k_2$. It is simple to check that

\begin{equation}
(\ast) =
\frac{(\slashed{p}+\slashed{\ell}-\slashed{q})}{(p-q)^2-\ell^2}(2\slashed{p}
+2\slashed{\ell}-\slashed{q})=1+\frac{\slashed{p}-\slashed{\ell}}{\slashed{p}
+\slashed{\ell}-\slashed{q}}+\frac{2\slashed{\ell}}{\slashed{p}+\slashed{\ell}
-\slashed{q}}\,,
\label{triangle5}
\end{equation}
and plugging it into eq.\eqref{triangle4}, one ends up with

\begin{equation}
T_{\mu\nu\mu'\nu'}(k_1,k_2) = T^{(1)}_{\mu\nu\mu'\nu'}(k_1,k_2) +
T^{(2)}_{\mu\nu\mu'\nu'}(k_1,k_2) + \tilde{T}_{\mu\nu\mu'\nu'}(k_1,k_2)\,,
\label{triangle6}
\end{equation}
with

\begin{eqnarray}
T^{(1)}_{\mu\nu\mu'\nu'}(k_1,k_2) &=& -\frac{1}{256}\int \frac{d^4p}{(2\pi)^4}
\int
\frac{d^{n-4}\ell}{(2\pi)^{n-4}}\mathrm{Tr}\left[\frac{\slashed{p}+\slashed{\ell
}}{p^2-\ell^2}(2p-k_1)_\mu\gamma_\nu\frac{(\slashed{p}+\slashed{\ell}-\slashed{k
}_1)}{(p-k_1)^2-\ell^2} \right.\nonumber\\
&\times&\left.(2p-2k_1-k_2)_{\mu'}\gamma_{\nu'}\frac{\gamma_5}{2}\right]\,,
\nonumber\\
T^{(2)}_{\mu\nu\mu'\nu'}(k_1,k_2) &=& -\frac{1}{256}\int \frac{d^4p}{(2\pi)^4}
\int
\frac{d^{n-4}\ell}{(2\pi)^{n-4}}\mathrm{Tr}\left[\frac{\slashed{p}+\slashed{\ell
}}{p^2-\ell^2}(2p-k_1)_\mu\gamma_\nu\frac{(\slashed{p}+\slashed{\ell}-\slashed{k
}_1)}{(p-k_1)^2-\ell^2} \right.\nonumber\\
&\times
&\left.(2p-2k_1-k_2)_{\mu'}\gamma_{\nu'}\frac{(\slashed{p}-\slashed{\ell})}{
\slashed{p}+\slashed{\ell}-\slashed{q}}\frac{\gamma_5}{2}\right]\,,\nonumber\\
\tilde{T}_{\mu\nu\mu'\nu'}(k_1,k_2) &=& -\frac{1}{256}\int \frac{d^4p}{(2\pi)^4}
\int
\frac{d^{n-4}\ell}{(2\pi)^{n-4}}\mathrm{Tr}\left[\frac{\slashed{p}+\slashed{\ell
}}{p^2-\ell^2}(2p-k_1)_\mu\gamma_\nu\frac{(\slashed{p}+\slashed{\ell}-\slashed{k
}_1)}{(p-k_1)^2-\ell^2} \right.\nonumber\\
&\times&\left.(2p-2k_1-k_2)_{\mu'}\gamma_{\nu'}\frac{\slashed{\ell}}{\slashed{p}
+\slashed
{\ell}-\slashed{q}}{\gamma_5}\right]\,.
\label{triangle7}
\end{eqnarray}
We detail the computation of each contribution $T^{(1)}$, $T^{(2)}$ and
$\tilde{T}$ in the following lines.

\subsection{$T^{(1)}_{\mu\nu\mu'\nu'}(k_1,k_2)$}

The contribution $T^{(1)}$ can be expressed as

\begin{eqnarray}
T^{(1)}_{\mu\nu\mu'\nu'}(k_1,k_2) &=& -\frac{1}{256}\int
\frac{d^4p}{(2\pi)^4}\int\frac{d^{n-4}\ell}{(2\pi)^{n-4}}\frac{(2p-k_1)_{\mu}
(2p-2k_1-k_2)_{\mu'}}{2(p^2-\ell^2)\left[(p-k_1)^2-\ell^2\right]}\nonumber\\
&\times&\underbrace{\mathrm{Tr}\left[(\slashed{p}+\slashed{\ell}
)\gamma_\nu(\slashed{p}+\slashed{\ell}-\slashed{k}_1)\gamma_{\nu'}\gamma_5\right
]}_{4ip^{\alpha}k^{\beta}_1\epsilon_{\alpha\nu\beta\nu'}}\,.
\label{triangle8}
\end{eqnarray}
Employing the Feynman parametrization, expression \eqref{triangle8} is written
as

\begin{eqnarray}
&&T^{(1)}_{\mu\nu\mu'\nu'}(k_1,k_2)\label{triangle9}\\
&=& -\frac{i}{128}\int
\frac{d^4p}{(2\pi)^4}\int\frac{d^{n-4}\ell}{(2\pi)^{n-4}}\int^{1}_{0}dx\frac{
(2p-k_1)_{\mu}(2p-2k_1-k_2)_{\mu'}}{\left\{\left[(p-k_1)^2-\ell^2\right]
x+(1-x)(p^2-\ell^2)\right\}^2}p^\alpha
k^{\beta}_1\epsilon_{\alpha\nu\beta\nu'}\,.
\0
\end{eqnarray}
Performing the shift $p\rightarrow p+xk_1$ and taking into account that just
even powers of $p$ in the numerator will result on non-vanishing contributions
to $T^{(1)}$, one obtains

\begin{eqnarray}
&&T^{(1)}_{\mu\nu\mu'\nu'}(k_1,k_2)\label{triangle10}\\
& =&
\frac{i}{128}\int^{1}_{0}dx\int\frac{d^{n-4}\ell}{(2\pi)^{n-4}}\int\frac{d^4p}{
(2\pi)^4}\frac{2p_{\mu'} (1-2x)k_{1\mu}+2p_\mu \left[2(1-x)k_1 +
k_2\right]_{\mu}}{\left[p^2+x(1-x)k^{2}_{1}-\ell^2\right]^2}p^\alpha
k^{\beta}_1\epsilon_{\alpha\nu\beta\nu'}\,.\0
\end{eqnarray} 
Making use of Lorentz symmetry, one can make the following replacement,

\begin{equation}
p^{\mu}p^{\nu}\,\longrightarrow\,\frac{1}{4}\eta^{\mu\nu}p^2\,,
\label{triangle11}
\end{equation}
which gives rise to

\begin{eqnarray}
&&T^{(1)}_{\mu\nu\mu'\nu'}(k_1,k_2)\label{triangle12}\\
& =&
\frac{i}{256}\int^{1}_{0}dx\int\frac{d^{n-4}\ell}{(2\pi)^{n-4}}\int\frac{d^4p}{
(2\pi)^4}\frac{\delta^{\alpha}_{\mu'} (1-2x)k_{1\mu}+\delta^{\alpha}_{\mu}
\left[2(1-x)k_1 +
k_2\right]_{\mu'}}{\left[p^2+x(1-x)k^{2}_{1}-\ell^2\right]^2}p^2
k^{\beta}_1\epsilon_{\alpha\nu\beta\nu'}\,.
\0
\end{eqnarray} 
After taking into account the contraction of the Kronecker deltas with the
$\epsilon$-tensor and imposing the symmetrization of $(\mu \leftrightarrow \nu)$
and $(\mu' \leftrightarrow \nu')$ one immediately sees that the contribution
from $T^{(1)}$ vanishes. 

\subsection{$T^{(2)}_{\mu\nu\mu'\nu'}(k_1,k_2)$}

\begin{eqnarray}
&& T^{(2)}_{\mu\nu\mu'\nu'}(k_1,k_2)\label{triangle13}\\
& =& \frac{1}{256}\int \frac{d^4p}{(2\pi)^4}\int
\frac{d^{n-4}\ell}{(2\pi)^{n-4}}\frac{(2p+k_1)_{\mu}(2p-k_2)_{\mu'}}{
2(p^2-\ell^2)\left[(p-k_2)^2-\ell^2\right]}\underbrace{\mathrm{Tr}\left[
\gamma_\nu(\slashed{p}+\slashed{\ell})\gamma_{\nu'}(\slashed{p}+\slashed{\ell}
-\slashed{k}_2)\gamma_{5}\right]}_{4ip^{\alpha}k^{\beta}_{2}\epsilon_{
\nu\alpha\nu'\beta}}\,.\0
\end{eqnarray}
As before, one employs the Feynman parametrization and in very strict analogy,
perform the shift $p\rightarrow p+xk_2$. This renders

\begin{eqnarray}
&&T^{(2)}_{\mu\nu\mu'\nu'}(k_1,k_2)\label{triangle14}\\
& =& \frac{i}{128}\int^{1}_{0}dx\int
\frac{d^{n-4}\ell}{(2\pi)^{n-4}}\int
\frac{d^4p}{(2\pi)^4}\frac{(2p+k_1+2xk_2)_{\mu}(2p-(1-x)k_2)_{\mu'}}{\left[
p^2-\ell^2-x(x-1)k^2_2\right]^2}p^{\alpha}k^{\beta}_{2}\epsilon_{
\nu\alpha\nu'\beta}\,.
\0
\end{eqnarray}
Collecting just the even power of $p$ in the numerator of \eqref{triangle14} and
applying the relation \eqref{triangle11}, one immediately obtains

\begin{equation}
T^{(2)}_{\mu\nu\mu'\nu'}(k_1,k_2) = \frac{i}{256}\int^{1}_{0}dx\int
\frac{d^{n-4}\ell}{(2\pi)^{n-4}}\int
\frac{d^4p}{(2\pi)^4}\frac{\delta^{\alpha}_{\mu}(x-1)k_{2\mu'}+\delta^{\alpha}_{
\mu'}(k_1+2xk_2)_\mu}{\left[p^2-\ell^2-x(x-1)k^2_2\right]^2}k^{\beta}_{2}
\epsilon_{\nu\alpha\nu'\beta}\,.
\label{triangle15}
\end{equation}
For the same reasons described in the previous subsection, after
symmetrizations, the contribution from $T^{(2)}$ vanishes.

\subsection{$\tilde{T}_{\mu\nu\mu'\nu'}(k_1,k_2)$}

\begin{eqnarray}
&&\tilde{T}_{\mu\nu\mu'\nu'}(k_1,k_2) =
-\frac{1}{256}\int\frac{d^4p}{(2\pi)^4}\int
\frac{d^{n-4}\ell}{(2\pi)^{n-4}}\frac{(2p-k_1)_{\mu}(2p-2k_1-k_2)_{\mu'}}{
(p^2-\ell^2)\left[(p-k_1)^2-\ell^2\right]\left[(p-q)^2-\ell^2\right]}\nonumber\\
&\times&~~~~~\underbrace{\mathrm{Tr}\left[(\slashed{p}+\slashed{\ell}
)\gamma_\nu(\slashed{p}+\slashed{\ell}-\slashed{k}_1)\gamma_{\nu'}(\slashed{p}
+\slashed{\ell}-\slashed{q})\slashed{\ell}\gamma_5\right]}_{4ik^{\alpha}_1
k^{\beta}_2\epsilon_{\nu\alpha\nu'\beta}}\,.
\label{triangle16}
\end{eqnarray}
The Feynman parametrization leads to

\begin{eqnarray}
\tilde{T}_{\mu\nu\mu'\nu'}(k_1,k_2) &=& -\frac{i}{32}k^{\alpha}_1
k^{\beta}_2\epsilon_{\nu\alpha\nu'\beta}\int \frac{d^4p}{(2\pi)^4}\int
\frac{d^{n-4}\ell}{(2\pi)^{n-4}}\int^{1}_{0}dx\nonumber\\
&\times&\int^{1-x}_{0}dy\frac{(2p-k_1)_{\mu}(2p-2k_1-k_2)_{\mu'}}{\left\{\left[
(p-k_1)^2-\ell^2\right]x+\left[(p-q)^2-\ell^2\right]y+(p^2-\ell^2)(1-x-y)
\right\}^3}\ell^2\,.\nonumber\\
\label{triangle17}
\end{eqnarray}
Making the shift $p\rightarrow p+xk_1+yq$ and few algebraic manipulations,
eq.\eqref{triangle17} becomes

\begin{eqnarray}
\tilde{T}_{\mu\nu\mu'\nu'}(k_1,k_2) &=& -\frac{i}{32}k^{\alpha}_1
k^{\beta}_2\epsilon_{\nu\alpha\nu'\beta}\int^{1}_{0}dx\int^{1-x}_{0}dy\int
\frac{d^4p}{(2\pi)^4}\nonumber\\
&\times&\int
\frac{d^{n-4}\ell}{(2\pi)^{n-4}}\frac{(2p+2xk_1+2yq-k_1)_{\mu}
(2p+2xk_1+2yq-2k_1-k_2)_{\mu'}}{\left[p^2-\ell^2+2k_1\cdot k_2
y(1-y-x)\right]^3}\ell^2\,.\nonumber\\
\label{triangle18}
\end{eqnarray}
Taking the numerator of \eqref{triangle18}, collecting just those terms which
contribute to the trace anomaly and employing the relation \eqref{triangle11},
eq.\eqref{triangle18} becomes

\begin{eqnarray}
\tilde{T}_{\mu\nu\mu'\nu'}(k_1,k_2) &=& -\frac{i}{32}k^{\alpha}_1
k^{\beta}_2\epsilon_{\nu\alpha\nu'\beta}\int^{1}_{0}dx\int^{1-x}_{0}dy\int
\frac{d^4p}{(2\pi)^4}\nonumber\\
&\times&\int
\frac{d^{n-4}\ell}{(2\pi)^{n-4}}\frac{p^2\eta_{\mu\mu'}+4y(x+y-1)k_{1\mu'}k_{
2\mu}}{\left[p^2-\ell^2+2k_1\cdot k_2 y(1-y-x)\right]^3}\ell^2\,.
\label{triangle19}
\end{eqnarray}
To make sense of the integrals present in \eqref{triangle19}, we make a Wick
rotation 
$k^0 \to i k^0_E$ for any momentum $k^\mu$: so, for instance, in the previous
integral $p^2 \to -p^2_E$, etc.
So \eqref{triangle19} is replaced by
\begin{eqnarray}
\tilde{T}_{\mu\nu\mu'\nu'}(k_1,k_2) &=& \frac{1}{32}k^{\alpha}_1
k^{\beta}_2\epsilon_{\nu\alpha\nu'\beta}\int^{1}_{0}dx\int^{1-x}_{0}dy\int
\frac{d^4p}{(2\pi)^4}\nonumber\\
&\times&\int
\frac{d^{n-4}\ell}{(2\pi)^{n-4}}\frac{p^2\eta_{\mu\mu'}-4y(x+y-1)k_{1\mu'}k_{
2\mu}}{\left[p^2+\ell^2+2k_1\cdot k_2 y(1-y-x)\right]^3}\ell^2\,.
\label{triangle19'}
\end{eqnarray}
and we dispense from explicitly indicating the Euclidean momenta whenever it is
not strictly necessary.
Now the integrals are well-defined and we can use the following results
\begin{eqnarray}
\int \frac{d^{n-4}\ell}{(2\pi)^{n-4}}\frac{\ell^2}{\left[p^2+2k_1\cdot k_2
y(1-y-x)+\ell^2\right]^3}&=& \frac 1 {(4\pi)^{(n-4)/2}}\frac{
n-4}{4}\nonumber\\
&\times&\frac{1}{\left[p^2+2k_1\cdot k_2
y(1-y-x)\right]^{4-\frac{n}{2}}}\Gamma\left(4-\frac{n}{2}\right)\,,\nonumber\\
\int \frac{d^4p}{(2\pi)^4}\frac{1}{\left[p^2+2k_1\cdot k_2
y(1-y-x)\right]^{4-\frac{n}{2}}}&=&
\frac{1}{(4\pi)^2}\frac{\Gamma\left(2-\frac{n}{2}\right)}{\Gamma\left(4-\frac{n}
{2}\right)}\left(\frac{1}{2k_1\cdot k_2
y(1-y-x)}\right)^{2-\frac{n}{2}}\,,\nonumber\\
\int \frac{d^4p}{(2\pi)^4}\frac{p^2}{\left[p^2+2 k_1\cdot k_2
y(1-x-y)\right]^{4-\frac{n}{2}}}&=&
\frac{2}{(4\pi)^{2}}\frac{\Gamma\left(1-\frac{n}{2}\right)}{\Gamma\left(4-\frac
{n}{2}\right)}\left(\frac{1}{2k_1\cdot k_2
y(1-y-x)}\right)^{1-\frac{n}{2}}\,.\nonumber\\
\label{triangle20}
\end{eqnarray}
Using \eqref{triangle20} and performing the integration over the Feynman
parameters $(x,y)$ and returning to the Lorentzian metric, one obtains

\begin{equation}
\tilde{T}_{\mu\nu\mu'\nu'}(k_1,k_2) = \frac{1}{6144\pi^2}k^{\alpha}_1
k^{\beta}_2
\epsilon_{\nu\nu'\alpha\beta}\left( \eta_{\mu\mu'}k_1 \cdot
k_2-k_{1\mu'}k_{2\mu}\right)\,.
\label{triangle21}
\end{equation}
Of course, as previously mentioned, one should symmetrize expression
\eqref{triangle21} with respect to $(\mu \leftrightarrow \nu)$ and $(\mu'
\leftrightarrow \nu')$. Then, \eqref{triangle21} becomes

\begin{equation}
\tilde{T}_{\mu\nu\mu'\nu'}(k_1,k_2) = \frac{1}{6144\pi^2}k^{\alpha}_1
k^{\beta}_2\left(k_1 \cdot k_2 t_{\mu\nu\mu'\nu'\alpha\beta}
-t^{(21)}_{\mu\nu\mu'\nu'\alpha\beta}\right)\,,
\label{triangle22}
\end{equation}
The tensors $t$ and $t^{(21)}$ have been defined in \eqref{t} and \eqref{t21}.

On top of that one should add the contribution from the ``cross diagram",
namely, the contribution coming from the simultaneous exchanges $(k_1
\leftrightarrow k_2, \mu \leftrightarrow \mu' , \nu \leftrightarrow \nu')$.
Hence, the sum of \eqref{triangle22} with the cross diagram contribution gives
rise to

\begin{equation}
\tilde{T}^\mathrm{(tot)}_{\mu\nu\mu'\nu'}(k_1,k_2) =
\frac{1}{3072\pi^2}k^{\alpha}_1 k^{\beta}_2\left(k_1 \cdot k_2
t_{\mu\nu\mu'\nu'\alpha\beta}-t^{(21)}_{\mu\nu\mu'\nu'\alpha\beta}\right)\,.
\label{triangle24}
\end{equation}

\section{Derivation of Feynman rules}
\label{s:feynmanrules}

\subsection{Ordinary gravity}
\label{ss:ordgravity}

Consider a free  theory coupled to ordinary gravity. We assume that the action
has the expansion
\be
S&=& \sum_{n=0}^\infty S_n \equiv S_0+  
\sum_{n=1}^\infty \int \prod_{i=1}^n dx_i\,\frac 1 {n!}\,\frac {\delta^n
S}{\delta h_{\mu_1\nu_1}(x_1) \ldots \delta h_{\mu_n\nu_n}(x_n)}
\Bigg{\vert}_{h=0}   h_{\mu_1\nu_1}(x_1) \ldots  h_{\mu_n\nu_n}(x_n)\0\\
&=& S_0 + \int dx \,\frac {\delta S}{\delta h_{\mu\nu}(x)}\Bigg{\vert}_{h=0}  
h_{\mu\nu}(x)\label{Sseries}\\
&& + \frac 12 \int dx_1dx_2\, 
\frac {\delta^2 S}{\delta h_{\mu_1\nu_1}(x_1)\delta h_{\mu_2\nu_2}(x_2)}
\Bigg{\vert}_{h=0}   h_{\mu_1\nu_1}(x_1)h_{\mu_2\nu_2}(x_2)+\ldots\0
\ee
The e.m. tensor is defined as
\be
T^{\mu\nu} = \frac 2{\sqrt {g}} \frac {\delta S}{\delta g_{\mu\nu}},\quad\quad 
T_{\mu\nu} =- \frac 2{\sqrt {g}} \frac {\delta S}{\delta g^{\mu\nu}}
\label{Tmunucl}
\ee
In the following we have in mind the free fermion theory in 4d defined by
({\ref{action}), and
set  $g_{\mu\nu}= \eta_{\mu\nu}+ h_{\mu\nu}$. 

We need the expansion 
\be
\sqrt{|g|}&=&  1+\frac 12 (\tr\, h)+\frac 18 (\tr\,  h)^2 -\frac 14 (\tr\,  h^2)
-\frac 18 (\tr\,  h)(\tr\,  h^2) +\frac 1{48}(\tr h)^3 +\frac 16 (\tr
h^3)+\ldots\0\\
&\equiv& \sum_{n=0} O_n(h)\label{sqrtg}\\
\frac 1{\sqrt g} &=& 1-\frac 12 (\tr\, h)+\frac 18 (\tr\,  h)^2 +\frac 14 (\tr\,
 h^2) -\frac 18 (\tr\,  h)(\tr\,  h^2) -\frac 1{48}(\tr h)^3 -\frac 16 (\tr
h^3)+\ldots\0\\
&\equiv& \sum_{n=0} \hat O_n(h)\label{sqrtg-1}
\ee
where by $h$ is meant the matrix $h_{\mu\nu}$, and $O_0(h)=1, \,\,O_1(h)=\frac
12 (\tr\, h), \ldots$, etc.
Next we consider the complete expansion of (\ref{action}) in powers of $h$, like
(\ref{approxaction3rdorder}).

Now, using (\ref{Tmunucl}), one can write
\be
 \,T^{\mu\nu}(x) &= &\frac 2{\sqrt g}\left(\frac {\delta S}{\delta
h_{\mu\nu}(x)}\Bigg{\vert}_{h=0} +  \int dx_2\, 
\frac {\delta^2 S}{\delta h_{\mu\nu}(x)\delta h_{\mu_2\nu_2}(x_2)}
\Bigg{\vert}_{h=0}   h_{\mu_2\nu_2}(x _2)\right.\0\\
&&+\left. \frac 12\int dx_2 dx_3\, 
\frac {\delta^3 S}{\delta h_{\mu\nu}(x)\delta h_{\mu_2\nu_2}(x_2)\delta
h_{\mu_3\nu_3}(x_3)}
\Bigg{\vert}_{h=0}   h_{\mu_2\nu_2}(x_2)  h_{\mu_3\nu_3}(x_3)+\ldots\right)\0\\
&&\equiv T_{(0)}^{\mu\nu}(x)+T_{(1)}^{\mu\nu}(x)+\ldots \label{Tmunu2}
\ee
which implies
\be
 T_{(n)}^{\mu\nu}(x)&=&\sum_{m=0}^n \hat O_{n-m}(h(x)) \frac 2
{m!}\label{Tmunun}\\
&&\cdot  \int \prod_{i=1}^m dx_i\, 
\frac {\delta^{m+1} S}{\delta h_{\mu\nu}(x)\delta h_{\mu_1\nu_1}(x_1) \ldots
\delta h_{\mu_m\nu_m}(x_m)}
\Bigg{\vert}_{h=0}   h_{\mu_1\nu_1}(x_1) \ldots  h_{\mu_m\nu_m}(x_m)\0
\ee
So we can rewrite
\be
S_n= \frac 1{2n} \int dx \,\left(\sum_{m=1}^n O_{n-m}(h(x)) 
T_{(m-1)}^{\mu\nu}(x)\right) h_{\mu\nu}(x) \label{Sn}
\ee
For instance
\be
S_1&=& \frac 12 \int dx \,   T_{(0)}^{\mu\nu}(x) \,h_{\mu\nu}(x),\label{S1}\\
S_2&=&   \frac 14 \int dx \,\left( T_{(1)}^{\mu\nu}(x) + \frac 12 (\tr\, h(x))
\, T_{(0)}^{\mu\nu}(x)\right) \,h_{\mu\nu}(x),\label{S2}\\
S_3&=&\frac 16 \int dx \,   \left(  T_{(2)}^{\mu\nu}(x) + \frac 12 (\tr\, h(x))
\, T_{(1)}^{\mu\nu}(x)+
\frac 18\left( (\tr\, h(x))^2 - 2 (\tr\, h^2(x))\right) T_{(0)}^{\mu\nu}(x)
\right)\,h_{\mu\nu}(x)\0\\
\label{S3}
\ee
and
\be
T_{(0)}^{\mu\nu}(x) &=&  2 \frac {\delta S}{\delta
h_{\mu\nu}(x)}\Bigg{\vert}_{h=0}\label{T0}\\
T_{(1)}^{\mu\nu}(x) &=&  -  (\tr\, h(x)) \frac {\delta S}{\delta
h_{\mu\nu}(x)}\Bigg{\vert}_{h=0}+ 2
 \int dx_1\,\frac {\delta^{2} S}{\delta h_{\mu\nu}(x)\delta h_{\mu_1\nu_1}(x_1)
}
\Bigg{\vert}_{h=0}   h_{\mu_1\nu_1}(x_1)  \label{T1}\\
T_{(2)}^{\mu\nu}(x) &=& \frac 14 \left( (\tr\, h(x))^2 + 2 (\tr
\,h(x)^2)\right)\frac {\delta S}{\delta h_{\mu\nu}(x)}\Bigg{\vert}_{h=0}\0\\
&&- (\tr\, h(x))
\int dx_1\,\frac {\delta^{2} S}{\delta h_{\mu\nu}(x)\delta h_{\mu_1\nu_1}(x_1) }
\Bigg{\vert}_{h=0}   h_{\mu_1\nu_1}(x_1)\label{T2}\\
&&+2 \int dx_1 dx_2 \frac {\delta^{3} S}{\delta h_{\mu\nu}(x)\delta
h_{\mu_1\nu_1}(x_1) \delta h_{\mu_2\nu_2}(x_2)}
\Bigg{\vert}_{h=0}   h_{\mu_1\nu_1}(x_1)h_{\mu_2\nu_2}(x_2)\0
\ee

{\bf Remark.} Since $S= \int \sqrt{|g|} {\cal L}$, the derivatives of $S$ in the
previous formulas, when applied to $\sqrt{|g|}$, will produce terms $\sim {\cal
L}$ which vanish on shell. These are contact terms. They produce contraction of
the Feynman diagrams
whereby a fermion internal line drops and the two endpoints collapse to a single
one.  These are contact terms. They are not the
only ones. Other contact terms are produced by seagull vertices, i.e. vertices
with two fermion legs and two or more graviton legs,
by contracting the fermion legs with a propagator, thus forming a fermion loop.

\subsection{One-loop one-point function}
\label{ss:oneloop}

Representing by $\phi$ the matter fields in the model, the one-loop 1pt function
of $T^{\mu\nu}$ in the presence of a metric $g_{\mu\nu}= \eta_{\mu\nu}+
h_{\mu\nu}$ is
\be
\langle\!\langle T^{\mu\nu}(x) \rangle \! \rangle &=& \int {\cal D}\phi 
T^{\mu\nu}(x) \, e^{i S[\phi,h]}\label{1pt1loop}\\
&=& \int {\cal D}\phi  \left( T_{(0)}^{\mu\nu}(x)+  T_{(1)}^{\mu\nu}(x) +
T_{(2)}^{\mu\nu}(x) +\ldots\right) e^{i \left(S_0+S_1+ S_2+\ldots\right)}\0\\
&=& \int {\cal D}\phi \left[ \left(  T_{(0)}^{\mu\nu}(x)+  T_{(1)}^{\mu\nu}(x) +
T_{(2)}^{\mu\nu}(x) +\ldots\right) e^{i \left(S_1+ S_2+\ldots\right)} \right] 
e^{i S_0}\0
\ee
$ e^{i S_0}$ has been singled out as the free part of the integration measure.
The rest of $S$ (the interaction) is treated perturbatively.

Rearranging (\ref{1pt1loop}) order by order in $h$:
\be
&&\langle\!\langle T^{\mu\nu}(x) \rangle \! \rangle =
\int {\cal D}\phi\,  T_{(0)}^{\mu\nu}(x) \, e^{i S_0}\label{1pt1loop2}\\
&&+ \int {\cal D}\phi \left(i S_1\, T_{(0)}^{\mu\nu}(x)+
T_{(1)}^{\mu\nu}(x)\right)  \, e^{i S_0}\0\\
&&+ \int {\cal D}\phi \left((i S_2-\frac 12 S_1^2)\, T_{(0)}^{\mu\nu}(x) + i
S_1\, T_{(1)}^{\mu\nu}(x)+ T_{(2)}^{\mu\nu}(x) \right)\, e^{i S_0}\0\\
&&+ \int {\cal D}\phi \left((i S_3-S_1 S_2- \frac i{3!}S_1^3)\,
T_{(0)}^{\mu\nu}(x) 
+ (i S_2-\frac 12 S_1^2)\, T_{(1)}^{\mu\nu}(x)+ i S_1 T_{(2)}^{\mu\nu}(x) +
T_{(3)}^{\mu\nu}(x)\right)\, e^{i S_0}\0\\
&&+\ldots\0
\ee

Next we introduce auxiliary external currents and couple them to the free field
in $S_0$. For instance
if the free fields are $\psi, \bar \psi$, we introduce  $j, \bar j$ and add a
term
\be
\langle\!\langle T^{\mu\nu}(x) \rangle \! \rangle[j,\bar j] =
\int  { {\cal D}\bar\psi{\cal D} \psi}  \, \Big{(}\ldots\ldots\ldots\Big{)}  \exp[iS_0 +i \int (\bar j
\psi+\overline \psi j)]\0
\ee 
and set at the end $j=\bar j=0$. At this point in 
$\Big{(}\ldots\ldots\ldots\Big{)}$  one can replace 
$\psi $ by $\frac {\delta}{\delta \bar j}$ and $\overline\psi $ by $-\frac
{\delta}{\delta j}$, so that the
only remaining dependence on $\psi$ and $\overline \psi$ is in the factor
$\exp[iS_0 + \int (\bar j \psi+\overline \psi j)]$. Since the exponent is a
quadratic expression, one can formally integrate over 
 $\psi$ and $\overline \psi$ by completing the square. This leads to an
irrelevant infinite constant times
\be
\exp[-i\int \bar j \, P \,j]\label{jpj}
\ee
where $P$ is the inverse of the kinetic differential operator in $S_0$, i. e.
the propagator in configuration space.
Finally
\be
\langle\!\langle T^{\mu\nu}(x) \rangle \! \rangle &=& 
 \Big{[} T_{(0)}^{\mu\nu}(x) \, \exp[-i\int \bar j \, P \,j] \label{1pt1loop3}\\
&&+  \left(i S_1\, T_{(0)}^{\mu\nu}(x)+ T_{(1)}^{\mu\nu}(x)\right) 
\,\exp[-i\int \bar j \, P \,j] \0\\
&&+  \left((i S_2-\frac 12 S_1^2)\, T_{(0)}^{\mu\nu}(x) + i S_1\,
T_{(1)}^{\mu\nu}(x)+ T_{(2)}^{\mu\nu}(x) \right)\, \exp[-i\int \bar j \, P \,j]
\0\\
&&+ \left((i S_3-S_1 S_2- \frac i{3!}S_1^3)\, T_{(0)}^{\mu\nu}(x) 
+ (i S_2-\frac 12 S_1^2)\, T_{(1)}^{\mu\nu}(x)+ i S_1 T_{(2)}^{\mu\nu}(x) +
T_{(3)}^{\mu\nu}(x)\right)\,\0\\
&&\cdot \exp[-i\int \bar j \, P \,j]\Big{]}\Big{\vert}_{j=\bar j=0} +\ldots\0
\ee
where all the $\psi, \overline\psi$ fields in $T_{(n)}, S_n$ are understood to
be replaced by
 $\frac {\delta}{\delta \bar j}$ and $-\frac {\delta}{\delta j}$, respectively.
This is the
final expression of the 1pt one-loop correlator from which the Feynman rules are
extracted. 
Eq.(\ref{1pt1loop3}) is thus rewritten as
\be
&&\langle\!\langle T^{\mu\nu}(x) \rangle \! \rangle =
  \langle 0| T_{(0)}^{\mu\nu}(x)|0\rangle  \,  \label{1pt1loop4}\\
&&+ \langle 0| {\cal T} \left(i S_1\, T_{(0)}^{\mu\nu}(x)+
T_{(1)}^{\mu\nu}(x)\right)  |0\rangle  \0\\
&&+\langle 0|  {\cal T}   \left((i S_2-\frac 12 S_1^2)\, T_{(0)}^{\mu\nu}(x) + i
S_1\, T_{(1)}^{\mu\nu}(x)+ T_{(2)}^{\mu\nu}(x) \right)|0\rangle,  \0\\
&&+\langle 0| {\cal T}   \left((i S_3-S_1 S_2- \frac i{3!}S_1^3)\,
T_{(0)}^{\mu\nu}(x) 
+ (i S_2-\frac 12 S_1^2)\, T_{(1)}^{\mu\nu}(x)+ i S_1 T_{(2)}^{\mu\nu}(x) +
T_{(3)}^{\mu\nu}(x)\right)|0\rangle\,\0\\
&& +\ldots\0
\ee
and the time-orderd amplitudes are computed by means of Feynman diagrams.

\subsection{MAT background}
\label{ss:MATbackground}

In this subsection the reference is to the expanded action
(\ref{2ndorderaction}). We rewrite it as
\be
S&=& S_0+ \sum_{n+m\geq 1} 
 \frac 1{n!} \frac 1{m!} \int \prod_{i=0}^n \prod_{j=0}^m dx_i \,dy_j\,\frac
{\delta^{i+j} S}{\delta h_{\mu_1\nu_1}(x_1) \ldots \delta
h_{\mu_i\nu_i}(x_i)\delta k_{\lambda_1\rho_1}(_1)\ldots \delta
k_{\lambda_j\rho_j}(y_j)}\Bigg{\vert}_{h,k=0}\0\\
&&\quad\quad \times   h_{\mu_1\nu_1}(x_1) \ldots 
h_{\mu_j\nu_j}(x_j)k_{\lambda_1\rho_1}(_1)\ldots  k_{\lambda_j\rho_j}(y_j) \0\\
&=& S_0 + \int dx \,\frac {\delta S}{\delta h_{\mu\nu}(x)}\Bigg{\vert}_{h,k=0}  
h_{\mu\nu}(x)+ \int dy \,\frac {\delta S}{\delta
k_{\lambda\rho}(x)}\Bigg{\vert}_{h,k=0}   k_{\lambda\rho}(y)\label{Sserieshk}\\
&& + \frac 12 \int dx_1 dx_2\, 
\frac {\delta^2 S}{\delta h_{\mu_1\nu_1}(x_1)\delta h_{\mu_2\nu_2}(x_2)}
\Bigg{\vert}_{h,k=0}   h_{\mu_1\nu_1}(x_1)h_{\mu_2\nu_2}(x_2)+\ldots \0\\
&&\equiv\sum_{n,m=0}^\infty S_{n,m} \0
\ee
where $S_0\equiv S_{0,0}$. {As long as we differentiate $S$ from
the right it functionally depends
on the axial-complex variable $g+\gamma_5 f$. So the functional derivatives with
respect to $h_{\mu\nu}$ and $k_{\mu\nu}$ have to be understood as 
\be
\frac {\delta}{\delta h_{\mu\nu}(x)}= \int d^4x' \frac {\delta
G_{\lambda\rho}(x')} {\delta  h_{\mu\nu}(x)}\,
\frac {\stackrel {\rightarrow}{\delta}}{ \delta
G_{\lambda\rho}(x')}=\frac{\stackrel {\rightarrow}{\delta}}{\delta
G_{\mu\nu}(x)}
\label{deltahG}\\
\frac {\delta}{\delta k_{\mu\nu}(x)}= \int d^4x' \frac {\delta
G_{\lambda\rho}(x')} {\delta  k_{\mu\nu}(x)}
\,\frac {\stackrel {\rightarrow}{\delta}}{
G_{\lambda\rho}(x')}=\gamma_5\frac{\stackrel {\rightarrow}{\delta}}{\delta
G_{\mu\nu}(x)}
\label{deltakG}
\ee
Now, going back to the definitions of $T^{\mu\nu}$ and $T_5^{\mu\nu}$, 
\eqref{fullem}, \eqref{Tmunu} and \eqref{T5munu},
one can see that, in the case when $\sqrt{|G|}$ is absorbed in $\psi$ we can
write}
\be
T^{\mu\nu}(x) &=& 2 \Bigg{(}\frac {\delta S}{\delta
h_{\mu\nu}(x)}\Bigg{\vert}_{h,k=0} +  \int dx_2\, 
\frac {\delta^2 S}{\delta h_{\mu\nu}(x)\delta h_{\mu_2\nu_2}(x_2)}
\Bigg{\vert}_{h,k=0}h_{\mu_2\nu_2}(x_2)\0\\
&&+  \int dy\, 
\frac {\delta^2 S}{\delta h_{\mu\nu}(x)\delta k_{\lambda\rho}(y)}
\Bigg{\vert}_{h,k=0}k_{\lambda\rho}(y)+ \ldots\Bigg{)}\label{TmunuS}\\
&=& T_{(0,0)}^{\mu\nu}(x) + T_{(1,0)}^{\mu\nu}(x) + T_{(0,1)}^{\mu\nu}(x)
+\ldots\0
\ee
and
\be
T_5^{\mu\nu}(x) &=& 2\Bigg{(}\frac {\delta S}{\delta
k_{\mu\nu}(x)}\Bigg{\vert}_{h,k=0} +  \int dy_2\, 
\frac {\delta^2 S}{\delta k_{\mu\nu}(x)\delta k_{\mu_2\nu_2}(y_2)}
\Bigg{\vert}_{h,k=0}k_{\mu_2\nu_2}(y_2)\0\\
&&+  \int dy\, 
\frac {\delta^2 S}{\delta k_{\mu\nu}(x)\delta h_{\lambda\rho}(y)}
\Bigg{\vert}_{h,k=0}h_{\lambda\rho}(y)+ \ldots\Bigg{)}\label{T5munuS}\\
&=&  T_{5(0,0)}^{\mu\nu}(x) + T_{5(0,1)}^{\mu\nu}(x) + T_{5(1,0)}^{\mu\nu}(x)+
\ldots\0 
\ee
Therefore
\be
T_{(n,m)}^{\mu\nu}(x)&=& \sum_{i=0}^n \sum_{j=0}^m 
\frac 2{i!j!}\int \prod_{i=1}^n \prod_{j=1}^m dx_i \,dy_j\,\label{Tnm}\\
&&\times\, \frac {\delta^{i+j+1} S}{\delta h_{\mu\nu}(x) \delta
h_{\mu_1\nu_1}(x_1) \ldots \delta h_{\mu_i\nu_i}(x_i)\delta
k_{\lambda_1\rho_1}(_1)\ldots \delta
k_{\lambda_j\rho_j}(y_j)}\Bigg{\vert}_{h,k=0}\0\\
&&\times\, h_{\mu_1\nu_1}(x_1) \ldots 
h_{\mu_j\nu_j}(x_j)k_{\lambda_1\rho_1}(_1)\ldots  k_{\lambda_j\rho_j}(y_j)\0
\ee
and
\be
T_{5(n,m)}^{\lambda\rho}(x)&=& \sum_{i=0}^n \sum_{j=0}^m 
\frac 2{i!j!}\int \prod_{i=1}^n \prod_{j=1}^m dx_i \,dy_j\,\label{T5nm}\\
&&\times\, \frac {\delta^{i+j+1} S}{ \delta h_{\mu_1\nu_1}(x_1) \ldots \delta
h_{\mu_i\nu_i}(x_i)\delta k_{\lambda\rho}(x)\delta
k_{\lambda_1\rho_1}(_1)\ldots \delta
k_{\lambda_j\rho_j}(y_j)}\Bigg{\vert}_{h,k=0}\0\\
&&\times\, h_{\mu_1\nu_1}(x_1) \ldots 
h_{\mu_j\nu_j}(x_j)k_{\lambda_1\rho_1}(_1)\ldots  k_{\lambda_j\rho_j}(y_j)\0
\ee
So,
\be
S_{n,m}=c_{n,m}\!&&\!\Bigg{(} \frac 1{2n} \, \sum_{i=0}^n \sum_{j=0}^m\int dx\,
T_{(i-1,j)}^{\mu\nu}(x)h_{\mu\nu}(x)\label{STT5}\\ 
&&+ \frac 1{2m}\,  \sum_{i=0}^n 
\sum_{j=0}^m \int dx\,
T_{5(i,j-1)}^{\mu\nu}(x) k_{\mu\nu}(x)\Bigg{)}\0
\ee
where $c_{n,m}= 1$ for either $n=0$ or $m=0$, $c_{n,m}= \frac 12$ otherwise. 
For instance
\be
S_{1,0}&=& \frac 12 \int dx \,   T_{(0,0)}^{\mu\nu}(x) \,h_{\mu\nu}(x)= \int
dx\, \frac {\delta S}{\delta h_{\mu\nu}(x)}\Bigg{\vert}_{h,k=0}h_{\mu\nu}(x)
\,\label{S10}\\
S_{0,1}&=& \frac 12 \int dy \,   T_{5(0,0)}^{\mu\nu}(y) \,k_{\mu\nu}(y)= \int
dy\, \frac {\delta S}{\delta k_{\mu\nu}(y)}\Bigg{\vert}_{h,k=0} k_{\mu\nu}(y)
\,\label{S01}\\
S_{2,0}&=&   \frac 18 \int dx \, T_{(1,0)}^{\mu\nu}(x)
\,h_{\mu\nu}(x),\label{S20}\\
S_{0,2}&=&   \frac 18 \int dy \, T_{5(0,1)}^{\mu\nu}(y) 
\,k_{\mu\nu}(y),\label{S02}\\
S_{1,1}&=&   \frac 14\int dx \,T_{(0,1)}^{\mu\nu}(x)h_{\mu\nu}(x) + \frac 14\int
dy \, T_{5(1,0)}^{\mu\nu}(y)
\,k_{\mu\nu}(y),\label{S11}
\ee
and
\be
T_{(0,0)}^{\mu\nu}(x) &=&  2 \frac {\delta S}{\delta
h_{\mu\nu}(x)}\Bigg{\vert}_{h,k-=0}\label{T00}\\
T_{(1,0)}^{\mu\nu}(x) &=&  2
 \int dx_1\,\frac {\delta^{2} S}{\delta h_{\mu\nu}(x)\delta h_{\mu_1\nu_1}(x_1)}
\Bigg{\vert}_{h,k=0}   h_{\mu_1\nu_1}(x_1)  \label{T10}\\
T_{(0,1)}^{\mu\nu}(x) &=& 2 \int dy\,\frac {\delta^{2} S}{\delta
h_{\mu\nu}(x)\delta k_{\lambda\rho}(y) }
\Bigg{\vert}_{h,k=0}   k_{\lambda\rho}(y)  \label{T01}
\ee
\be
T_{(2,0)}^{\mu\nu}(x) &=&  \int dx_1 dx_2 \frac {\delta^{3} S}{\delta
h_{\mu\nu}(x)\delta
h_{\mu_1\nu_1}(x_1) \delta h_{\mu_2\nu_2}(x_2)}
\Bigg{\vert}_{h,k=0}   h_{\mu_1\nu_1}(x_1)h_{\mu_2\nu_2}(x_2)\label{T20}
\ee
\be
T_{(0,2)}^{\mu\nu}(x) &=& \int dy_1 dy_2 \frac {\delta^{3} S}{\delta
h_{\mu\nu}(x)\delta
k_{\lambda_1\rho_1}(_1) \delta k_{\lambda_2\rho_2}(y_2)}\Bigg{\vert}_{h,k=0}  
k_{\lambda_1\rho_1}(_1)k_{\lambda_2\rho_2}(y_2)\label{T02}\\
T_{(1,1)}^{\mu\nu}(x) &=&2 \int dx_1 dy \frac {\delta^{3} S}{\delta
h_{\mu\nu}(x)\delta
h_{\mu_1\nu_1}(x_1) \delta k_{\lambda\rho}(y)}\Bigg{\vert}_{h,k=0}  
k_{\mu_1\nu_1}(x_1)k_{\lambda\rho}(y)\\label{T11}
\ee
Similarly
\be
T_{5(0,0)}^{\mu\nu}(x) &=&  2 \frac {\delta S}{\delta
k_{\mu\nu}(x)}\Bigg{\vert}_{h,k-=0}\label{T500}\\
T_{5(1,0)}^{\mu\nu}(x) &=&  2
 \int dx_1\,\frac {\delta^{2} S}{\delta k_{\mu\nu}(x)\delta h_{\mu_1\nu_1}(x_1)
}
\Bigg{\vert}_{h,k=0}   h_{\mu_1\nu_1}(x_1)  \label{T510}\\
T_{5(0,1)}^{\mu\nu}(x) &=& 2 \int dy\,\frac {\delta^{2} S}{\delta
k_{\mu\nu}(x)\delta k_{\lambda\rho}(y) }
\Bigg{\vert}_{h,k=0}   k_{\lambda\rho}(y)  \label{T501}
\ee
\be
T_{5(2,0)}^{\mu\nu}(x) &=&  \int dx_1 dx_2 \frac {\delta^{3} S}{\delta
k_{\mu\nu}(x)\delta
h_{\mu_1\nu_1}(x_1) \delta h_{\mu_2\nu_2}(x_2)}
\Bigg{\vert}_{h,k=0}   h_{\mu_1\nu_1}(x_1)h_{\mu_2\nu_2}(x_2)\label{T520}
\ee
\be
T_{5(0,2)}^{\mu\nu}(x) &=& \int dy_1 dy_2 \frac {\delta^{3} S}{\delta
k_{\mu\nu}(x)\delta
k_{\lambda_1\rho_1}(_1) \delta k_{\lambda_2\rho_2}(y_2)}\Bigg{\vert}_{h,k=0}  
k_{\lambda_1\rho_1}(_1)k_{\lambda_2\rho_2}(y_2)\label{T502}\\
T_{5(1,1)}^{\mu\nu}(x) &=& 2 \int dx_1 dy \frac {\delta^{3} S}{\delta
k_{\mu\nu}(x)\delta
h_{\mu_1\nu_1}(x_1) \delta k_{\lambda\rho}(y)}\Bigg{\vert}_{h,k=0}  
h_{\mu_1\nu_1}(x_1)k_{\lambda\rho}(y)\label{T511}
\ee
The explicit expression of $T_{(0,0)}^{\mu\nu}(x)$ and $T_{5(0,0)}^{\mu\nu}(x)$
are given in eqs.(\ref{Tmunu0},\ref{T5munu0}).

\subsection{The one-loop one-point functions}
\label{ss:oneloopaxial}

The one-loop one-point functions of $T^{\mu\nu}$ and $T_5^{\mu\nu}$ are defined
in path integral terms as follows.
\be
\langle\!\langle {\rm T}^{\mu\nu}(x) \rangle \! \rangle &=& \int {\cal D}\phi 
{\rm T}^{\mu\nu}(x) \, e^{i S[\phi,h]}\label{1pt1loopT}\\
 &=& \int {\cal D}\phi \left[ \left(  {\rm T}_{(0,0)}^{\mu\nu}(x)+  {\rm
T}_{(1,0)}^{\mu\nu}(x) + {\rm T}_{(0,1)}^{\mu\nu}(x) +\ldots\right) e^{i
\left(S_{10}+ S_{01}+\ldots\right)} \right]  e^{i S_0}\0
\ee
where ${\rm T}$ can be either $T$ or $T_5$. Expanding the exponential:
\be
&&\langle\!\langle {\rm T}^{\mu\nu}(x) \rangle \! \rangle =
\int {\cal D}\phi\, {\rm  T}_{(0,0)}^{\mu\nu}(x) \, e^{i
S_0}\label{1pt1loopT2}\\
&&+ \int {\cal D}\phi \left(i S_{10}\, {\rm T}_{(0,0)}^{\mu\nu}(x)+ {\rm
T}_{(1,0)}^{\mu\nu}(x)\right)  \, e^{i S_0}\0\\
&&+ \int {\cal D}\phi \left(i S_{01}\, {\rm T}_{(0,0)}^{\mu\nu}(x)+ {\rm
T}_{(0,1)}^{\mu\nu}(x)\right)  \, e^{i S_0}\0\\
&&+ \int {\cal D}\phi \left((i S_{20}-\frac 12 S_{10}^2)\, {\rm
T}_{(0,0)}^{\mu\nu}(x) + i S_{10}\, {\rm T}_{(1,0)}^{\mu\nu}(x)+ {\rm
T}_{(2,0)}^{\mu\nu}(x) \right)\, e^{i S_0}\0\\
&&+ \int {\cal D}\phi \left((i S_{02}-\frac 12 S_{01}^2)\, {\rm
T}_{(0,0)}^{\mu\nu}(x) + i S_{01}\, {\rm T}_{(0,1)}^{\mu\nu}(x)+ {\rm
T}_{(0,2)}^{\mu\nu}(x) \right)\, e^{i S_0}\0\\
&&+ \int {\cal D}\phi \left((i S_{11}- S_{01} S_{01})\, {\rm
T}_{(0,0)}^{\mu\nu}(x) + i S_{01}\, {\rm T}_{(1,0)}^{\mu\nu}(x)+  i S_{10}\,
{\rm T}_{(0,1)}^{\mu\nu}(x)+{\rm T}_{(1,1)}^{\mu\nu}(x) \right)\, e^{i S_0}\0\\
&&+\ldots\0
\ee

Next we introduce auxiliary external currents $J$ and $\bar J$  and couple them
to the free field $\bar \Psi, \Psi$ in $S_0$. 
\be
\langle\!\langle T^{\mu\nu}(x) \rangle \! \rangle[J,\bar J] =
\int { {\cal D}\bar\Psi{\cal D} \Psi} \, \Big{(}\ldots\ldots\ldots\Big{)}  \exp[iS_0 +i \int (\bar J
\Psi+\overline \Psi J)]\0
\ee 
and set at the end $J=\bar J=0$. At this point in 
$\Big{(}\ldots\ldots\ldots\Big{)}$  one can replace 
$\Psi $ by $\frac {\delta}{\delta \bar J}$ and $\overline\Psi $ by $-\frac
{\delta}{\delta J}$, so that the
only remaining dependence on $\Psi$ and $\overline \Psi$ is in the factor
$\exp[iS_0 + \int (\bar J\Psi+\overline \Psi J)]$.  Formally integrating over 
$\Psi$ and $\overline \Psi$   leads to an irrelevant infinite constant times
\be
\exp[-i\int \bar J\, P \,J]\label{JPJ}
\ee
where $P$ is the inverse of the kinetic differential operator in $S_0$, i. e.
the propagator in configuration space. The final expression is the same as
(\ref{1pt1loopT2}) with $ e^{i S_0}$ replaced by $\exp[-i\int \bar j \, P \,j]$,
from which the Feynman rules can be extracted. This is interpreted as
\be
&&\langle\!\langle {\rm T}^{\mu\nu}(x) \rangle \! \rangle =
\langle 0|{\rm  T}_{(0,0)}^{\mu\nu}(x)|0 \rangle\, \label{1pt1loopVEV}\\
&&+ \langle 0|{\cal T}  \left(i S_{10}\, {\rm T}_{(0,0)}^{\mu\nu}(x)+ {\rm
T}_{(1,0)}^{\mu\nu}(x)\right) |0 \rangle\  \0\\
&&+ \langle 0|{\cal T}   \left(i S_{01}\, {\rm T}_{(0,0)}^{\mu\nu}(x)+ {\rm
T}_{(0,1)}^{\mu\nu}(x)\right)|0 \rangle\   \0\\
&&+\langle 0| {\cal T}    \left((i S_{20}-\frac 12 S_{10}^2)\, {\rm
T}_{(0,0)}^{\mu\nu}(x) + i S_{10}\, {\rm T}_{(1,0)}^{\mu\nu}(x)+ {\rm
T}_{(2,0)}^{\mu\nu}(x) \right)|0 \rangle\ \0\\
&&+ \langle 0|{\cal T}   \left((i S_{02}-\frac 12 S_{01}^2)\, {\rm
T}_{(0,0)}^{\mu\nu}(x) + i S_{01}\, {\rm T}_{(0,1)}^{\mu\nu}(x)+ {\rm
T}_{(0,2)}^{\mu\nu}(x) \right) |0 \rangle\ \0\\
&&+ \langle 0|{\cal T}  m \left((i S_{11}- S_{01} S_{01})\, {\rm
T}_{(0,0)}^{\mu\nu}(x) + i S_{01}\, {\rm T}_{(1,0)}^{\mu\nu}(x)+  i S_{10}\,
{\rm T}_{(0,1)}^{\mu\nu}(x)+{\rm T}_{(1,1)}^{\mu\nu}(x) \right)|0 \rangle \0\\
&&+\ldots\0\\
&&\equiv   \langle 0|{\rm T}_{(0,0)}^{\mu\nu}(x)|0\rangle\label{Tseriesrm}\\
&&+\sum_{n+m\geq 1}^\infty \frac 1{2^{n+m} n!m!}  
\int \prod_{i,j,i+j\geq 1} dx_i\,dy_j h_{\mu_1\nu_1}(x_1)\ldots
h_{\mu_i\nu_i}(x_i) k_{\lambda_1\rho_1}(_1)\ldots k_{\lambda_j\rho_j}(y_j)\0\\
&&\times \, {\bf T}^{\mu\nu\mu_1\nu_1\ldots
\mu_n\nu_n,\lambda_1\rho_1\ldots\lambda_m\rho_m}(x,x_1,\ldots,x_n,_1,\ldots,
y_m),\0
\ee
The expansion coefficients ${\bf T}^{\mu\nu\mu_1\nu_1\ldots
\mu_n\nu_n}(x,x_1,\ldots,x_n)$, where
${\bf T}$ stands both for ${\cal  T}$ and ${\cal  T}_5$, are introduced for
convenience.
 
\subsection{Trace Ward indentities}
\label{ss:WIaxial}

The quantum Ward identities for the Weyl and axial Weyl symmetry are given by
(\ref{quantumWardWeyl1})  
and (\ref{quantumWardWeyl2}).
We need to expand them in series of $h$ and $k$. With reference to
(\ref{Tseriesrm}) we get
\be
\ET_{(0,0)} (x) &\equiv& \langle 0| T_{(0,0)}{}_\mu^\mu
(x)|0\rangle=0\label{ET00}\\
\ET_{(1,0)}(x,x_1) &\equiv& {\cal T}_{(1,0)}{}_{\mu}^{\mu\mu_1\nu_1}(x,x_1)
+2\delta(x-x_1) \langle 0|T_{(0,0)}^{\mu_1\nu_1}(x_1)|0\rangle=0\label{ET10}\\
 \ET_{(0,1)}(x,{y}_1) &\equiv& {\cal T}_{(0,1)}{}_{\mu}^{\mu\mu_1\nu_1}(x,{y}_1)
+2\delta(x-{y}_1) \langle
0|{T_{5(0,0)}^{\mu_1\nu_1}}({y}_1)|0\rangle=0\label{ET01}\\
 \ET_{(1,1)}(x,x_1,{y}_1) &\equiv&
{\cal T}_{(1,1)}{}_{\mu}^{\mu\mu_1\nu_1\lambda_1\rho_1}(x,x_1,{y}_1)
+2\delta(x-x_1)
{\cal T}_{(0,1)}^{\mu_1\nu_1\lambda_1\rho_1}(x_1,{y}_1)\0\\
&&+ 2\delta(x-{y}_1){\cal T}_{5(1,0)}^{\mu_1\nu_1\lambda_1\rho_1}(x_1,{y}_1) =0
\label{ET11}\\
{\ET}_{(2,0)}(x,x_1,x_2) &\equiv&{\cal T}_{(2,0)}{}_\mu^{\mu
\mu_1\nu_1\mu_2\nu_2}(x,x_1,x_2)\0\\
&& +2(\delta(x-x_1) +\delta(x-x_2)) {\cal
T}_{(1,0)}^{\mu_1\nu_1\mu_2\nu_2}(x_1,x_2)=0
\label{ET20}\\
\ET_{(0,2)}(x,{y}_1,y_2) &\equiv& {\cal T}_{(0,2)}{}_\mu^{\mu
\lambda_1\rho_1\lambda_2\rho_2}(x,{y}_1,y_2) \0\\
&&+2(\delta(x-{y}_1) +\delta(x-y_2))
{\cal T}_{5(0,1)}^{\lambda_1\rho_1\lambda_2\rho_2}({y}_1,y_2)=0 \label{ET02}\\
&&\ldots\0
\ee
 and
\be
\ET_{5(0,0)} (x) &\equiv& \langle 0| T_{5(0,0)}{}_\lambda^\lambda
(x)|0\rangle=0\label{ET500}\\
\ET_{5(1,0)}(x,x_1) &\equiv& {\cal
T}_{5(1,0)}{}_\lambda^{\lambda\mu_1\nu_1}(x,x_1)
+2\delta(x-x_1) \langle 0|T_{5(0,0)}^{\mu_1\nu_1}(x_1)|0\rangle=0\label{ET510}\\
 \ET_{5(0,1)}(x,{y}_1) &\equiv& {\cal
T}_{5(0,1)}{}_\lambda^{\lambda\mu_1\nu_1}(x,{y}_1)
+2\delta(x-{y}_1) \langle
0|T_{(0,0)}^{\lambda_1\rho_1}({y}_1)|0\rangle=0\label{ET501}\\
 \ET_{5(1,1)}(x,x_1,{y}_1) &\equiv&
{\cal T}_{5(1,1)}{}_\lambda^{\lambda\mu_1\nu_1\lambda_1\rho_1}(x,x_1,{y}_1)
+2\delta(x-x_1){\cal T}_{5(1,0)}^{\mu_1\nu_1\lambda_1\rho_1}(x_1,{y}_1)\0\\
&&+ 2\delta(x-{y}_1)  {\cal T}_{(0,1)}^{\mu_1\nu_1\lambda_1\rho_1}(x_1,{y}_1)
=0\label{ET511}\\
 \ET_{5(2,0)}(x,x_1,x_2) &\equiv& {\cal T}_{5(2,0)}{}_\lambda^{\lambda
\mu_1\nu_1\mu_2\nu_2}(x,x_1,x_2)\0\\
&& {+2(\delta(x-x_1) +\delta(x-x_2)) {\cal
T}_{(1,0)}^{\mu_1\nu_1\mu_2\nu_2}(x_1,x_2)=0}\label{ET520}\\
\ET_{5(0,2)}(x,{y}_1,y_2) &\equiv& {\cal T}_{5(0,2)\lambda}{}^{\lambda
\lambda_1\rho_1\lambda_2\rho_2}(x,{y}_1,y_2)\0\\
&&{+2(\delta(x-y_1) +\delta(x-y_2))
{\cal T}_{(0,1)}^{\lambda_1\rho_1\lambda_2\rho_2}(y_1,y_2)=0}  \label{ET502}\\
&&\ldots\0
\ee
where
\be
 {\cal T}_{(1,0)}^{\mu\nu\mu_1\nu_1}(x,x_1) &=& 
i \langle 0|{\cal T}T_{(0,0)}^{\mu\nu}(x)T_{(0,0)}^{\mu_1\nu_1}(x_1)|0 \rangle+4
\langle 0| \frac {\delta^{2} S}{\delta h_{\mu\nu}(x)\delta h_{\mu_1\nu_1}(x_1)
}|0 \rangle\label{TTmunu10}\\
 {\cal T}_{(0,1)}^{\mu\nu\lambda_1\rho_1}(x,{y}_1) &=& 
i \langle 0|{\cal T}T_{{5}(0,0)}^{\mu\nu}(x)T_{(0,0)}^{\lambda_1\rho_1}({y}_1)|0
\rangle +4
\langle 0| \frac {\delta^{2} S}{\delta h_{\mu\nu}(x)\delta
{k}_{\lambda_1\rho_1}({y}_1) }|0 \rangle\label{TTmunu01} 
\ee
and
\be
&&{\cal T}_{(2,0)}^{\mu\nu\mu_1\nu_1\mu_2\nu_2}(x,x_1,x_2) = 
- \langle 0|{\cal
T}T_{(0,0)}^{\mu\nu}(x)T_{(0,0)}^{\mu_1\nu_1}(x_1)T_{(0,0)}^{\mu_2\nu_2}(x_2)|0
\rangle\0\\
&&+4
 i  \langle 0|{\cal T}T_{(0,0)}^{\mu_1\nu_1}(x_1)\frac {\delta^{2} S}{\delta
h_{\mu\nu}(x)\delta h_{\mu_2\nu_2}(x_2) }|0 \rangle+4i  \langle 0|{\cal
T}T_{(0,0)}^{\mu_2\nu_2}(x_2)\frac {\delta^{2} S}{\delta
h_{\mu_1\nu_1}(x_1)\delta h_{\mu\nu}(x) }|0 \rangle\0\\
&&+4
 i  \langle 0|{\cal T}T_{(0,0)}^{\mu\nu}(x)\frac {\delta^{2} S}{\delta
h_{\mu_1\nu_1}(x_1)\delta h_{\mu_2\nu_2}(x_2) }|0 \rangle + 8 \langle 0| \frac
{\delta^{3} S}{\delta h_{\mu\nu}(x)\delta h_{\mu_1\nu_1}(x_1)
h_{\mu_2\nu_2}(x_2) }|0 \rangle\label{TTTmunu20}
\ee
\be
&&{\cal T}_{(0,2)}^{\mu\nu\lambda_1\rho_1\lambda_2\rho_2}(x,{y}_1,y_2) = 
- \langle 0|{\cal
T}T_{(0,0)}^{\mu\nu}(x)T_{5(0,0)}^{\lambda_1\rho_1}(_1)T_{5(0,0)}^{
\lambda_2\rho_2}(y_2)|0 \rangle\0\\
&&+4
 i  \langle 0|{\cal T}T_{5(0,0)}^{\lambda_1\rho_1}({y}_1)\frac {\delta^{2}
S}{\delta h_{\mu\nu}(x)\delta k_{\lambda_2\rho_2}(y_2) }|0 \rangle+4i  \langle
0|{\cal T}T_{5(0,0)}^{\lambda_2\rho_2}(y_2)\frac {\delta^{2} S}{\delta
k_{\lambda_1\rho_1}({y}_1)\delta h_{\mu\nu}(x) }|0 \rangle\0\\
&&+4
 i  \langle 0|{\cal T}T_{(0,0)}^{\mu\nu}(x)\frac {\delta^{2} S}{\delta
k_{\lambda_1\rho_1}({y}_1)\delta k_{\lambda_2\rho_2}(y_2) }|0 \rangle+ 8 \langle
0| \frac {\delta^{3} S}{\delta h_{\mu\nu}(x)\delta k_{\lambda_1\rho_1}({y}_1)
h_{\lambda_2\rho_2}(y_2) }|0 \rangle \label{TTTmunu02}
\ee
and
\be
&&{\cal T}_{(1,1)}^{\mu\nu\mu_1\nu_1\lambda_1\rho_1}(x,x_1,{y}_1) = 
- \langle 0|{\cal
T}T_{(0,0)}^{\mu\nu}(x)T_{(0,0)}^{\mu_1\nu_1}(x_1)T_{5(0,0)}^{\lambda_1\rho_1}
({y}_1)|0 \rangle\0\\
&&+4
 i  \langle 0|{\cal T}T_{5(0,0)}^{\lambda\rho_1}({y}_1)\frac {\delta^{2}
S}{\delta
h_{\mu\nu}(x)\delta h_{\mu_1\nu_1}(x_1) }|0 \rangle+4i  \langle 0|{\cal
T}T_{(0,0)}^{\mu_1\nu_1}(x_1)\frac {\delta^{2} S}{\delta
k_{\lambda_1\rho_1}({y}_1)\delta h_{\mu\nu}(x) }|0 \rangle\0\\
&&+4
 i  \langle 0|{\cal T}T_{(0,0)}^{\mu\nu}(x)\frac {\delta^{2} S}{\delta
k_{\lambda_1\rho_1}({y}_1)\delta h_{\mu_1\nu_1}(x_1) }|0 \rangle+ 8 \langle 0|
\frac {\delta^{3} S}{\delta h_{\mu\nu}(x)\delta h_{\mu_1\nu_1}(x_1)
{k}_{\lambda_1\rho_1}({y}_1) }|0 \rangle \label{TTTmunu11}
\ee
and for the axial tensors
\be
{ {\cal T}_{5(1,0)}^{\lambda\rho\mu_1\nu_1}(x,x_1)} &=& 
i \langle 0|{\cal T}T_{5(0,0)}^{\lambda\rho}(x)T_{(0,0)}^{\mu_1\nu_1}(x_1)|0
\rangle+4
\langle 0| \frac {\delta^{2} S}{\delta k_{\lambda\rho}(x)\delta
h_{\mu_1\nu_1}(x_1) }|0 \rangle\label{TT5munu10}\\
 {\cal T}_{5(0,1)}^{\lambda\rho\lambda_1\rho_1}(x,{y}_1) &=& 
i \langle 0|{\cal
T}T_{5(0,0)}^{\lambda\rho}(x)T_{(0,0)}^{\lambda_1\rho_1}({y}_1)|0 \rangle+4
\langle 0| \frac {\delta^{2} S}{\delta k_{\lambda\rho}(x)\delta
k_{\lambda_1\rho_1}({y}_1) }|0 \rangle\label{TT5munu01} 
\ee

\be
&&{\cal T}_{5(1,1)}^{\lambda\rho\mu_1\nu_1\lambda_1\rho_1}(x,x_1,{y}_1) = 
- \langle 0|{\cal
T}T_{5(0,0)}^{\lambda\rho}(x)T_{(0,0)}^{\mu_1\nu_1}(x_1)T_{5(0,0)}^{
\lambda_1\rho_1}({y}_1)|0 \rangle\0\\
&&+4
 i  \langle 0|{\cal T}T_{5(0,0)}^{\lambda\rho}(x)\frac {\delta^{2} S}{\delta
h_{\mu_1\nu_1}(x_1)\delta k_{\lambda_1\rho_1}({y}_1) }|0 \rangle+4i  \langle
0|{\cal T}T_{(0,0)}^{\mu_1\nu_1}(x_1)\frac {\delta^{2} S}{\delta
k_{\lambda_1\rho_1}({y}_1)\delta {k}_{\lambda\rho}(x) }|0 \rangle\0\\
&&+4
 i  \langle 0|{\cal T}T_{5(0,0)}^{\lambda\rho}(x)\frac {\delta^{2} S}{\delta
k_{\lambda_1\rho_1}({y}_1)\delta h_{\mu_1\nu_1}(x_1) }|0 \rangle+ 8 \langle 0|
\frac {\delta^{3} S}{\delta h_{\lambda\rho}(x)\delta h_{\mu_1\nu_1}(x_1)
h_{\lambda_1\rho_1}({y}_1) }|0 \rangle \label{TTT5munu11}
\ee
and
\be
&&{\cal T}_{5(2,0)}^{\lambda\rho\mu_1\nu_1\mu_2\nu_2}(x,x_1,x_2) = 
- \langle 0|{\cal
T}T_{5(0,0)}^{\lambda\rho}(x)T_{(0,0)}^{\mu_1\nu_1}(x_1)T_{(0,0)}^{\mu_2\nu_2}
(x_2)|0 \rangle\0\\
&&+{4}
 i  \langle 0|{\cal T}T_{(0,0)}^{\mu_1\nu_1}(x_1)\frac {\delta^{2} S}{\delta
k_{\lambda\rho}(x)\delta h_{\mu_2\nu_2}(x_2) }|0 \rangle+{4}i  \langle 0|{\cal
T}T_{(0,0)}^{\mu_2\nu_2}(x_2)\frac {\delta^{2} S}{\delta
h_{\mu_1\nu_1}(x_1)\delta k_{\lambda\rho}(x) }|0 \rangle\0\\
&&+{4}
 i  \langle 0|{\cal T}T_{{5}(0,0)}^{\lambda\rho}(x)\frac {\delta^{2} S}{\delta
h_{\mu_1\nu_1}(x_1)\delta h_{\mu_2\nu_2}(x_2) }|0 \rangle + 8 \langle 0| \frac
{\delta^{3} S}{\delta k_{\lambda\rho}(x)\delta h_{\mu_1\nu_1}(x_1)
h_{\mu_2\nu_2}(x_2) }|0 \rangle\0\label{TTT5munu20}
\0
\ee
and
\be
&&{\cal T}_{5(0,2)}^{\lambda\rho\lambda_1\rho_1\lambda_2\rho_2}(x,{y}_1,y_2) = 
- \langle 0|{\cal
T}T_{5(0,0)}^{\lambda\rho}(x)T_{5(0,0)}^{\lambda_1\rho_1}({y}_1)T_{5(0,0)}^{
\lambda_2\rho_2}(y_2)|0 \rangle\0\\
&&+4
 i  \langle 0|{\cal T}T_{5(0,0)}^{\lambda_1\rho_1}({y}_1)\frac {\delta^{2}
S}{\delta k_{\lambda\rho}(x)\delta k_{\lambda_2\rho_2}(y_2) }|0 \rangle+4i 
\langle 0|{\cal T}T_{5(0,0)}^{\lambda_2\rho_2}(y_2)\frac {\delta^{2} S}{\delta
k_{\lambda_1\rho_1}({y}_1)\delta k_{\lambda\rho}(x) }|0 \rangle\0\\
&&+{4}
 i  \langle 0|{\cal T}T_{5(0,0)}^{\lambda\rho}(x)\frac {\delta^{2} S}{\delta
k_{\lambda_1\rho_1}({y}_1)\delta k_{\lambda_2\rho_2}(y_2) }|0 \rangle+ 4 \langle
0| \frac {\delta^{3} S}{\delta k_{\lambda\rho}(x)\delta
k_{\lambda_1\rho_1}({y}_1)
{k}_{\lambda_2\rho_2}(y_2) }|0 \rangle\label{TTT5munu02}
\ee

\section{Samples of Feynman diagram calculations}
\label{s:samples}

In this Appendix we give more details on some of the Feynman diagram computed in
section \ref{s:2}. 

\subsection{$T_{(0)}$ two-point function}

Let us start
from a very simple one, the calculation of $ \langle 0|{\cal
T}T_{(0)}^{\mu\nu}(x)T_{(0)}^{\lambda\rho}(y)|0 \rangle$. In momentum space
there corresponds:
\be
-\frac 1{64} \int \frac {d^4p}{(2\pi)^4} \tr \left(\frac 1{\slashed {p}}
(2p+k)^\mu \gamma^\nu \frac 1{\slashed{p}+\slashed{k}} (2p+k)^\lambda
\gamma^\rho \frac {1+\gamma_5}2+ \left(\begin{matrix} \mu\leftrightarrow \nu\\
 \lambda \leftrightarrow \rho \end{matrix} \right)\right)\label{T0T01}
\ee
whose odd parity part is
\be
-\frac i{36} \int \frac {d^4p}{(2\pi)^4}\left(\epsilon^{\sigma \nu\tau\rho}\frac
{p_\sigma k_\tau (2p+k)^\mu (2p+k)^\lambda}{p^2(p+k)^2}+ \left(\begin{matrix}
\mu\leftrightarrow \nu\\
 \lambda \leftrightarrow \rho \end{matrix} \right)\right) \label{T0T02}
\ee
The corresponding regulated expression is
\be
-\frac i{36} \int \frac {d^4p}{(2\pi)^4}\int\frac {d^\delta
\ell}{(2\pi)^\delta}\left(\epsilon^{\sigma \nu\tau\rho}\frac {p_\sigma k_\tau
(2p+k)^\mu (2p+k)^\lambda}{(p^2-\ell^2)((p+k)^2-\ell^2)}+ \left(\begin{matrix}
\mu\leftrightarrow \nu\\
 \lambda \leftrightarrow \rho \end{matrix} \right)\right) \label{T0T03}
\ee
Only the terms quadratic in $p$ in the numerator may survive for symmetry
reasons, but for the same reason they give rise to $\delta_\sigma^\mu$ and
$\delta_\sigma^\lambda$, which leads to the vanishing of (\ref{T0T03}). If we
contract (\ref{T0T03}) with $\eta_{\mu\nu}$ its vanishing is even more evident.

\subsubsection{Terms $P-V_{ffh}-P-V'_{ffhh}$ and similar}
\label{ss:PVPV'}

We wish to evaluate the terms contained in $\langle 0|{\cal
T}T_{(0)}^{\mu\nu}(x)\frac {\delta^{2} S}{\delta
h_{\mu_1\nu_1}(x_1)\delta h_{\mu_2\nu_2}(x_2) }|0 \rangle$. They are diagram
with an incoming graviton line of momentum $q$ and two outgoing ones of momentum
$k_1, k_2$. The first is the diagram $P-V_{ffh}-P-V'_{ffhh}$, whose odd part is
\be
\frac 3{ 512} \int \frac {d^4p}{(2\pi)^4}\!&&\!\tr\left[\left( \left(\frac
1{\slashed {p}} (2p-q)^\mu \gamma^\nu \frac 1{\slashed{p}-\slashed{q}}
(2p-q)^{\mu_1} \gamma^{\nu_1}\eta^{\nu_1\nu_2} + \left(\begin{matrix} \mu
\leftrightarrow \nu\\ \mu_1\leftrightarrow \nu_1\\  \mu_2 \leftrightarrow \nu_2
\end{matrix} \right)\right)\right.\right.\0\\
&&+\left.\left. \begin{matrix}{}\\{}\\{}\end{matrix}(\mu_1,\nu_1)\leftrightarrow
(\mu_2,\nu_2)\right) \frac {\gamma_5}2\right]\label{PVffhPV'ffhh}
\ee
Saturating it with $\eta_{\mu\nu}$ one gets
\be
\frac 3{ 512} \int \frac {d^4p}{(2\pi)^4}\!&&\!\tr\left[\left( \left(\frac
1{\slashed {p}} (2\slashed{p}-\slashed {q})\frac 1{\slashed{p}-\slashed{q}}
(2p-q)^{\mu_1} \gamma^{\nu_1}\eta^{\nu_1\nu_2} + \left(\begin{matrix}
\mu_1\leftrightarrow \nu_1\\  \mu_2 \leftrightarrow \nu_2 \end{matrix}
\right)\right)\right.\right.\0\\
&&+\left.\left. \begin{matrix}{}\\{}\\{}\end{matrix} 
(\mu_1,\nu_1)\leftrightarrow (\mu_2,\nu_2)\right) \frac
{\gamma_5}2\right]\label{PVffhPV'ffhh2}
\ee
which clearly vanishes because of the $\gamma$ trace. It follows that also the
odd part of the diagram
$P-V'_{ffh}-P-V'_{ffhh}$ vanishes.

The same conclusion holds if in these previous diagrams we replace  $V'_{ffhh}$
with $V''_{ffhh}$ and $V'''_{ffhh}$.

Proceeding in the same way we can prove that also the odd part of 
\be
\eta_{\mu\nu} \langle 0|{\cal T}T_{(0)}^{\mu_1\nu_1}(x_1)\frac {\delta^{2}
S}{\delta h_{\mu\nu}(x)\delta h_{\mu_2\nu_2}(x_2) }|0 \rangle
\ee
vanishes. But there is a simpler way to get rid of the terms containing one
$T_{(0)}$ factor and one second derivative of $S$ and it is to prove that their
odd parity part vanishes before taking the trace. 

Let us consider again (\ref{PVffhPV'ffhh}), that is the untraced 
$P-V_{ffh}-P-V'_{ffhh}$. Introducing a dimensional regulator $\delta$ we can
rewrite it as
\be
\frac 3{1024} \int \frac {d^4p}{(2\pi)^4}\int\frac {d^\delta
\ell}{(2\pi)^\delta}\,\left[ \tr\left(\frac {\slashed{p}}{p^2-\ell^2}
\gamma_\nu\frac {\slashed {p}-\slashed {q}}{(p-q)^2-\ell^2}\gamma_{\mu_2}
\gamma_5 \right) (2p-q)^\mu (2p-q)^{\mu_1}  \eta^{\nu_1\nu_2} +\ldots\right]\0\\
\label{PVffhPV'ffhh3}
\ee
where the dots denote the symmetrizations indicated in (\ref{PVffhPV'ffhh}).
Taking the $\gamma$ trace:
\be
 \frac {3i}{256} \int \frac {d^4p}{(2\pi)^4}\int\frac {d^\delta
\ell}{(2\pi)^\delta}\, \epsilon^{\sigma\nu\tau\mu_2} p_\sigma q_\tau 
\frac{ (2p-q)^\mu (2p-q)^{\mu_1} 
\eta^{\nu_1\nu_2}}{(p^2-\ell^2)((p-q)^2-\ell^2)}+\ldots 
\label{PVffhPV'ffhh4}
\ee
The integrand has two $p^2$ terms in the numerator. They are proportional
respectively to $\epsilon^{\mu\nu\tau\mu_2} q_\tau q^{\mu_1}  
\eta^{\nu_1\nu_2}$ and $\epsilon^{\mu_1\nu\tau\mu_2}  q_\tau q^{\mu}  
\eta^{\nu_1\nu_2}$. The first vanishes under the $\mu\leftrightarrow \nu$
symmetrization, the other under  the symmetrization
$(\mu_1,\nu_1)\leftrightarrow (\mu_2,\nu_2)$.

Next we do the same for the untraced $P-V_{ffh}-P-V'''_{ffhh}$. The relevant
integral is
\be
\frac 1{128}\int \frac {d^4p}{(2\pi)^4}\tr\left[  \frac 1{\slashed {p}}
(2p-q)^\mu \gamma^\nu \frac 1{\slashed{p}-\slashed{q}} (2\slashed{p}-\slashed{q}
)\frac {\gamma_5}2 (\eta^{\mu_1\nu_1}\eta^{\mu_2\nu_2}-
\eta^{\mu_1\nu_2}\eta^{\mu_2\nu_1}-\eta^{\mu_1\mu_2}\eta^{\nu_1\nu_2})\right]
\label{PVffhPVffhh'''}
\ee
symmetrized in $\mu\leftrightarrow \nu$. Writing $ 2\slashed{p}-\slashed{q}
=\slashed{p}+ \slashed{p}-\slashed{q}$ and simplifying with the denominators, we
get two terms each with a  trace of two $\gamma$'s with $\gamma_5$, which
vanishes.

\subsubsection{The term $P-V_{ffh}-P-V^\epsilon_{ffhh}$}
\label{ss:PVPVe}

This term requires a bit more elaboration. The starting point is  the integral
\be
\frac i{512}\int \frac {d^4p}{(2\pi)^4}\tr\left[  \frac 1{\slashed {p}}
(2p-q)^\mu \gamma^\nu \frac 1{\slashed{p}-\slashed{q}} t^{\mu_1\nu_1\mu_2\nu_2
\kappa\lambda} (k_1-k_2)_\lambda \gamma_\kappa \frac
{1+\gamma_5}2\right]\label{PVPVepsilon1}
\ee
which has to be symmetrized in $\mu\leftrightarrow \nu$. The odd part is
\be
\frac i{1024} \int \frac {d^4p}{(2\pi)^4}\left[\tr\left( \frac 1{\slashed
{p}}\gamma^\nu \frac 1{\slashed{p}-\slashed{q}} \gamma_\kappa\right) (2p-q)^\mu
t^{\mu_1\nu_1\mu_2\nu_2 \kappa\lambda} (k_1-k_2)_\lambda + (\mu\rightarrow
\nu)\right]\label{PVPVepsilon2}
\ee
Next we introduce the dimensional regulator and use Lorentz covariance to obtain
\be
&&\frac i{256} \int \frac {d^4p}{(2\pi)^4}\int\frac {d^\delta
\ell}{(2\pi)^\delta}\,\frac
1{(p^2-\ell^2)((p-q)^2-\ell^2)}\times\label{PVPVepsilon3}\\
&&\left[ \left(p^\nu(p-q)_\kappa- (p\!\cdot\!(p-q)+\ell^2) \delta^\nu_\kappa
+p_\kappa(p-q)^\nu \right)(2p-q)^\mu t^{\mu_1\nu_1\mu_2\nu_2 \kappa\lambda}
(k_1-k_2)_\lambda + (\mu\rightarrow \nu)\right]\0
\ee
Next we introduce a Feynman parameter $x$, $0\leq x\leq 1$ and represent
\be
\frac 1{(p^2-\ell^2)((p-q)^2-\ell^2)}= \int_0^1 dx\, \frac 1{((p-xq)^2-\ell^2
+x(1-x)q^2)^2},\0
\ee
then we change variable $p\to p'= p-xq$. The result is
\be 
&&\frac i{256} \int \frac {d^4p}{(2\pi)^4}\int\frac {d^\delta
\ell}{(2\pi)^\delta}\int_0^1 dx\, \frac{t^{\mu_1\nu_1\mu_2\nu_2 \kappa\lambda}
(k_1-k_2)_\lambda}{(p^2-\ell^2 +x(1-x)q^2)^2}\left[ \frac 12 \left(
\eta^{\mu\nu} q_\kappa + \delta^\mu_\kappa q^\nu+  \delta^\nu_\kappa
q^\mu\right) p^2 (2x-1) \right.\0\\
&& +(2 q^\mu q^\nu q_\kappa - q^\mu q^2 \delta_\kappa^\nu ) x(1-x)(1-2x) \left.
-\frac 32 q^\mu p^2\delta^\nu_\kappa (2x-1) -\ell^2 q^\mu\delta^\nu_\kappa
(2x-1) \right]\label{PVPVepsilon4}
\ee
which must be symmetrized under $\mu \leftrightarrow \nu$. All the terms vanish
because of the $x$ integration.


\end{document}